\documentclass{jpp}
\usepackage{graphicx}
\usepackage{tensor}
\usepackage{bm}

\usepackage[utf8]{inputenc}
\usepackage[T1]{fontenc}
\usepackage{amsmath}

\newcommand{\diffIL}[2]{\mathrm{d}#1/\mathrm{d}#2}	
\newcommand{\diff}[2]{\frac{\mathrm{d}#1}{\mathrm{d}#2}}	
\renewcommand{\tt}{\tilde{\theta}}							
\renewcommand{\d}{\textrm{d}}								

\shorttitle{Reflection-Driven MHD Turbulence}

\title{Reflection-driven magnetohydrodynamic turbulence in the solar
  atmosphere and solar wind}
 
\author{Benjamin D. G. Chandran\aff{1}\corresp{\email{benjamin.chandran@unh.edu}} \and Jean
C. Perez\aff{2}}

\affiliation{
\aff{1} Department of Physics and Astronomy, University of New Hampshire, Durham,
New Hampshire 03824,  USA 
\aff{2} Department of Aerospace, Physics and Space Sciences, Florida Institute
of Technology, Melbourne, Florida, 32901, USA
}

\begin{document}

\maketitle

\begin{abstract}
  We present three-dimensional direct numerical simulations and an
  analytic model of reflection-driven magnetohydrodynamic (MHD)
  turbulence in the solar wind. Our simulations describe transverse,
  non-compressive MHD fluctuations within a narrow magnetic flux tube
  that extends from the photosphere, through the chromosphere and
  corona, and out to a heliocentric distance~$r$ of 21 solar
  radii~$(R_{\odot})$. We launch outward-propagating ``$\bm{z}^+$
  fluctuations'' into the simulation domain by imposing a randomly
  evolving photospheric velocity field. As these fluctuations
  propagate away from the Sun, they undergo partial reflection,
  producing inward-propagating ``$\bm{z}^-$ fluctuations.''
  Counter-propagating fluctuations subsequently interact, causing
  fluctuation energy to cascade to small scales and dissipate. Our
  analytic model incorporates dynamic alignment, allows for strongly
  or weakly turbulent nonlinear interactions, and divides the
  $\bm{z}^+$ fluctuations into two populations with different
  characteristic radial correlation lengths. The inertial-range power
  spectra of $\bm{z}^+$ and $\bm{z}^-$ fluctuations in our simulations
  evolve toward a $k_\perp^{-3/2}$ scaling at $r>10 R_{\odot}$, where
  $k_\perp$ is the wave-vector component perpendicular to the
  background magnetic field. In two of our simulations, the $\bm{z}^+$ power
  spectra are much flatter between the coronal base and
  $r \simeq 4 R_{\odot}$. We argue that these spectral scalings are
  caused by: (1) high-pass filtering in the upper chromosphere; (2) the
  anomalous coherence of
  inertial-range $\bm{z}^-$ fluctuations in a reference frame propagating outwards with the $\bm{z}^+$
  fluctuations; and (3) the change in the sign of the radial derivative of
  the Alfv\'en speed at $r=r_{\rm m} \simeq 1.7 R_{\odot}$, which
  disrupts this anomalous coherence between $r=r_{\rm m}$ and
  $r\simeq 2r_{\rm m}$.  At $r>1.3 R_{\odot}$, the turbulent-heating
  rate in our simulations is comparable to the turbulent-heating rate
  in a previously developed solar-wind model that agreed with a number
  of observational constraints, consistent with the hypothesis that
  MHD turbulence accounts for much of the heating of the fast solar
  wind.
 \end{abstract}


\maketitle

\vspace{0.2cm} 
\section{Introduction}
\label{sec:intro}
\vspace{0.2cm} 

One model for the origin of the solar wind relies upon Alfv\'en waves
(AWs) with wavelengths much larger than the proton gyroradius and
frequencies much smaller than the proton cyclotron frequency. In this
model, photospheric motions and/or magnetic reconnection in the solar
atmosphere launch AWs into the corona and solar wind, where the AWs
undergo partial non-WKB (Wentzel-Kramers-Brillouin) reflection
\citep{velli89,zhou89}. Subsequent interactions between
counter-propagating AW packets transfer fluctuation energy from large
scales to small scales. At sufficiently small scales, the fluctuation
energy dissipates. Large-scale AWs also exert an outward force on the
plasma. Several studies have found that this dissipation and momentum
deposition can account for much of the heating and acceleration of the
solar wind
\citep[e.g.,][]{cranmer07,verdini10,chandran11,vanderholst14}.

A number of authors have investigated different aspects of
reflection-driven magnetohydrodynamic (MHD) turbulence. For example, \cite{heinemann80},
\cite{velli93}, and \cite{hollweg07} investigated the linear AW
propagation problem, accounting for radial variations in the density,
outflow velocity, and magnetic-field strength.  \cite{dmitruk02},
\cite{cranmer05}, \cite{verdini07},  \cite{chandran09c}, and \cite{zank18}
investigated the radial evolution of MHD turbulence in the solar
atmosphere and solar wind accounting for reflection and nonlinear interactions. \cite{cranmer07}, \cite{verdini10},
\cite{chandran11}, \cite{vanderholst14}, and \cite{usmanov14} incorporated
reflection-driven MHD turbulence into one-dimensional (1-D) and 3-D solar-wind models.
\cite{verdini09a} and \cite{verdini12} carried out numerical
simulations of reflection-driven MHD turbulence, in which they
approximated the nonlinear terms in the governing equations using a
shell model. \cite{dmitruk03} carried out direct numerical simulations
of reflection-driven MHD turbulence (i.e., without approximating the
nonlinear terms) in the corona in the absence of a background
flow. \cite{vanballegooijen11} carried out direct numerical
simulations of reflection-driven MHD turbulence in the chromosphere
and corona without a background flow. \cite{perez13},
\cite{vanballegooijen16}, and \cite{vanballegooijen17} carried out
direct numerical simulations of reflection-driven MHD turbulence from
the low corona to the Alfv\'en critical point (at a heliocentric
distance~$r$ of $r_{\rm A} \sim 10 R_{\odot}$) and beyond, taking into
account the solar-wind outflow velocity.

In Section~\ref{sec:DNS} of this paper, we present three new direct numerical simulations of
reflection-driven MHD turbulence extending from the photosphere, through the chromosphere,
through a coronal hole, and out to $r=21 R_{\odot}$.  These simulations
go beyond previous simulations extending to $r\gtrsim r_{\rm A}$ by
incorporating the chromosphere. This enables us to account, at least
in an approximate way, for the strong turbulence that develops in
the chromosphere, which launches a broad spectrum of fluctuations into the
corona \citep{vanballegooijen11}. Our simulations also reach larger~$r$
than the simulations of \cite{perez13} and contain 16 times
as many grid points in the field-perpendicular plane as the
simulations of \cite{vanballegooijen17}. 

To offer some insight into the physical processes at work in our
simulations, we present an analytic model of reflection-driven MHD
turbulence in Section~\ref{sec:model}. This model accounts for the
generation of inward-propagating AWs by non-WKB reflection, nonlinear
interactions between counter-propagating AW packets, and the
development of alignment between outward-propagating and
inward-propagating fluctuations. For reasons that
we describe in Sections~\ref{sec:DNS} and~\ref{sec:model}, we divide
the outward-propagating fluctuations into two populations with
different characteristic radial correlation lengths. Our model
reproduces our numerical results reasonably well.

The power-law scalings of the inertial-range power spectra in our simulations vary with radius. We discuss the causes of these variations in Section~\ref{sec:124}, after reviewing several relevant studies 
 in Section~\ref{sec:homogeneous}. We briefly discuss other
 wave-launching parameter regimes in Section~\ref{sec:universality}
 and phase mixing in Section~\ref{sec:phase}, and we present our
 conclusions in Section~\ref{sec:conclusion}.

\section{Transverse, non-compressive fluctuations in a radially
  stratified corona and solar wind}
\label{sec:nlho} 

We focus exclusively on non-compressive fluctuations, which are
observed to dominate the energy density of solar-wind
turbulence~\citep{tumarsch95}, and which carry an energy flux in the
low corona that is sufficient to power the solar wind
\citep{depontieu07}. A disadvantage of our approach is that we neglect
nonlinear couplings between compressive and non-compressive
fluctuations \citep[see, e.g.][]{cho03a,chandran05a,luo06,chandran08b,yoon09,shoda19}, which are likely
important in the solar atmosphere and solar wind. For example, the
plasma density varies by a factor of $\sim 6$ over a distance of a few
thousand km perpendicular to the background magnetic field~$\bm{B}_0$
in the low corona \citep{raymond14}, which suggests that phase mixing
\citep{heyvaerts83} is an efficient mechanism for cascading AW energy
to small scales measured perpendicular to~$\bm{B}_0$ near the
Sun.\footnote{In contrast, {\em Helios} radio occultation observations
  show that the fractional density variations drop to $0.1 - 0.2$ at
  $r \in (5R_{\odot}, 20R_{\odot})$ \citep{hollweg10}.}  We also
neglect the parametric decay of AWs into slow magnetosonic waves and
counter-propagating AWs
\citep[e.g.,][]{galeev63,sagdeev69,cohen74,tenerani17}, which may
cause outward-propagating AWs in the fast solar wind to acquire a
$k_\parallel^{-1}$ spectrum by the time these fluctuations reach
$r=0.3 \mbox{ au}$ \citep{chandran18a}, where $k_\parallel$ is the
wave-vector component parallel to the background magnetic field, and
1~au is the mean Earth-Sun distance. Nevertheless, the simulations
that we report in Section~\ref{sec:DNS} describe an important subset
of the full turbulent dynamics.

Our analysis begins with the continuity, momentum, and induction equations of ideal MHD,
\begin{equation}
\frac{\partial \rho}{\partial t} + \nabla \cdot (\rho \bm{v}) = 0,
\label{eq:cont} 
\end{equation} 
\begin{equation}
\rho \left(\frac{\partial \bm{v}}{\partial t} + \bm{v} \cdot \nabla \bm{v}\right) = 
- \nabla p_{\rm tot} + \frac{\bm{B} \cdot \nabla \bm{B}}{4\pi}  - \rho \nabla \Phi,
\label{eq:momentum} 
\end{equation} 
and
\begin{equation}
\frac{\partial \bm{B}}{\partial t} = \nabla \times (\bm{v} \times \bm{B}),
\label{eq:induction} 
\end{equation} 
where $\rho$, $\bm{v}$, and $\bm{B}$ are the mass density, velocity,
and magnetic field, $ \Phi$ is the gravitational potential, $p_{\rm
  tot} = p + B^2/8\pi$ is the total pressure, and $p$ is the plasma
pressure. We set
\begin{equation} 
\bm{v}   =  \bm{U} + \delta \bm{v}
\qquad \bm{B}   =  \bm{B}_0 + \delta \bm{B}
\end{equation} 
and take the background flow velocity $\bm{U}$ to be aligned with~$\bm{B}_0$. We neglect density fluctuations, setting
\begin{equation}
\delta \rho = 0.
\label{eq:drhozero}  
\end{equation} 
We assume that  the fluctuations are transverse and non-compressive, i.e.,
\begin{equation}
\delta \bm{v} \cdot \bm{B}_0 = 0 \qquad
\delta \bm{B} \cdot \bm{B}_0 = 0 \qquad
\nabla \cdot \delta \bm{v}= 0,
\label{eq:RMHD0}
\end{equation} 
and we take
$\rho$, $\bm{U}$, and $\bm{B}_0$ to be steady-state solutions of
equations~(\ref{eq:cont}) through  (\ref{eq:induction}) (as well as the
MHD energy equation).
The Alfv\'en velocity and Elsasser variables are given by
\begin{equation}
\bm{v}_{\rm A} = \frac{\bm{B}_0}{\sqrt{4\pi \rho}}
\qquad \bm{z}^\pm = \delta \bm{v} \mp \delta \bm{b},
\label{eq:Elsasser} 
\end{equation} 
where
$\delta \bm{b} = \delta \bm{B}/\sqrt{4\pi \rho}$.
Rewriting equations~(\ref{eq:momentum}) and (\ref{eq:induction}) in terms of $\bm{z}^\pm$, we obtain 
\citep{velli89,zhou90}
\[
\frac{\partial \bm{z}^\pm}{\partial t}  
+\left(\bm{U} \pm \bm{v}_{\rm A}\right)\cdot \nabla\bm{z}^\pm 
+ \bm{z}^\mp \cdot \nabla \left(\bm{U} \mp \bm{v}_A\right)
+ \frac{1}{2} \left(\bm{z}^- - \bm{z}^+\right) \left(\nabla \cdot\bm{v}_{\rm A} 
\mp \frac{1}{2}\nabla \cdot \bm{U} \right)
\]
\begin{equation}
\hspace{3cm}  = - \left( \bm{z}^\mp \cdot \nabla \bm{z}^\pm
+ \frac{\nabla p_{\rm tot}}{\rho}\right).
\label{eq:velli} 
\end{equation} 
As in homogeneous MHD turbulence, 
the $\rho^{-1} \nabla p_{\rm tot}$ term in~(\ref{eq:velli}) cancels
the compressive part of the $\bm{z}^\mp \cdot \nabla \bm{z}^\pm$ term
to maintain the condition $\nabla \cdot \bm{z}^\pm =0$.

We assume that the background magnetic field~$\bm{B}_0$ possesses a field line
that is purely radial. Working, temporarily, in spherical coordinates
$(r, \theta, \phi)$, with $\theta=0$ coinciding with this radial field line, 
we restrict our analysis to 
\begin{equation}
\theta \ll 1.
\label{eq:narrow} 
\end{equation} 
We further assume that
\begin{equation}
v_{A\phi} = U_\phi = \partial U/\partial \phi = \partial v_{\rm A}/\partial \phi =
0
\label{eq:notwist} 
\end{equation} 
and 
\begin{equation}
\frac{1}{B_0} \frac{\partial B_0}{\partial r} \sim O(r^{-1}).
\label{eq:lB} 
\end{equation} 
Since $\bm{z}^\mp \cdot \bm{B}_0 = 0$, these assumptions imply that to
leading order in~$\theta$ \citep{chandran15b}
\begin{equation}
\bm{\hat{b}}_0 \cdot \nabla = \frac{\partial}{\partial r},
\label{eq:defb0} 
\end{equation} 
and 
\begin{equation}
\bm{z}^\mp \cdot \nabla \left(\bm{U} \mp \bm{v}_{\rm A}\right)
 = \bm{z}^\mp (U \mp v_{\rm A}) (\nabla \cdot \bm{\hat{b}}_0/2),
\label{eq:HOapprox} 
\end{equation} 
where
\begin{equation}
\bm{\hat{b}}_0 = \frac{\bm{B}_0}{B_0}.
\label{eq:b0} 
\end{equation} 
We take $\bm{B}_0$ to be directed away from the Sun, so that
$\bm{z}^+$ ($\bm{z}^-$) corresponds to outward-propagating
(inward-propagating) fluctuations (when viewed in the local plasma
frame), and we define vector versions of the variables introduced by
Heinemann \& Olbert (1980),
\begin{equation}
\bm{g} = \frac{(1 + \eta^{1/2}) \bm{z}^+}{\eta^{1/4}} \qquad
\bm{f} = \frac{(1 - \eta^{1/2}) \bm{z}^-}{\eta^{1/4}},
\label{eq:deffg} 
\end{equation} 
where
\begin{equation}
\eta = \rho/\rho_a,
\label{eq:defeta} 
\end{equation} 
and $\rho_a$ is the value of $\rho$ at the Alfv\'en critical point, at
which $U=v_{\rm A}$.
Mass conservation and flux conservation imply that 
\begin{equation}
\frac{\rho U}{B_0} = \mbox{ constant},
\label{eq:cons} 
\end{equation} 
which in turn implies that
\begin{equation}
v_{\rm A} = \eta^{1/2} U.
\label{eq:vAU} 
\end{equation} 
With the use of~(\ref{eq:deffg}) and (\ref{eq:vAU}), we rewrite $\bm{z}^\pm$ in~(\ref{eq:velli}) in terms of $\bm{g}$ and $\bm{f}$,
obtaining the nonlinear
Heinemann-Olbert equations \citep{heinemann80,chandran09c},
\begin{equation} 
 \frac{\partial \bm{g}}{\partial t} + (U + v_{\rm A}) \frac{\partial \bm{g}}{\partial r} -\left(\frac{U+v_{\rm A}}{2v_{\rm A}}\right) \diff{v_{\rm A}}{r} \bm{f}
 =   - \bm{z}^- \cdot \nabla \bm{g} -
\left(\frac{1+\eta^{1/2}}{\eta^{1/4}}\right)\frac{\nabla p_{\rm tot}}{\rho}
\label{eq:Hg} 
\end{equation} 
\begin{equation} 
\frac{\partial \bm{f}}{\partial t} + (U - v_{\rm A}) \frac{\partial \bm{f}}{\partial r}
-  \left(\frac{U-v_{\rm A}}{2v_{\rm A}}\right)\diff{v_{\rm A}}{r} \bm{g}  
 =  -\bm{z}^+ \cdot \nabla \bm{f} -   \left(\frac{1 -\eta^{1/2}}{\eta^{1/4}}\right)
\frac{\nabla p_{\rm tot}}{\rho}.
\label{eq:Hf} 
\end{equation} 
Equations~(\ref{eq:Hg}) and (\ref{eq:Hf}) are equivalent to the
equations solved by \cite{perez13} and
\cite{vanballegooijen16,vanballegooijen17}.\footnote{Equations~(\ref{eq:cont}),
  (\ref{eq:momentum}), (\ref{eq:induction}), (\ref{eq:Hg}),
  (\ref{eq:Hf}), and the plasma internal-energy equation possess two
  conservation laws involving~$\bm{f}$ and~$\bm{g}$. The first is total-energy conservation, and the
  second is some times referred to as ``non-WKB wave-action
  conservation''
  \citep{heinemann80,cranmer05,verdini07,chandran15b}. This second
  conservation relation can be derived from the equation of
  cross-helicity conservation \citep{chandran15b}.}

Because~(\ref{eq:RMHD0}) is also satisfied by non-compressive
fluctuations in reduced MHD (RMHD), (\ref{eq:Hg}) and (\ref{eq:Hf}) could be viewed
as an inhomogeneous version of RMHD. However, the way in which we have
arrived at (\ref{eq:Hg}) and~(\ref{eq:Hf}) --- in particular, starting
with (\ref{eq:drhozero}) and~(\ref{eq:RMHD0}) as assumptions ---
differs from the usual derivation of the RMHD equations \citep[see,
e.g.,][]{schekochihin09}, which begins by assuming that $\delta B \ll
B_0$ and $\lambda \ll l$, where $\lambda$ ($l$) is the characteristic
length scale of the fluctuations measured perpendicular (parallel)
to~$\bm{B}_0$. We conjecture that (\ref{eq:Hg}) and~(\ref{eq:Hf}) may
provide a reasonable description of transverse, non-compressive
fluctuations and their mutual interactions even when the assumptions
$\delta B \ll B_0$ and $\lambda \ll l$ fail. For example, if
collisionless damping \citep{barnes66} or passive-scalar mixing
\citep{schekochihin16,meyrand19} removes compressive and longitudinal
fluctuations, then (\ref{eq:drhozero}) and~(\ref{eq:RMHD0}) may be
reasonable approximations even if $\delta B \sim B_0$ and $\lambda
\sim l$.  We note that neither our derivation of (\ref{eq:Hg}) and (\ref{eq:Hf}), nor the derivation of RMHD as a limit of the Vlasov equation \citep{schekochihin09}, requires that $\beta = 8\pi p/B^2$ be ordered as either large or small.

\vspace{0.2cm} 
\section{Direct Numerical Simulations}
\label{sec:DNS} 

\begin{figure}
\centerline{
\includegraphics[width=8cm,trim={0 8cm 0 4cm}]{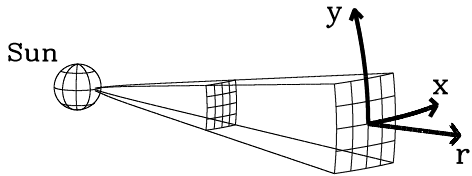}
} \caption{ \label{fig:clebsch}
Numerical domain of the REFLECT Code.}
\end{figure}

We have carried out three direct numerical simulations of
(\ref{eq:Hg}) and (\ref{eq:Hf}) using the
pseudo-spectral/Chebyshev REFLECT code \citep{perez13}. 
In each simulation, the numerical domain is a narrow
magnetic flux tube with a square cross section, as illustrated in
Fig.~\ref{fig:clebsch}. 
This flux tube extends from the photosphere at $r=r_{\rm min} =
1 R_{\odot}$, through the chromosphere,  the ``transition region'' (the
narrow layer at the top of the chromosphere), and
a coronal hole, and then out to a heliocentric distance of
\begin{equation}
r_{\rm max} = 21 R_{\odot}.
\label{eq:rmax} 
\end{equation} 
We model the transition region in our simulations as a
  discontinuity in the density at
\begin{equation}
r_{\rm tr} = 1.0026 R_{\odot},
\label{eq:rtr} 
\end{equation} 
a distance of roughly 1800~km above the photosphere. (We have
collected in table~\ref{tab:r}
several heliocentric distances that we refer to repeatedly in the
discussion to follow.)
The walls of the simulation domain are parallel to
the background magnetic field~$\bm{B}_0$. As $r$
increases and $\bm{B}_0(r)$ decreases, the width~$L_{\rm box}$ of the
simulation domain perpendicular to~$\bm{B}_0$ grows according
to the relation
\begin{equation}
L_{\rm box}(r) = L_{\rm box}(1 R_{\odot}) \left[\frac{B_0(1
      R_{\odot})}{B_0(r)}\right]^{1/2}.
\label{eq:Lbox} 
\end{equation} 
Because $B_0$ drops
sharply between the photosphere and the transition region (see
(\ref{eq:Bch}) below), $L_{\rm box}(r_{\rm tr}) \simeq 10 L_{\rm box}(1 R_{\odot})$. The values of $L_{\rm box}(1 R_{\odot})$
and $L_{\rm box}(r_{\rm tr})$ in our three simulations are listed in table~\ref{tab:parameters}.
We discuss why we choose these values for $L_{\rm box}(1 R_{\odot})$ in Section~\ref{sec:bc}.

\begin{table*}
\caption{\vspace{0.1cm} Simulation Parameters
\label{tab:parameters} }
\begin{center}
\begin{tabular}{lcccc}
\vspace{-0.7cm} 
&&  \\ 
\hline \hline 
\vspace{-0.25cm} 
& & \\
Quantity & & Run 1 & Run 2 & Run 3 \\
\vspace{-0.25cm} 
&& \\
\hline
\vspace{-0.25cm} && \\
$\delta v_{\rm ph, rms}$ & $\dots\dots\dots\dots$ & $1.3 \mbox{ km/s}$
  & $1.3 \mbox{ km/s}$ & $1.3 \mbox{ km/s}$\\
$\tau_{v}^{\rm(ph)}$ & $\dots\dots\dots\dots$ & $3.3 \mbox{ min}$ &
                                                                     $9.6 \mbox{ min}$ & $9.3 \mbox{ min}$\\
$L_{\rm box}(1 R_{\odot})$& $\dots\dots\dots\dots$ & $ 4.1 \times
                                                         10^2 \mbox{
                                                         km}$  & $ 4.1
                                                                 \times
                                                                 10^2
                                                                 \mbox{
                                                                 km}$  & $ 1.6 \times 10^3 \mbox{ km}$ \\

$L_{\rm box}(1.0026 R_{\odot})$& $\dots\dots\dots\dots$ & $ 4.1 \times
                                                         10^3 \mbox{
                                                         km}$  & $ 4.1
                                                                 \times
                                                                 10^3
                                                                 \mbox{
                                                                 km}$  & $ 1.6 \times 10^4 \mbox{ km}$ \\
Number of grid points & $\dots\dots\dots\dots$ & \hspace{0.3cm} $256^2
                                                 \times 16385$
                                                 \hspace{0.3cm} 
  & \hspace{0.3cm} $256^2 \times 16385$ \hspace{0.3cm} &
                                                         \hspace{0.3cm}
                                                         $256^2 \times
                                                         16385$
                                                         \hspace{0.3cm}
  \\
\vspace{-0.2cm} 
\\
\hline
\end{tabular}
\end{center}
\medskip
{\footnotesize $\delta v_{\rm ph, rms}$ is the r.m.s. amplitude of the velocity
  fluctuation at the photosphere, $\tau_v^{\rm (ph)}$ is the correlation
  time of the photospheric velocity,  and $L_{\rm box}$ is the
  perpendicular dimension (along either the $x$ or $y$ directions) of
  the numerical domain.
}
\vspace{0.7cm} 
\end{table*}

\begin{table*}
\caption{\vspace{0.1cm} Glossary of Heliocentric Distances
\label{tab:r} }
\begin{center}
\begin{tabular}{ccc}
\vspace{-0.4cm} 
&&  \\ 
\hline \hline 
\vspace{-0.25cm} 
& & \\
Symbol & Numerical Value & Corresponding location \\
\vspace{-0.25cm} 
&& \\
\hline
\vspace{-0.25cm} && \\
$r_{\rm tr}$ & $1.0026 R_{\odot}$ &   transition region \\
$r_{\rm b}$ & $1.0027 R_{\odot}$ &   coronal base \\
$r_{\rm m}$ & $1.71 R_{\odot}$ &   Alfv\'en-speed maximum \\
$r_{\rm A}$ & $11.1 R_{\odot}$ &   Alfv\'en critical point
  \\
$r_{\rm max}$ & $21 R_{\odot}$ & maximum~$r$ in simulation domain 
  \\
\vspace{-0.2cm} 
\\
\hline
\end{tabular}
\end{center}
\medskip
\end{table*}

At $r>r_{\rm tr}$, 
the field lines of~$\bm{B}_0$ are nearly
radial, even though we allow for super-radial expansion of the
magnetic field.
This is because the flux-tube width is much smaller than the
characteristic radial distance over which $B_0$ varies by a factor of
order unity. Because the flux tube is narrow and $\bm{B}_0$ is nearly
radial, we can ignore the
curvature of the field-perpendicular surfaces 
to a good approximation at $r>r_{\rm tr}$.
 We thus use Cartesian coordinates, $x$
and~$y$, to denote position in the plane perpendicular to the
radial line that runs down the centre of the simulation
domain.

At $r<r_{\rm tr}$, our assumption in Section~\ref{sec:nlho} that
$\bm{B}_0$ is nearly radial breaks down, because the flux tube expands
so rapidly with height above the photosphere. 
 Because of this, and because we neglect compressive fluctuations,
our simulations provide only a crude approximation of chromospheric
turbulence. Nevertheless, we retain the chromosphere in our
simulations, because turbulence in the actual chromosphere
launches a broad spectrum of AWs into the corona
\citep{vanballegooijen11}, and our model chromosphere gives us a way
of approximating this turbulent wave-launching process.

\subsection{Radial Profiles of $\rho$, $B_0$, and $U$}
\label{sec:profiles} 

We choose the radial profiles of $\rho$,
$U$, and~$B_0$ to approximate the conditions found in coronal holes and the
fast solar wind. Above the transition region, at $r> r_{\rm tr}$, we set
\begin{equation} 
\rho  =  \left(10^9 s^{-15.6} +
2.51\times10^6 s^{-3.76}  + 1.85\times
10^5 s^{-2}\right)m_{\rm p} \,\mbox{cm}^{-3},
\label{eq:rho0} 
\end{equation} 
\begin{equation} 
B_0  =  1.5 \left[s^{-6}(f_{\rm max} - 1) 
+ s^{-2}\right]\, \mbox{G},
\label{eq:B0} 
\end{equation} 
and
\begin{equation} 
U =  9.25 \times 10^{12}\left(\frac{B_0}{1
    \,\mbox{G}}\right) \left(\frac{\rho}{m_{\rm p}\mbox{
      cm}^{-3}}\right)^{-1}
\mbox{cm}\,\mbox{s}^{-1},
\label{eq:U0} 
\end{equation} 
where 
\begin{equation}
s = \frac{r}{R_\odot},
\label{eq:defs}
\end{equation}
\begin{equation}
f_{\rm max} = 9
\label{eq:fmax} 
\end{equation} 
is the super-radial expansion
factor, and $m_{\rm p}$ is the proton mass.  Equation~(\ref{eq:rho0})
is adapted from the coronal-hole electron-density measurements of
\cite{feldman97}. We have modified those authors' density profile by
adding the $s^{-2}$ term in~(\ref{eq:rho0}) so that the model
extrapolates to a reasonable density at large~$r$ and by increasing the
coefficient of the $s^{-15.6}$ term in order to match the 
low-corona density in the model of \cite{cranmer05}.  Equation~(\ref{eq:B0}) is taken
from \cite{hollweg02}.  The general form of~(\ref{eq:U0})  follows from~(\ref{eq:cons}).
 The numerical coefficient on the right-hand side of~(\ref{eq:U0}) is chosen so that 
\begin{equation}
U(r_{\rm b}) = 1.2 \mbox{ km/s} \qquad U(1 \mbox{ au}) = 750 \mbox{ km/s},
\label{eq:Uvals} 
\end{equation} 
where 
\begin{equation}
r_{\rm b} = 1.0027R_{\odot}
\label{eq:defrb} 
\end{equation} 
is a heliocentric distance just larger than~$r_{\rm tr}$ that we take
to correspond to the base of the corona. Given the radial profiles in
Equations~(\ref{eq:rho0}) through (\ref{eq:U0}), the Alfv\'en critical
point is at
\begin{equation}
r_{\rm A} = 11.1 R_{\odot},
\label{eq:rA} 
\end{equation} 
the Alfv\'en speed reaches its maximum value at 
\begin{equation}
r_{\rm m} = 1.71 R_{\odot},
\label{eq:rm} 
\end{equation} 
and 
\begin{equation}
v_{\rm A}(r_{\rm b}) = 935 \mbox{ km/s} \qquad v_{\rm A}(r_{\rm m}) =
2730 \mbox{ km/s} \qquad v_{\rm A}(r_{\rm A}) = U(r_{\rm A}) =  627
\mbox{ km/s}.
\label{eq:vAvals} 
\end{equation} 

Below the transition region, we set
\begin{equation}
\rho = \rho_{\rm ph} e^{c(1-s)/s},
\label{eq:rhoch} 
\end{equation} 
where
\begin{equation}
\rho_{\rm ph} = 4.78 \times 10^{16} m_{\rm p}\mbox{ cm}^{-3}
\label{eq:nph} 
\end{equation} 
is the photospheric density,
 $c = [s_{\rm tr}/(1-s_{\rm tr})]\ln(\rho_{\rm
   tr,<}/\rho_{\rm ph})$,
$s_{\rm tr} = r_{\rm tr}/R_{\odot}$, and $\rho_{\rm
  tr,<}$ is the density just below the transition region,
which we take to be 100 times greater than the value of the
density at $r=r_{\rm tr}$ from~(\ref{eq:rho0}). 
We then set \citep[cf.][]{vanballegooijen11}
\begin{equation}
B = \left[ \frac{(B_{\rm ph}^2 - B_{\rm tr}^2) (\rho -
    \rho_{\rm tr,<})}{\rho_{\rm ph} - \rho_{\rm tr,<}} +
  B_{\rm tr}^2\right]^{1/2},
\label{eq:Bch} 
\end{equation} 
at $r< r_{\rm tr}$, where 
\begin{equation}
B_{\rm ph} = 1400 \mbox{ G}
\label{eq:Bph} 
\end{equation} 
is the assumed magnetic-field strength in the photospheric
footpoint of the simulated flux tube, and $B_{\rm tr}$ is the
value of~$B$ at $r=r_{\rm tr}$ from~(\ref{eq:B0}).

\begin{figure*}
\centerline{
\includegraphics[trim = 0cm 4cm 0cm 0cm, width=6cm]{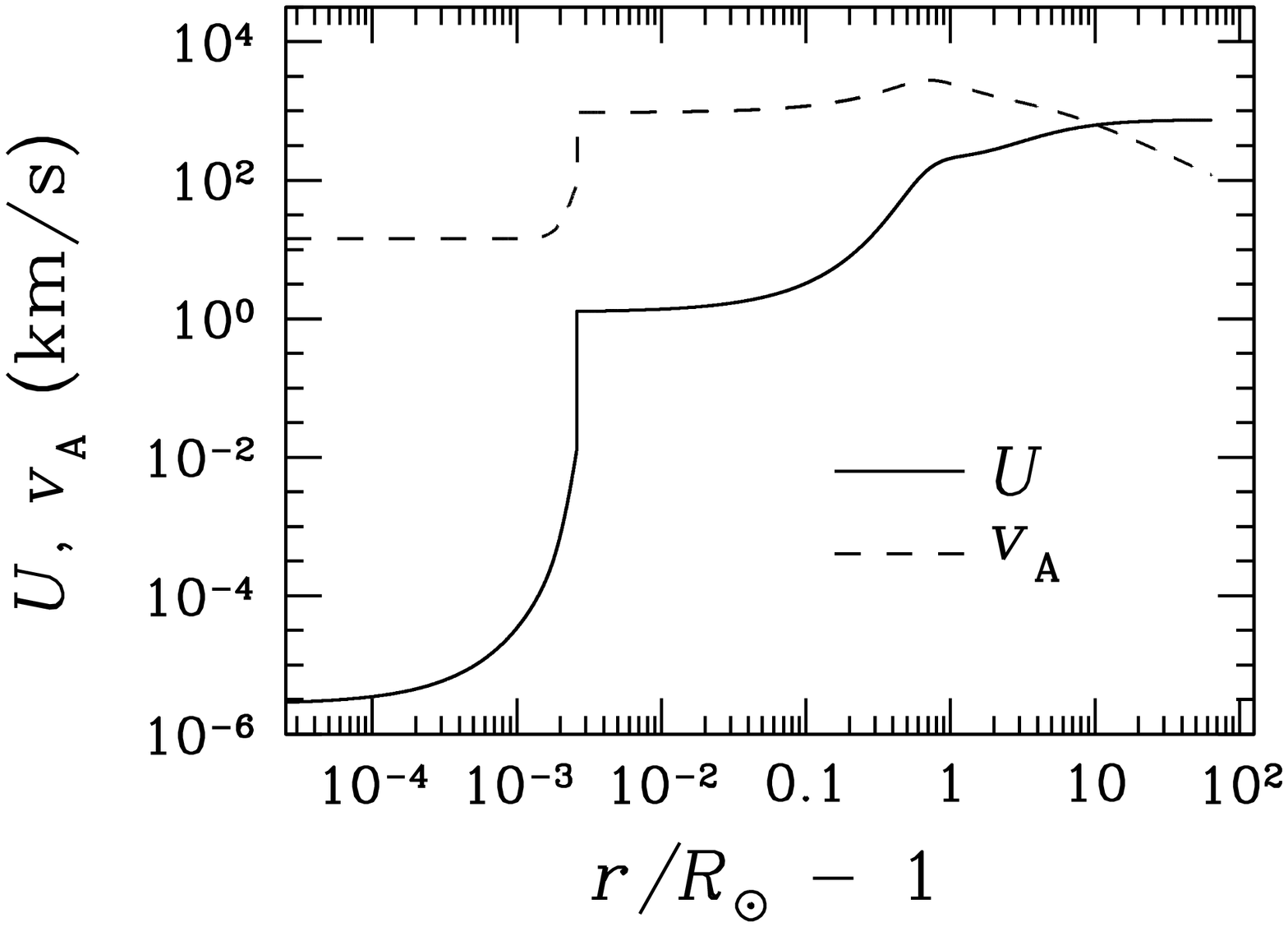}
\hspace{0.3cm} 
\includegraphics[trim = 0cm 4cm 0cm 0cm, width=6cm]{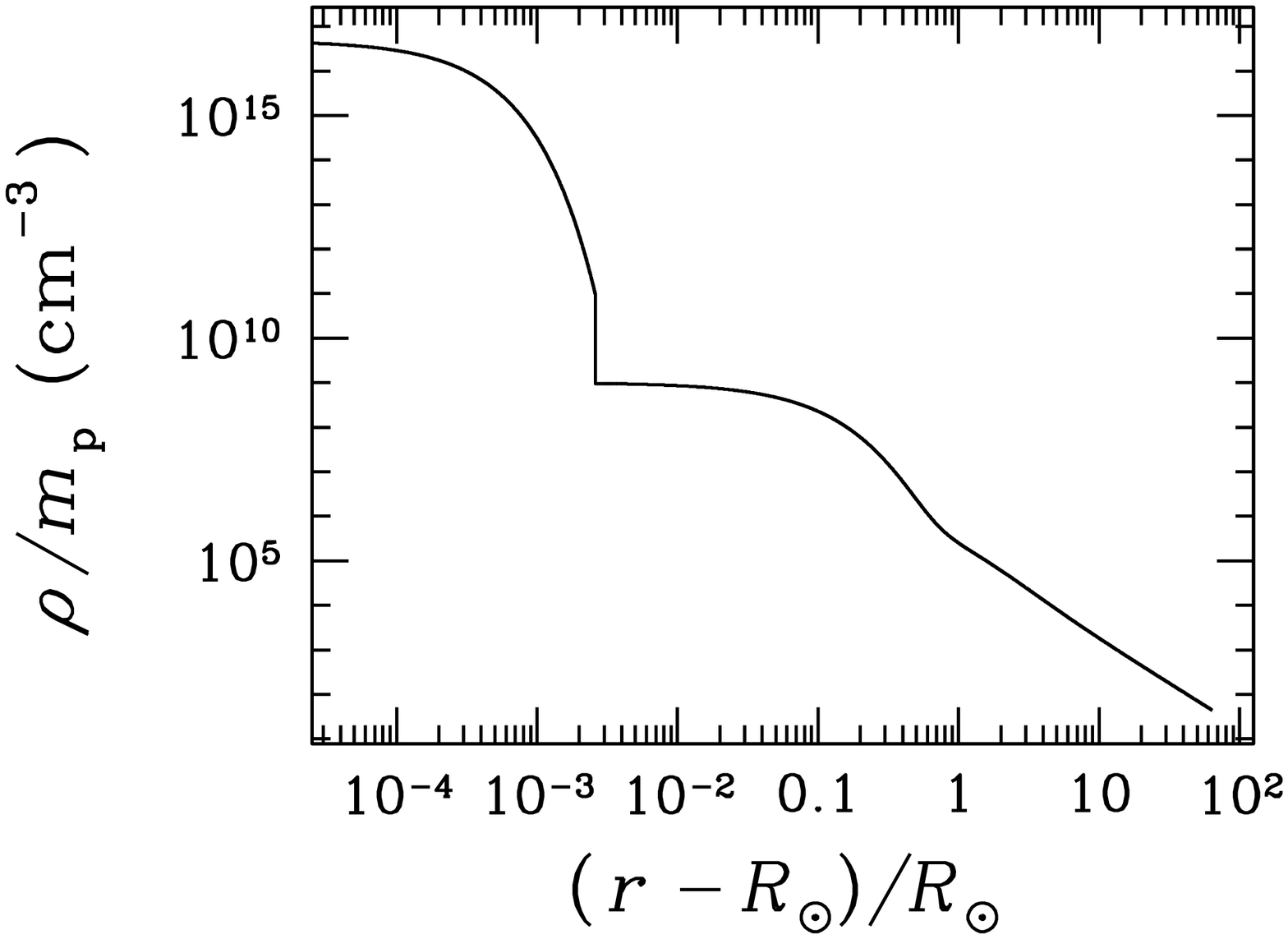}
}
\centerline{
\includegraphics[trim = 0cm 4cm 0cm 0cm, width=6cm]{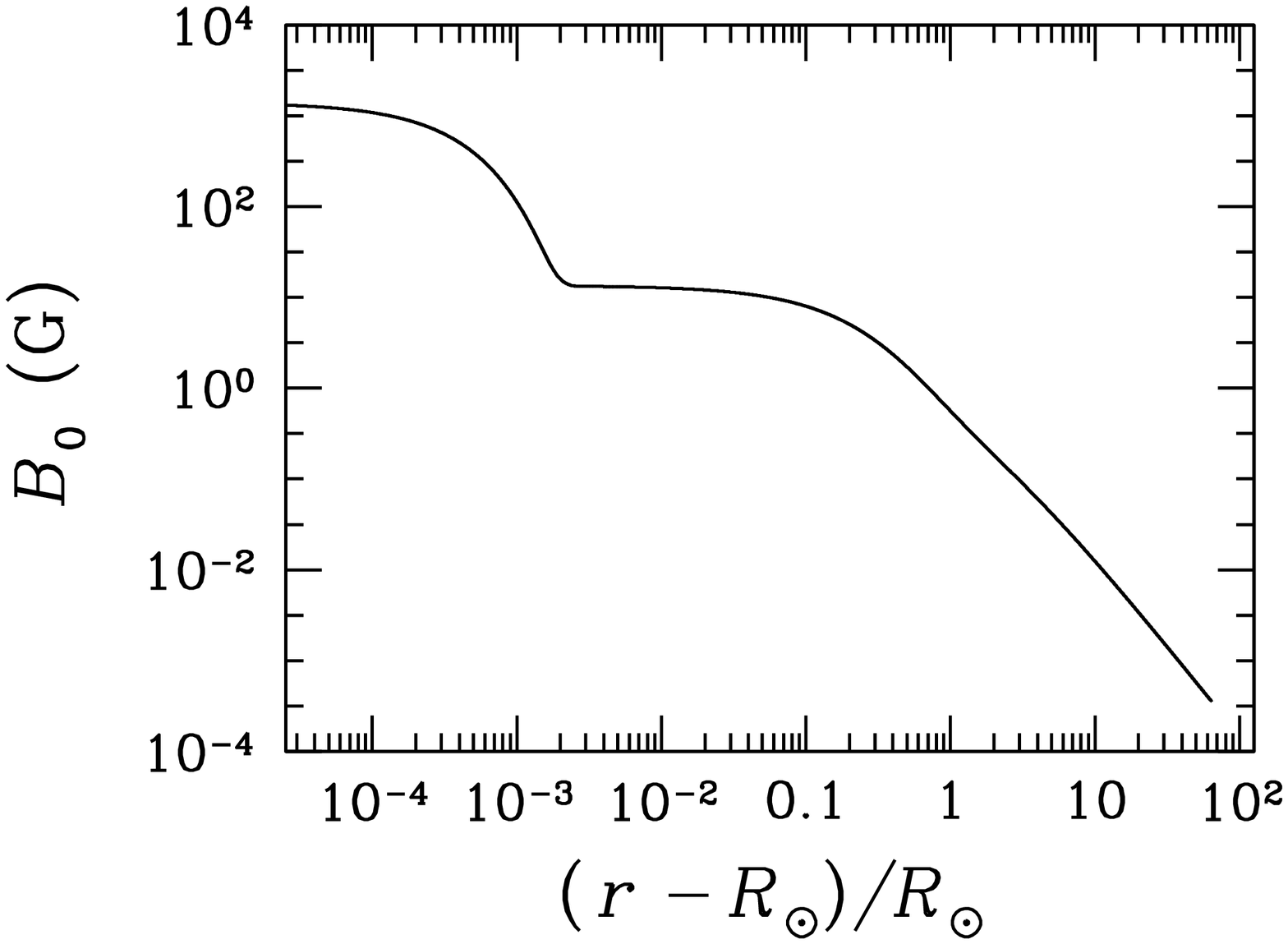}
\hspace{0.3cm} 
\includegraphics[trim = 0cm 4cm 0cm 0cm, width=6cm]{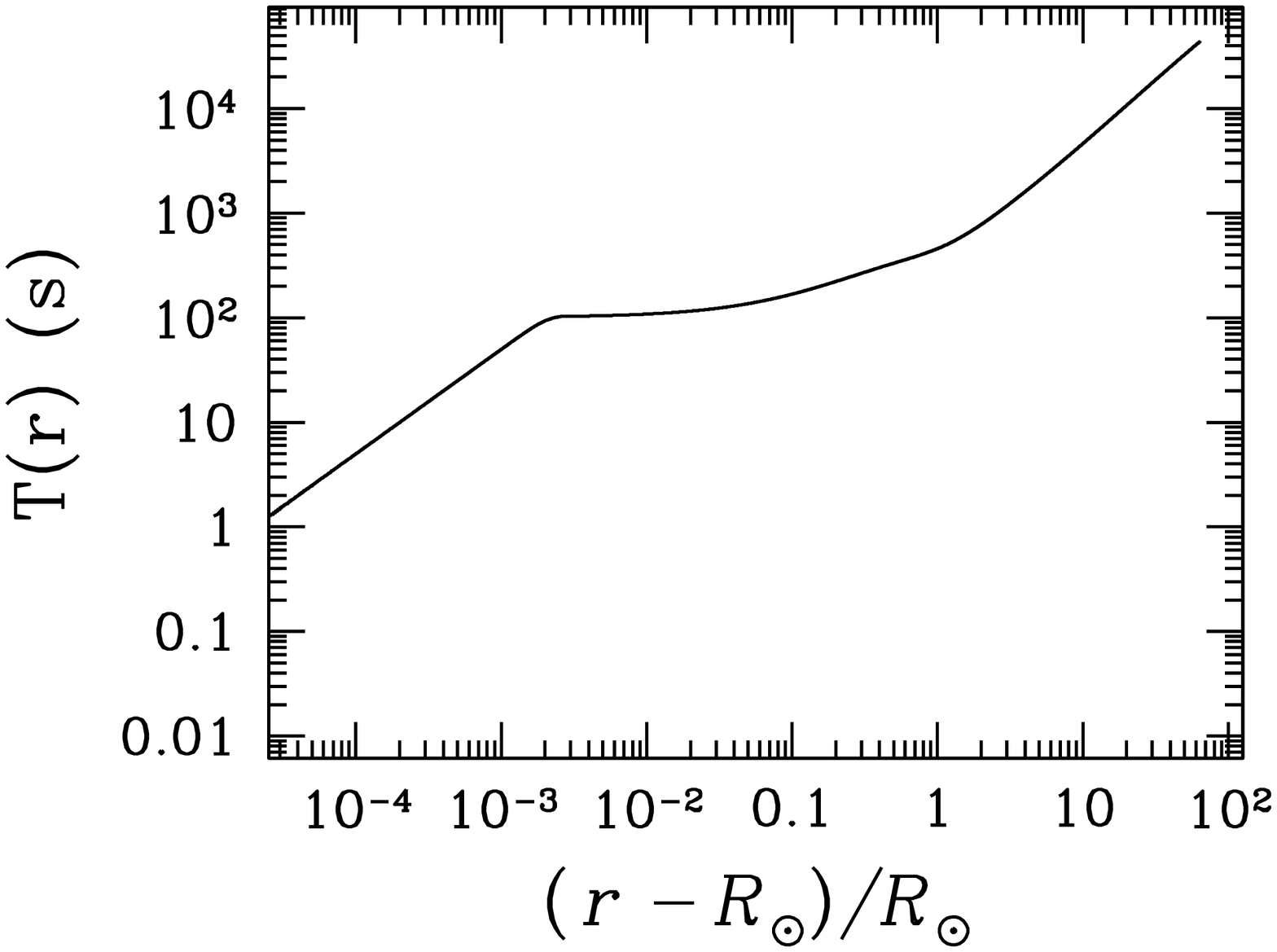}
}
\caption{The radial profiles of the solar-wind outflow
  velocity~$U$, Alfv\'en speed~$v_{\rm A}$, plasma
  density~$\rho$ divided by the proton mass~$m_{\rm p}$, background magnetic-field
  strength~$B_0$, and $\bm{z}^+$ travel time from the transition
  region~$T(r)$ in our direct numerical simulations. We
  use the same profiles when evaluating quantities in the analytic
  model that we present in Section~\ref{sec:model}.
\label{fig:profiles} 
}
\end{figure*}

We plot the radial profiles of $\rho$, $B$, $U$, and $v_{\rm
  A}$ in figure~\ref{fig:profiles}.
We also plot the $\bm{z}^+$ travel time between the photosphere
and radius~$r$,
\begin{equation}
T(r) = \int_{R_{\odot}}^r \frac{\d r}{U + v_{\rm A}}.
\label{eq:defT} 
\end{equation}

\subsection{Boundary Conditions}
\label{sec:bc} 

We take the $\bm{z}^\pm$ fluctuations to satisfy periodic boundary
conditions in the $xy$-plane.
At the photosphere,
we impose a time-dependent velocity field.
We set the velocity Fourier components at the photosphere equal to zero when 
$k_\perp > 3\times 2\pi/L_{\rm box}(R_{\odot})$, where
\begin{equation}
k_\perp = \sqrt{k_x^2 + k_y^2},
\label{eq:kperp} 
\end{equation} 
and $k_x$ and $k_y$ are the $x$ and $y$ components of the wave
vector~$\bm{k}$. We set the amplitudes of the velocity Fourier
components at $k_\perp \leq 3 \times 2\pi/L_{\rm box}(R_{\odot})$
equal to a constant, which we choose so that the root-mean-square (r.m.s.) amplitude of the
fluctuating velocity at the photosphere is
\begin{equation}
\delta v_{\rm ph, rms} = 1.3 \mbox{ km/s},
\label{eq:dvprms} 
\end{equation} 
consistent with observational constraints  on the velocities of solar
granules \citep{richardson50}. We then assign random values to the phases of these velocity
Fourier components at the discrete set of times
$t_{\rm n} = n \, \tau_0,$ where
$\tau_0 = 5 \mbox{ min}$ in Run~1 and $\tau_0 = 20
\mbox{ min}$ in Runs~2 and~3. To determine the
phases at times between successive $t_{\rm n}$, we use cubic
interpolation in time. We define the correlation time of the
photospheric velocity~$\tau_{v}^{\rm (ph)}$ to be the time lag over which the normalized
velocity autocorrelation function decreases from 1 to 0.5. The
resulting
velocity correlation times are listed in table~\ref{tab:parameters}.

Our choices of $\tau_0$ and $L_{\rm box}(R_{\odot})$ determine (at
least in part --- see Section~\ref{sec:twocomp}) the correlation
time~$\tau_{\rm c}$ and perpendicular correlation length~$L_\perp$ of
the AWs launched by the Sun. (Since we only drive photospheric
velocity modes with $k_\perp \leq 3 \times 2\pi/L_{\rm
  box}(R_{\odot})$, $L_\perp$ is a few times smaller than~$L_{\rm box}$.)
Estimates of
$L_{\perp}(r_{\rm b})$ range from~$\simeq 10^3 \mbox{ km}$
\citep{cranmer07,hollweg10,vanballegooijen16,vanballegooijen17} to more than
$10^4 \mbox{ km}$ \citep{dmitruk02,verdini07,verdini12}, and estimates
of $\tau_{\rm c}(r_{\rm b})$ range from $\simeq 1-5$
minutes~\citep{cranmer05,vanballegooijen16,vanballegooijen17} to one
or more hours~\citep{dmitruk03}. Given the uncertainty in 
$L_\perp(r_{\rm b})$ and $\tau_{\rm c}(r_{\rm b})$, we
vary $L_{\rm box}(R_{\odot})$ and $\tau_0$ by factors of 4 and 5,
respectively, in our different simulations in order to investigate how the values of
$L_\perp(r_{\rm b})$ and $\tau_{\rm c}(r_{\rm b})$ influence the
properties of the turbulence at larger~$r$.

No information flows into the simulation domain through 
the outer boundary at $r=r_{\rm max}$, because $r_{\rm max} > r_{\rm A}$. We thus do not impose an
additional boundary condition at the outer boundary.

\subsection{Hyper-Dissipation}
\label{sec:hyper} 

To dissipate the fluctuation energy that cascades to small
wavelengths, we add a hyper-dissipation term of the form
\begin{equation}
D_g = - \nu_g \left(\frac{\partial^2}{\partial x^2} +
  \frac{\partial^2}{\partial y^2}\right)^4 \bm{g}
\label{eq:Dg} 
\end{equation} 
to the right-hand side of~(\ref{eq:Hg}), and a
hyper-dissipation term of the form
\begin{equation}
D_f = - \nu_f \left(\frac{\partial^2}{\partial x^2} +
  \frac{\partial^2}{\partial y^2}\right)^4 \bm{f}
\label{eq:Df} 
\end{equation} 
to the right-hand side of~(\ref{eq:Hf}).  We choose the magnitude and
radial dependence of the hyper-dissipation coefficients~$\nu_g$ and
$\nu_f$ so that dissipation becomes important near the grid scale at
all radii in each simulation.  In particular, we take $\nu_g$ and
$\nu_f$ to be proportional
to~$[L_{\rm box}(r)/L_{\rm box}(R_{\odot})]^8$.

\subsection{Numerical Algorithm}
\label{sec:algorithm} 

The REFLECT Code solves (\ref{eq:Hg}) and~(\ref{eq:Hf}) using
a spectral element method based on a Chebyshev-Fourier
basis~\citep{CanHus88}. In each of our three simulations, we split the numerical
domain into 1024 subdomains. Each subdomain covers the full flux-tube
cross section pictured in figure~\ref{fig:clebsch} using 256 grid
points along both the $x$ and $y$ directions, but only part of
the flux tube's radial extent. 
Along the~$r$ axis, each subdomain
contains 17 grid points, 
two of which are boundary grid points. The total number of radial grid
points is 16385. Except at $r_{\rm min}$ and $r_{\rm
  max}$, these boundary grid points are shared by neighboring
subdomains. Eight of the subdomains are
in the chromosphere.

A Chebyshev/Fourier
transform of (\ref{eq:Hg}) and~(\ref{eq:Hf}) leads to a
system of ordinary differential equations for the Chebyshev-Fourier
coefficients in each subdomain.  These equations are coupled through
matching conditions (continuity of $\delta \bm{v}$ and $\delta \bm{B}$) at the boundaries
between neighboring subdomains.  The REFLECT code advances the solution
forward in time using a third-order Runge-Kutta method, with an
integrating factor to handle the hyper-dissipation terms.  Within each
subdomain, the REFLECT code discretizes the radial interval using a
Gauss-Lobatto grid, which makes it possible to compute the Chebyshev
transform using a fast cosine transform.

\subsection{Duration of the Simulations}
\label{sec:duration} 

We run each simulation from $t=0$ until $t= 13.2 \mbox{ hr}$.  Between $t=0$ and $t= 4 \mbox{ hr}$, the magnetic and kinetic
energies in the simulations fluctuate while trending upwards. For
reference, it takes 1.3~hours for an outward-propagating AW to travel
from the photosphere to the Alfv\'en critical point at
$r_{\rm A} = 11.1 R_{\odot}$, and 3~hours for an outward-propagating AW
to travel from the photosphere to~$r_{\rm max} = 21 R_{\odot}$ (see
figure~\ref{fig:profiles}).  After $t \simeq 4 \mbox{ hr}$, the
magnetic and kinetic energies fluctuate around a steady value.
We regard the turbulence as being in a statistical steady state at $t>
6 \mbox{ hr}$. All
the numerical results that we present are calculated from time
averages between $t= 6 \mbox{ hr}$ and $t = 13.2 \mbox{ hr}$, except
for the 
$z^+_{\rm HF, rms}$ and $z^+_{\rm LF, rms}$ profiles in Run~2; those
profiles,
because of technical difficulties,
were only computed from averages between
$t= 12 \mbox{ hr}$ and $t = 13 \mbox{ hr}$.

\subsection{Radial profiles of the fluctuation amplitudes}
\label{sec:z_profiles}

\begin{figure}
\centerline{
\includegraphics[trim = 0cm 4cm 0cm 0cm, width=6cm]{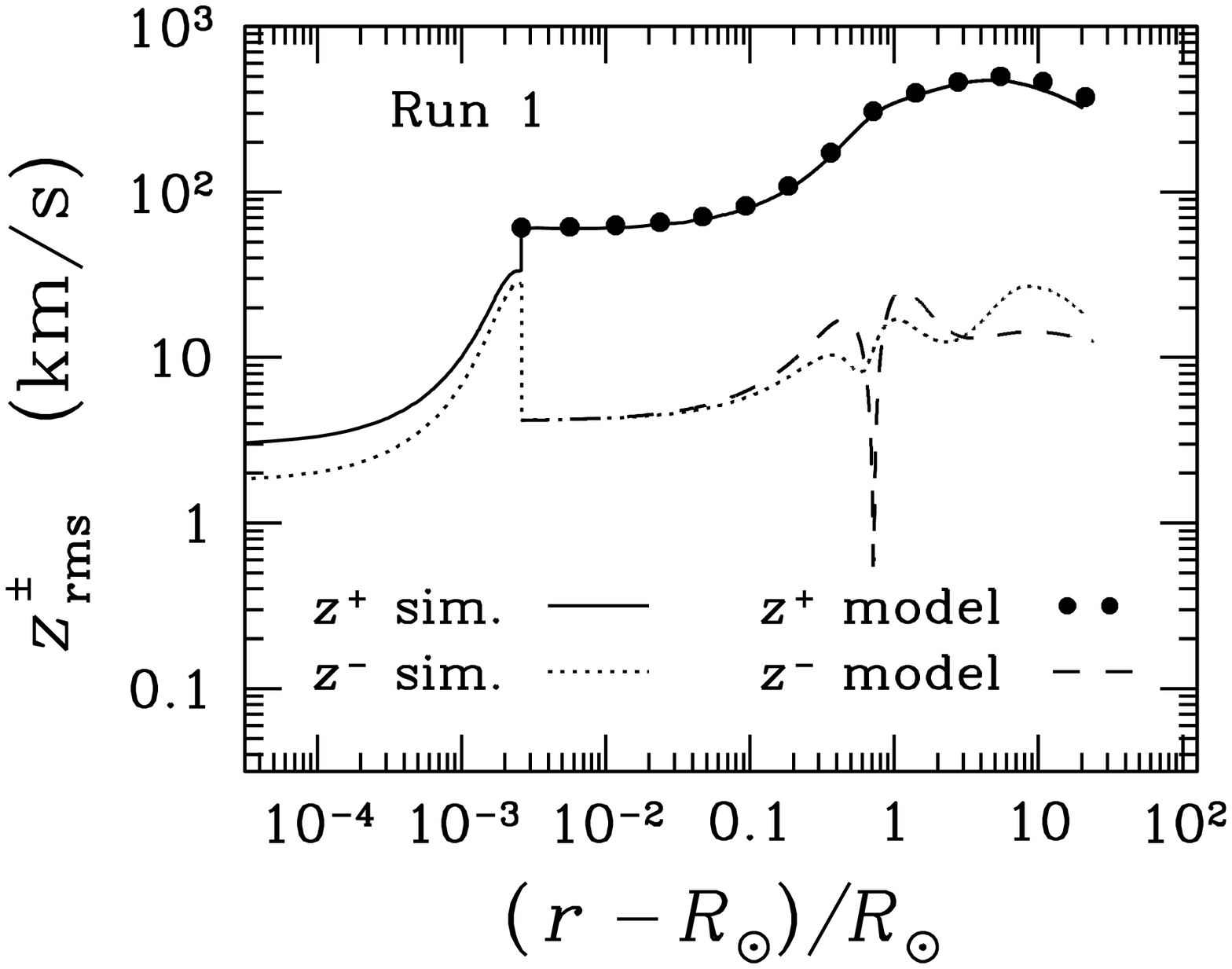}
\hspace{0.3cm} 
\includegraphics[trim = 0cm 4cm 0cm 0cm, width=6cm]{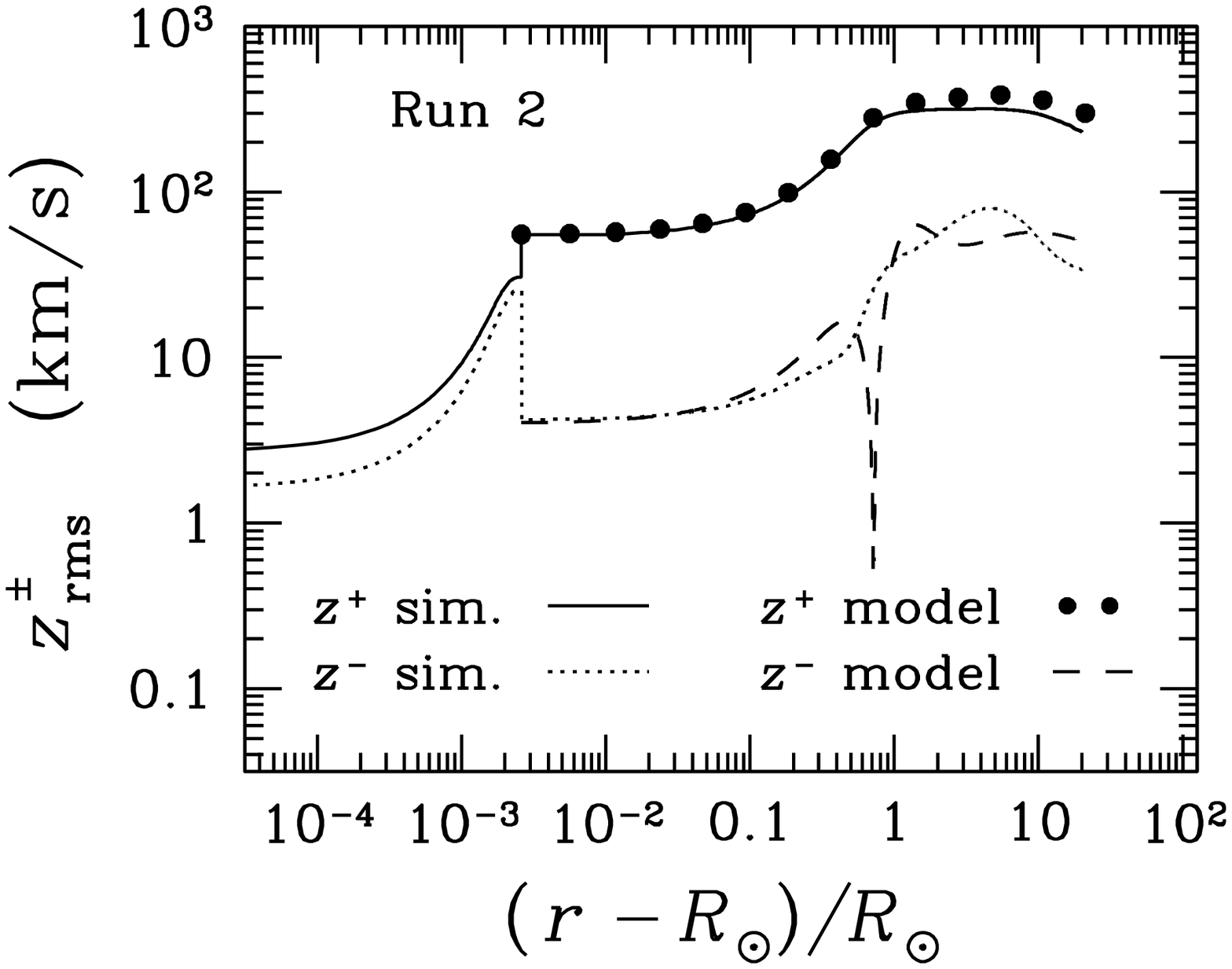}
}
\centerline{
\includegraphics[trim = 0cm 4cm 0cm 0cm, width=6cm]{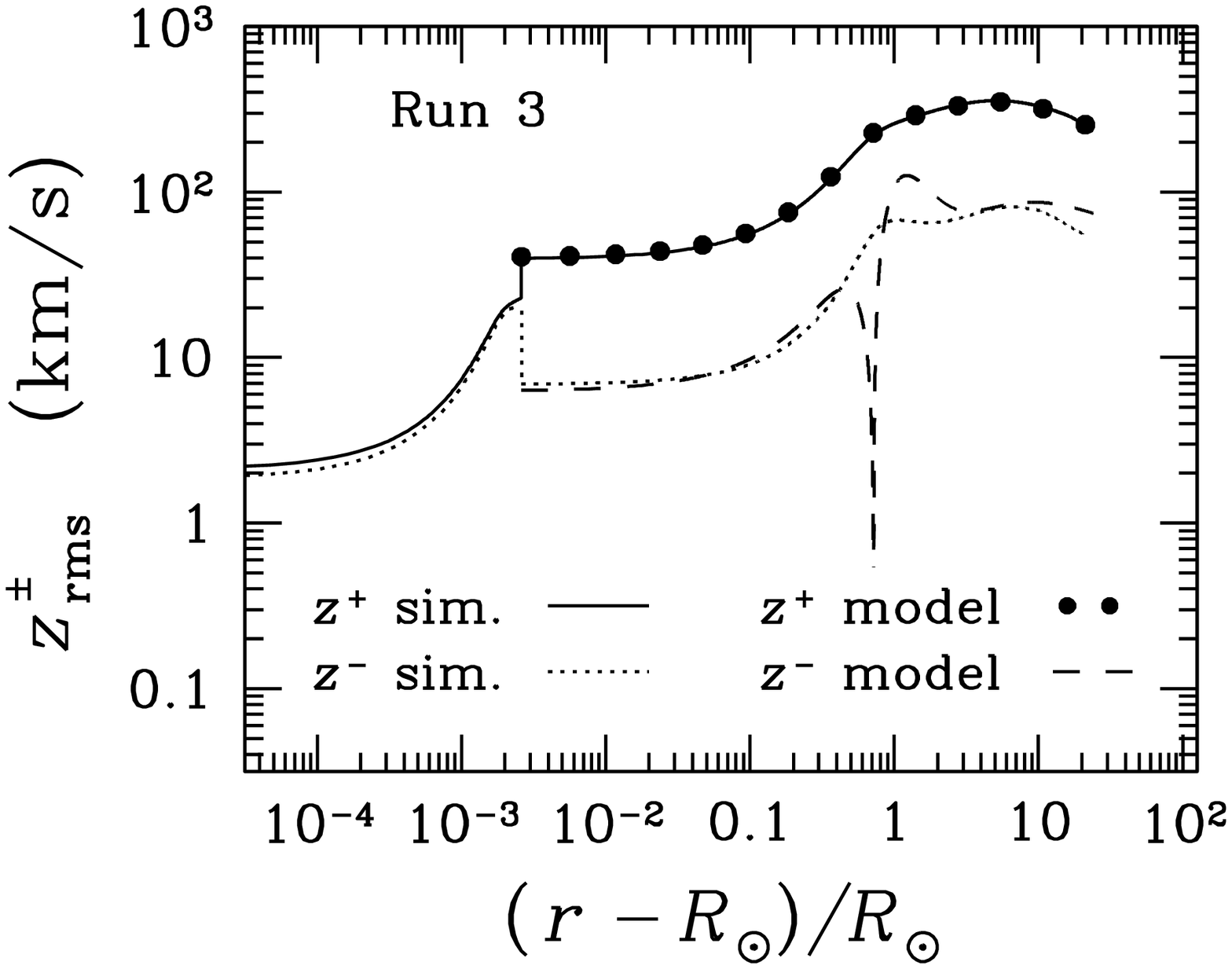}
\hspace{0.3cm} 
\includegraphics[trim = 0cm 4cm 0cm 0cm, width=6cm]{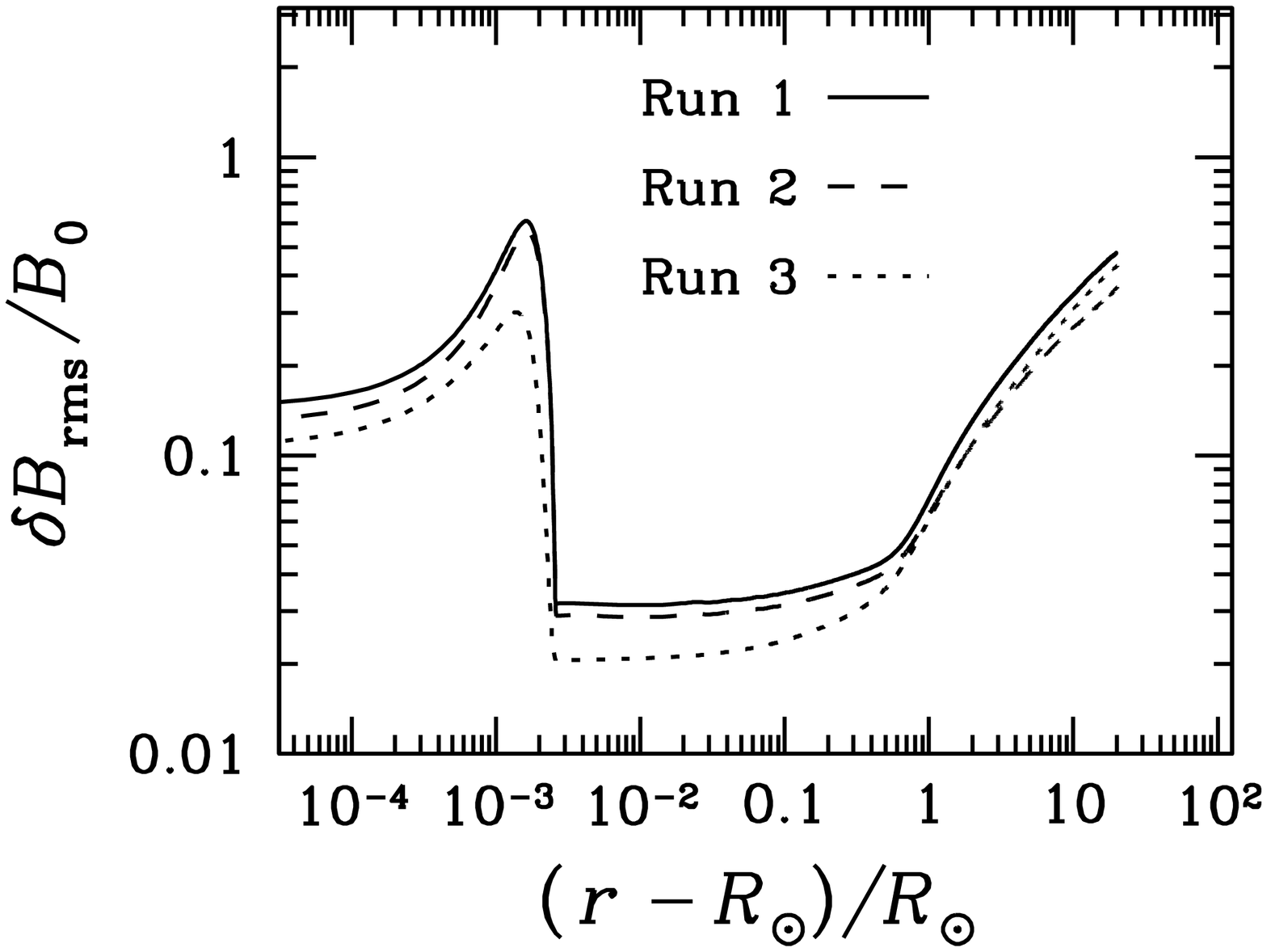}
}
\caption{
The top panels and bottom-left panel show the r.m.s. amplitudes of $\bm{z}^\pm$ in Runs~1
  through~3 and in the analytic model described in Section~\ref{sec:model}.
The lower-right panel shows $\delta B_{\rm rms}/B_0$ in Runs~1 through~3, where
$\delta B_{\rm rms}$ is the r.m.s. amplitude of the magnetic-field
fluctuation.
\label{fig:zpm_profiles} 
}
\end{figure}

In figure~\ref{fig:zpm_profiles}, we plot the r.m.s. amplitudes of
$\bm{z}^\pm$, denoted $z^\pm_{\rm rms}$, as a function of~$r$ in
Runs~1 through~3 and in the analytic model discussed in
Section~\ref{sec:model}. The lower-right panel of
figure~\ref{fig:zpm_profiles} shows the fractional variation in the
magnetic-field strength as a function of~$r$ in our three numerical
simulations. In all three simulations,
$z^+_{\rm rms} \simeq z^-_{\rm rms}$ in the chromosphere, because of
strong AW reflection at the transition region and photosphere. On the
other hand, $z^+_{\rm rms} \gg z^-_{\rm rms}$ in the corona and solar
wind because of the limited efficiency of reflection in these regions
and because $\bm{z}^-$ fluctuations are rapidly cascaded to small scales by
the large-amplitude $\bm{z}^+$ fluctuations.  

The value of $z^+_{\rm rms}$ increases between $r=R_{\odot}$ and
$r=5 R_{\odot}$ because of the radially decreasing density profile.
Equation~(\ref{eq:Hg}) implies that the r.m.s. amplitude of $\bm{g}$
($g_{\rm rms}$) is independent of~$r$ when (i) the fluctuations are in
a statistical steady state, (ii) $z^-_{\rm rms} \ll z^+_{\rm rms}$, and
(iii) nonlinear interactions can be ignored. At $r< 5 R_{\odot}$,
$\rho(r) \gg \rho(r_{\rm A})$, and it follows from (\ref{eq:deffg})
that $z^+_{\rm rms} \propto g_{\rm rms}
\rho^{-1/4}$.
Equations~(\ref{eq:deffg}) and (\ref{eq:Hg}) thus imply that the
linear physics of AW propagation causes $z^+_{\rm rms}$ to increase
rapidly with increasing~$r$ at $r< 5 R_{\odot}$, since
$z^-_{\rm rms} \ll z^+_{\rm rms}$ in this region.  When nonlinear
interactions are taken into account, $g_{\rm rms}$ becomes a
decreasing function of~$r$, but the linear physics ``wins out'' at
$r<5 R_{\odot}$, in the sense that
$z^+_{\rm rms} \propto g_{\rm rms} \rho^{-1/4}$ remains an increasing
function of~$r$. Since the rate of non-WKB reflection vanishes at
$r=r_{\rm m} = 1.71 R_{\odot}$, the $z^-$ fluctuations seen at
$r=r_{\rm m}$ in all three simulations must be generated elsewhere. At
$r<r_{\rm A}$, $z^-$ fluctuations propagate with a negative radial velocity once they
are produced, and thus the $z^-$ fluctuations seen at $r=r_{\rm m}$ in
the simulations originate at $r>r_{\rm m}$.

\begin{figure}
\centerline{
\includegraphics[trim = 0cm 4cm 0cm 0cm, width=6cm]{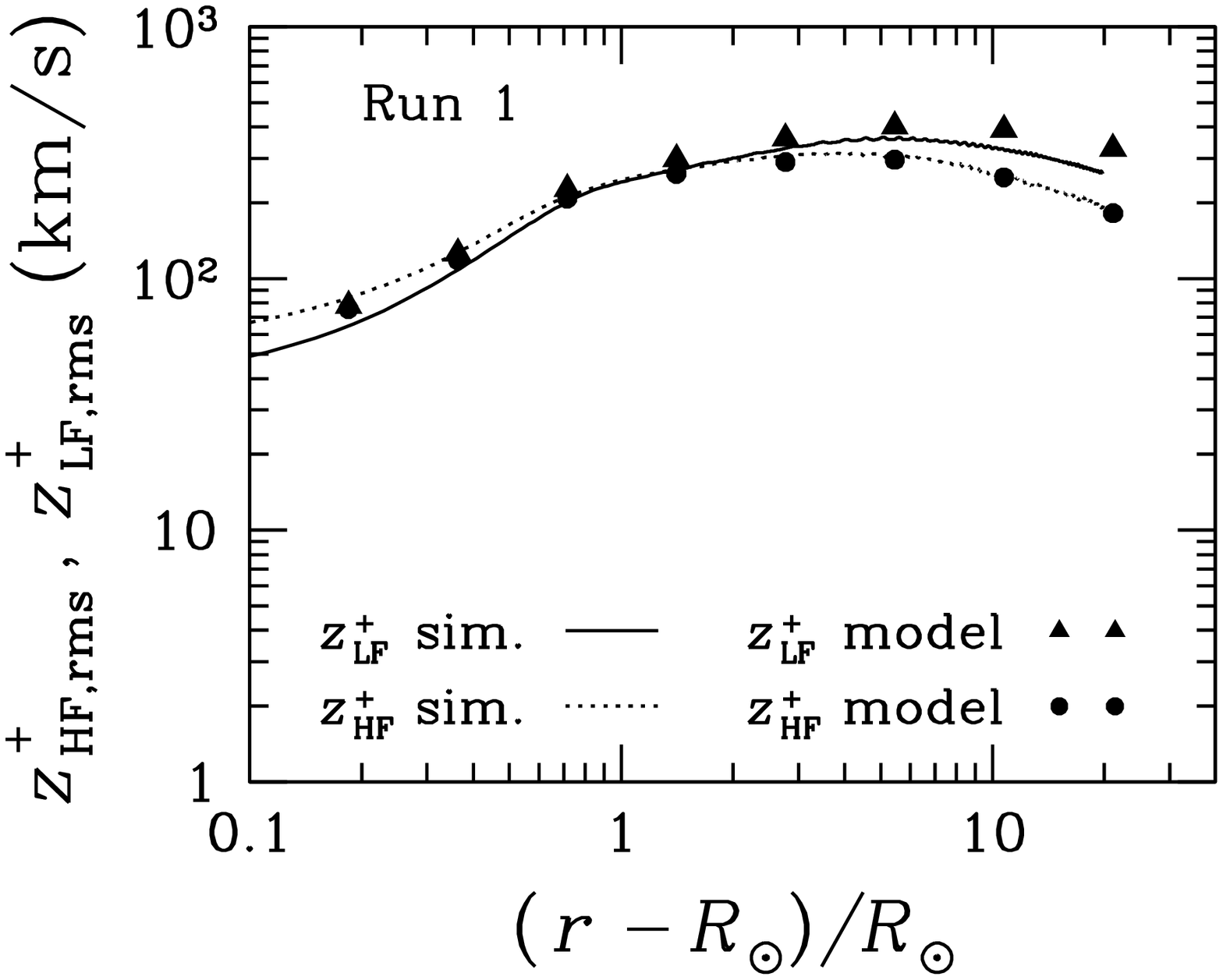}
\hspace{0.3cm} 
\includegraphics[trim = 0cm 4cm 0cm 0cm, width=6cm]{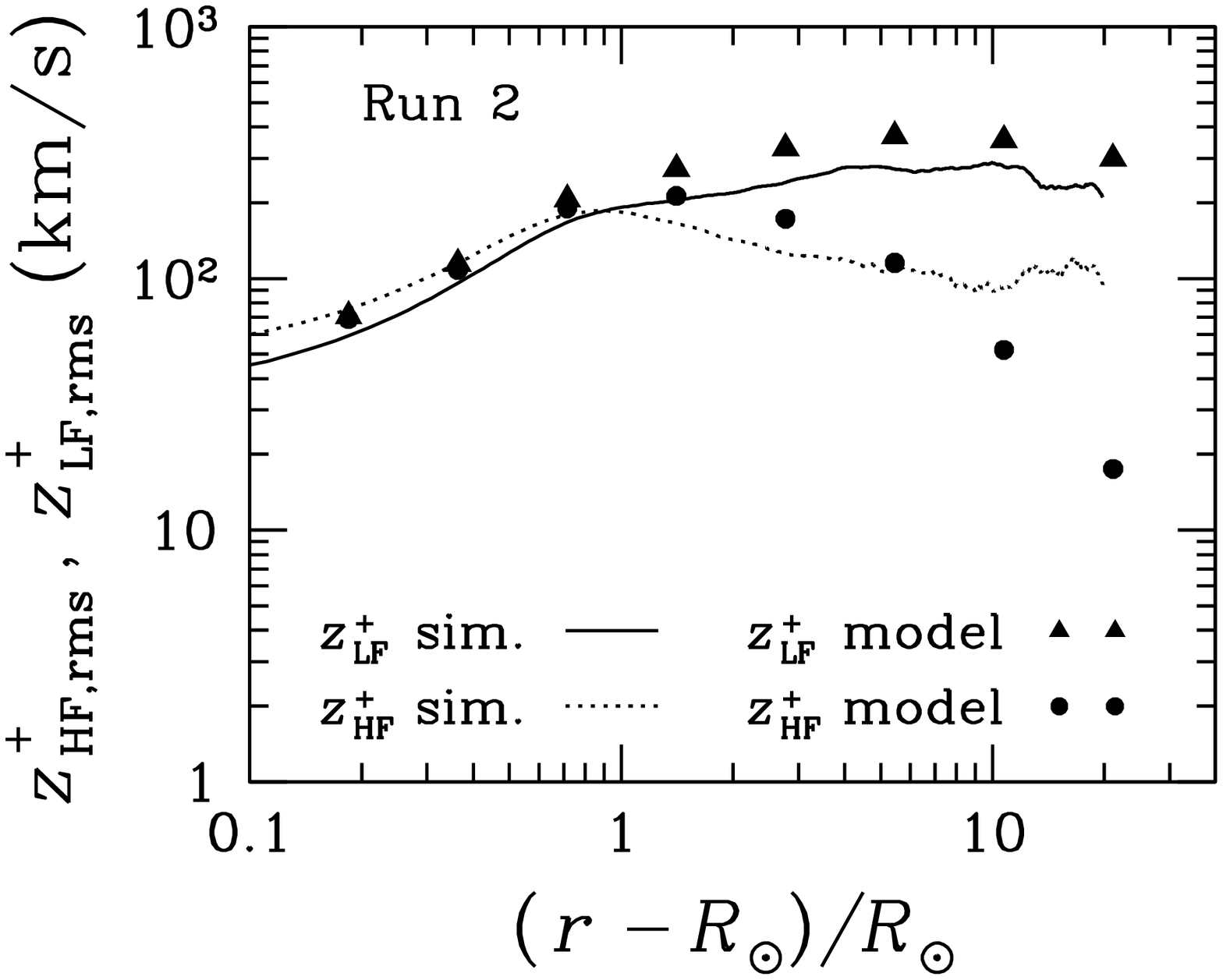}
}
\centerline{
\includegraphics[trim = 0cm 4cm 0cm 0cm, width=6cm]{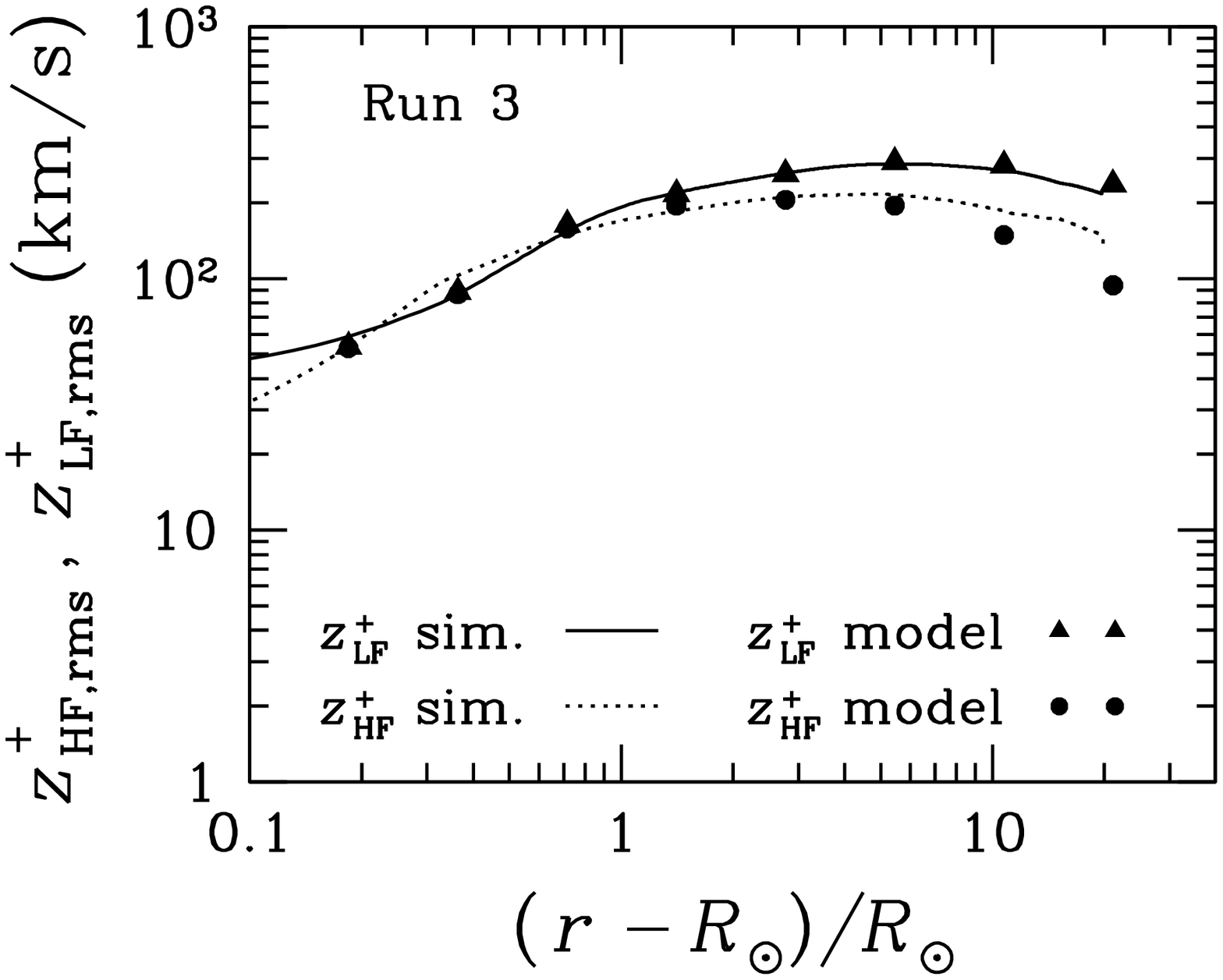}
}
\caption{
R.m.s. amplitudes of $\bm{z}^+_{\rm HF}$ and
$\bm{z}^+_{\rm LF}$ (defined in (\ref{eq:defgLF}) through
(\ref{eq:defgHFrms}) and (\ref{eq:zplusLFHF}))
in Runs~1
  through~3 and in the analytic model described in Section~\ref{sec:model}.
\label{fig:HFLF} 
}
\end{figure}

\subsection{Two components of outward-propagating fluctuations}
\label{sec:twocomp}

In our simulations, the transition region, which acts like an AW antenna, is
characterized by two time scales at the perpendicular outer scale of
the turbulence, which we take to be
\begin{equation}
L_\perp = \frac{1}{3} L_{\rm box}.
\label{eq:defLperp} 
\end{equation} 
The first time scale is the
correlation time of the photospheric velocity field, $\tau_{v}^{\rm (ph)}$,
which we define as the time increment required for the normalized velocity
autocorrelation function at the photosphere to decrease from 1 to 0.5.
This time increment is 3.3~min, 9.6~min, and
9.3~min in Runs 1, 2, and~3, respectively, as displayed in
table~\ref{tab:parameters}.
The second time scale is the nonlinear time scale 
\begin{equation}
\tau_{\rm nl} = \frac{L_{\perp}}{z^\pm_{\rm rms}}
\label{eq:taunl0} 
\end{equation} 
of the balanced
turbulence (``balanced'' meaning that $z^+_{\rm rms} \simeq z^-_{\rm rms}$) just below the
transition region at $r = r_{\rm tr, <} = r_{\rm tr} - \epsilon$,
where $\epsilon$ is an infinitesimal distance,
and $z^\pm_{\rm rms}(r_{\rm tr, <}) \simeq  30 \mbox{ km/s}$. (Section~\ref{sec:powerspectra}  discusses an effect that shortens this second time scale relative to the estimate in~(\ref{eq:taunl0}) in Runs~1 and~2.)
Although the right-hand side of~(\ref{eq:taunl0})  contains a
$\pm$ sign, we do not include a $\pm$ sign on the left-hand side, because we
will only evaluate~(\ref{eq:taunl0}) at locations at which
$z^+_{\rm rms} \simeq z^-_{\rm rms}$. 
We define
\begin{equation}
\tau_{\rm nl}^{\rm (tr)} = \tau_{\rm nl}(r_{\rm tr, <}).
\label{eq:taunltr} 
\end{equation} 
Given the values of $L_{\rm box}(r_{\rm tr})$ listed in
table~\ref{tab:parameters}, $\tau_{\rm nl}^{\rm (tr)}$ is 0.8~min, 0.8~min,
and 3~min in Runs~1, 2, and~3, respectively, values that are several
times smaller than~$\tau_{v}^{\rm (ph)}$. This suggests
that the transition region in our simulations launches two populations of $\bm{z}^+$
fluctuations characterized by different time scales and hence different
radial correlation lengths.

To investigate this possibility, we define
\begin{equation} 
\bm{g}_{\rm LF}(\tilde{x},\tilde{y},r,t) = 
\frac{1}{2\Delta}\int_{r_{\rm i}}^{r_{\rm i} +2\Delta} \d r^\prime \bm{g}\left(\tilde{x}, \tilde{y},
r^\prime,t\right) ,
\label{eq:defgLF} 
\end{equation} 
\begin{equation}
g_{\rm LF, rms} = \left\langle|\bm{g}_{\rm LF}|^2\right\rangle^{1/2} ,
\label{eq:defgLFrms} 
\end{equation} 
and 
\begin{equation}
g_{\rm HF, rms} = \sqrt{g_{\rm rms}^2 - g_{\rm LF, rms}^2},
\label{eq:defgHFrms} 
\end{equation} 
where $\tilde{x} = x/L_{\rm box}$, $\tilde{y} = y/L_{\rm box}$, and
$\langle\dots\rangle$ denotes an average over $x$, $y$, and~$t$. The quantity
\begin{equation}
\Delta = c_{\rm av} \tau_{\rm nl}^{\rm (tr)} v_{\rm A}(r_{\rm b})
\label{eq:defDelta} 
\end{equation} 
is the approximate radial correlation length in the low corona of a $\bm{z}^+$ fluctuation
that is generated by a disturbance at the transition region 
whose correlation time is~$\tau_{\rm nl}^{\rm (tr)}$,
\begin{equation}
r_{\rm i} = \left\{ \begin{array}{ll}
r_{\rm min} & \mbox{ if $r< r_{\rm min} + \Delta$} \\
r - \Delta & \mbox{ if $r_{\rm min} + \Delta \leq r \leq r_{\rm max} -
             \Delta $} \\
r_{\rm max} - 2\Delta &  \mbox{ if $r> r_{\rm max} - \Delta$}
\end{array}
\right.
\label{eq:ri} 
\end{equation} 
and $c_{\rm av}$ is a
dimensionless constant of order unity. We set 
\begin{equation}
c_{\rm av} \simeq 0.6,
\label{eq:cav} 
\end{equation} 
which enables us to carry out the radial average in~(\ref{eq:defgLF}) in a computationally efficient way,
using an integer number of subdomains. Given the above definitions, 
$\Delta =  0.08 R_{\odot}$ in Runs~1 and~2, and $\Delta = 0.32 R_{\odot}$
in Run~3. We define
\begin{equation}
z^+_{\rm LF, rms} = \frac{\eta^{1/4} g_{\rm LF, rms}}{1 + \eta^{1/2}} \qquad 
z^+_{\rm HF, rms} = \frac{\eta^{1/4} g_{\rm HF, rms}}{1 + \eta^{1/2}} .
\label{eq:zplusLFHF}
\end{equation}
We emphasize that, although we use the subscripts ``LF'' and ``HF'' as shorthand for ``low-frequency'' and ``high-frequency,'' the defining difference between $z^+_{\rm LF,rms}$ and $z^+_{\rm HF, rms}$ is the difference in their radial correlation lengths.

In figure~\ref{fig:HFLF} we plot the radial profiles of $z^+_{\rm LF, rms}$ and $z^+_{\rm HF, rms}$ in our numerical simulations and the analytic model of Section~\ref{sec:model}.
As this figure shows, all three
simulations contain both $\bm{z}^+_{\rm LF}$ and $\bm{z}^+_{\rm HF}$
fluctuations, and these fluctuations evolve in different ways as
they propagate away from the Sun.
In all three runs, $z^+_{\rm  HF,rms} \simeq z^+_{\rm LF,rms}$ in the low corona.
As $r$ increases, $z^+_{\rm HF,rms}/z^+_{\rm LF,rms}$ decreases,
particularly in Run~2,
suggesting that the high-frequency component of $\bm{z}^+$
cascades and dissipates more rapidly than the low-frequency component.

\subsection{Alignment}
\label{sec:alignment}

\begin{figure}
\centerline{
\includegraphics[trim = 0cm 4cm 0cm 0cm, width=6cm]{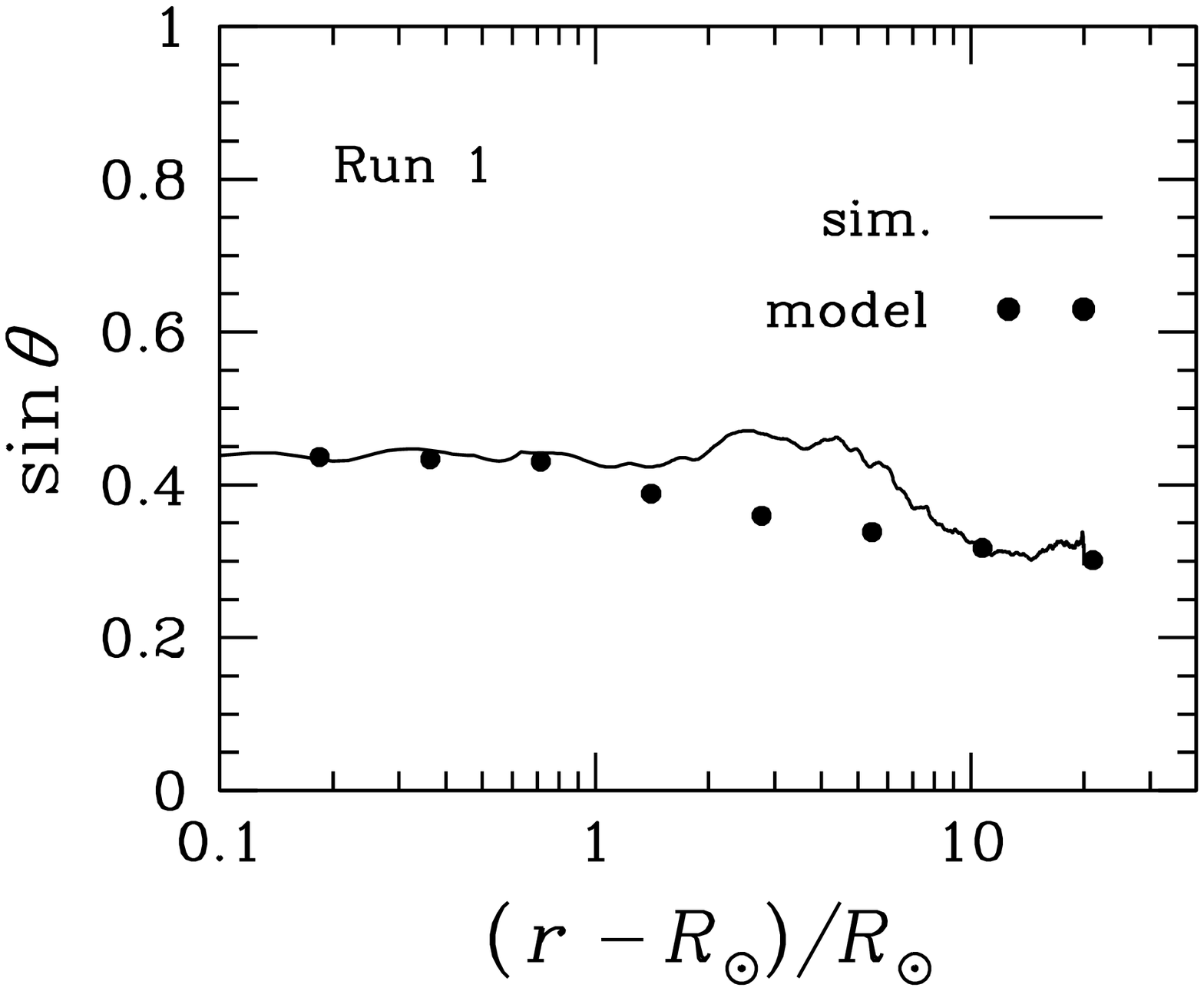}
\hspace{0.3cm} 
\includegraphics[trim = 0cm 4cm 0cm 0cm, width=6cm]{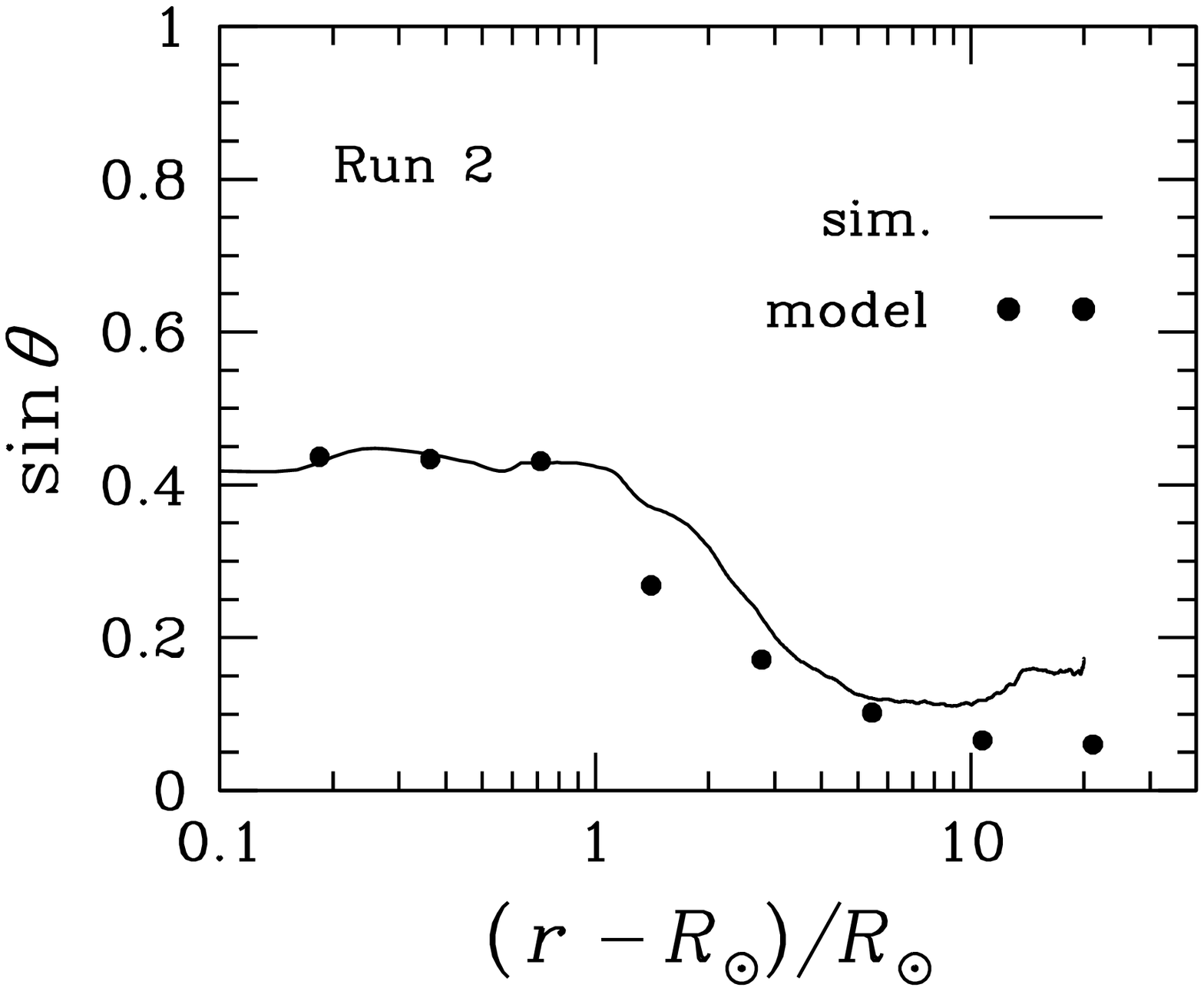}
}
\centerline{
\includegraphics[trim = 0cm 4cm 0cm 0cm, width=6cm]{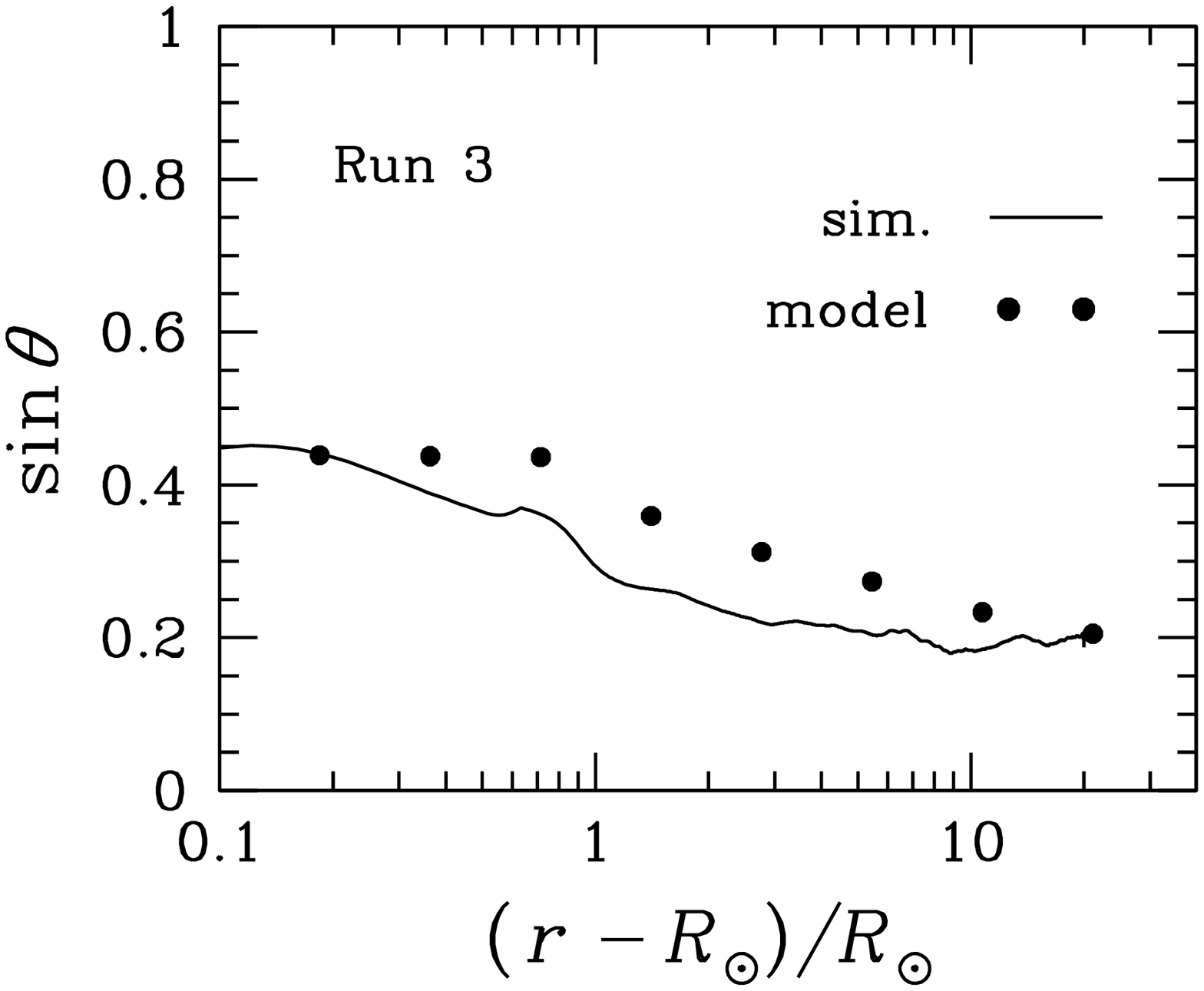}
}
\caption{The characteristic value of the sine of the alignment angle~$\theta$ between
$\bm{z}^+$ and $\bm{z}^-$, defined in (\ref{eq:defsinthetarms}), in Runs 1
through~3 and in the analytic model of Section~\ref{sec:model} (using (\ref{eq:sinthetaA})).
\label{fig:Theta_2comp} 
}
\end{figure}

Figure~\ref{fig:Theta_2comp}  shows the characteristic value of the
sine of the angle
between $\bm{z}^+$ and $\bm{z}^-$, 
\begin{equation}
\sin\theta = \frac{\langle |\bm{z}^+ \times
  \bm{z}^-|\rangle}{\langle |\bm{z}^+|\rangle \langle |
  \bm{z}^-|\rangle},
\label{eq:defsinthetarms} 
\end{equation} 
in both our numerical simulations and the model we present in
Section~\ref{sec:model}.  As $r$ increases, $\sin\theta$ decreases,
particularly in Run~2, causing nonlinear interactions between
$\bm{z}^+$ and $\bm{z}^-$ to weaken \citep[see,
e.g.,][]{boldyrev05,boldyrev06,perez13,chandran15}.\footnote{We note
  that a different alignment angle, between $\delta\bm{v}$ and
  $\delta \bm{b}$ fluctuations, was the basis for Boldyrev's~(2006)
  theory of scale-dependent dynamic alignment.\nocite{boldyrev06}}

\subsection{Turbulent heating}
\label{sec:TH} 

\begin{figure}
\centerline{
\includegraphics[trim = 0cm 4cm 0cm 0cm, width=6cm]{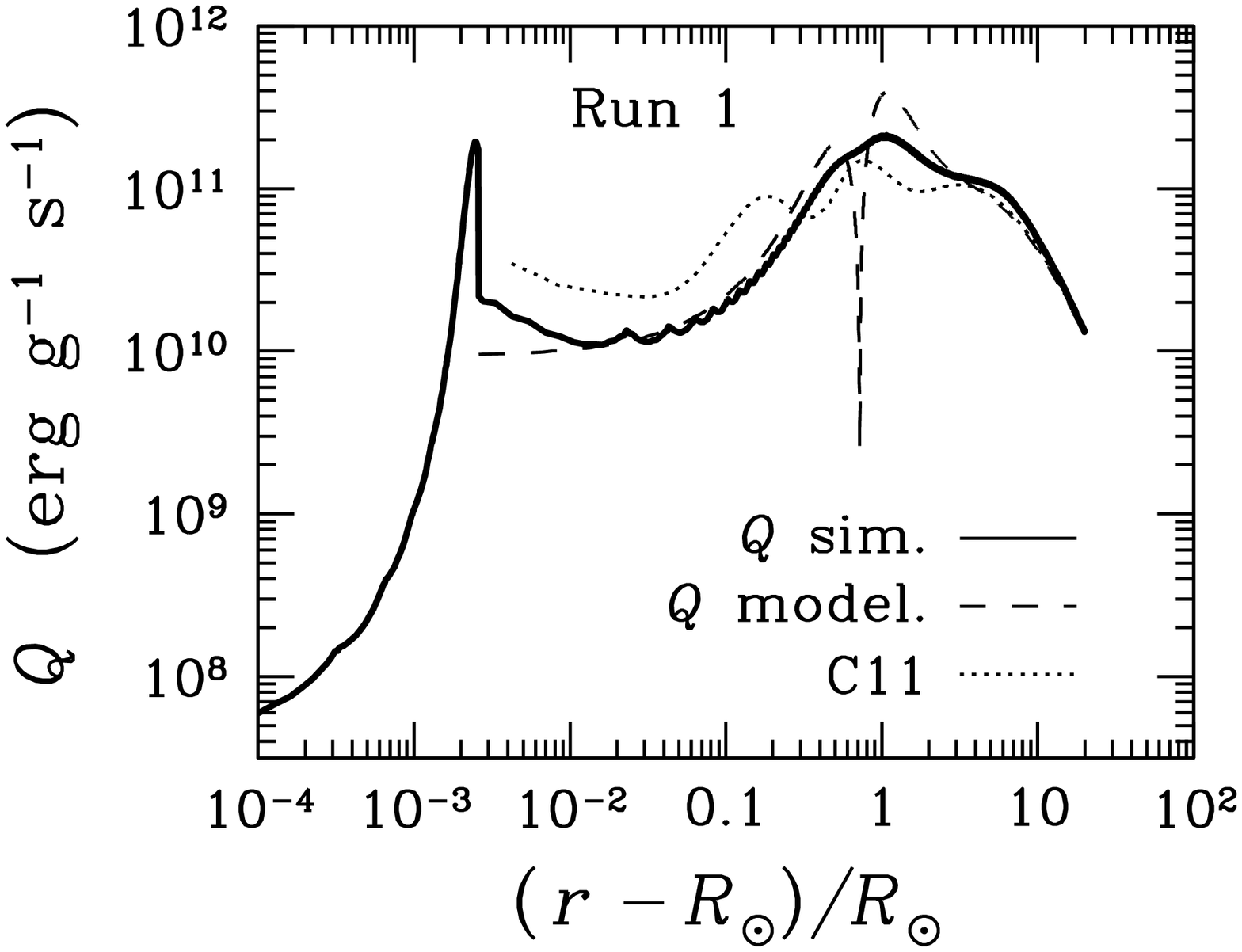}
\hspace{0.3cm} 
\includegraphics[trim = 0cm 4cm 0cm 0cm, width=6cm]{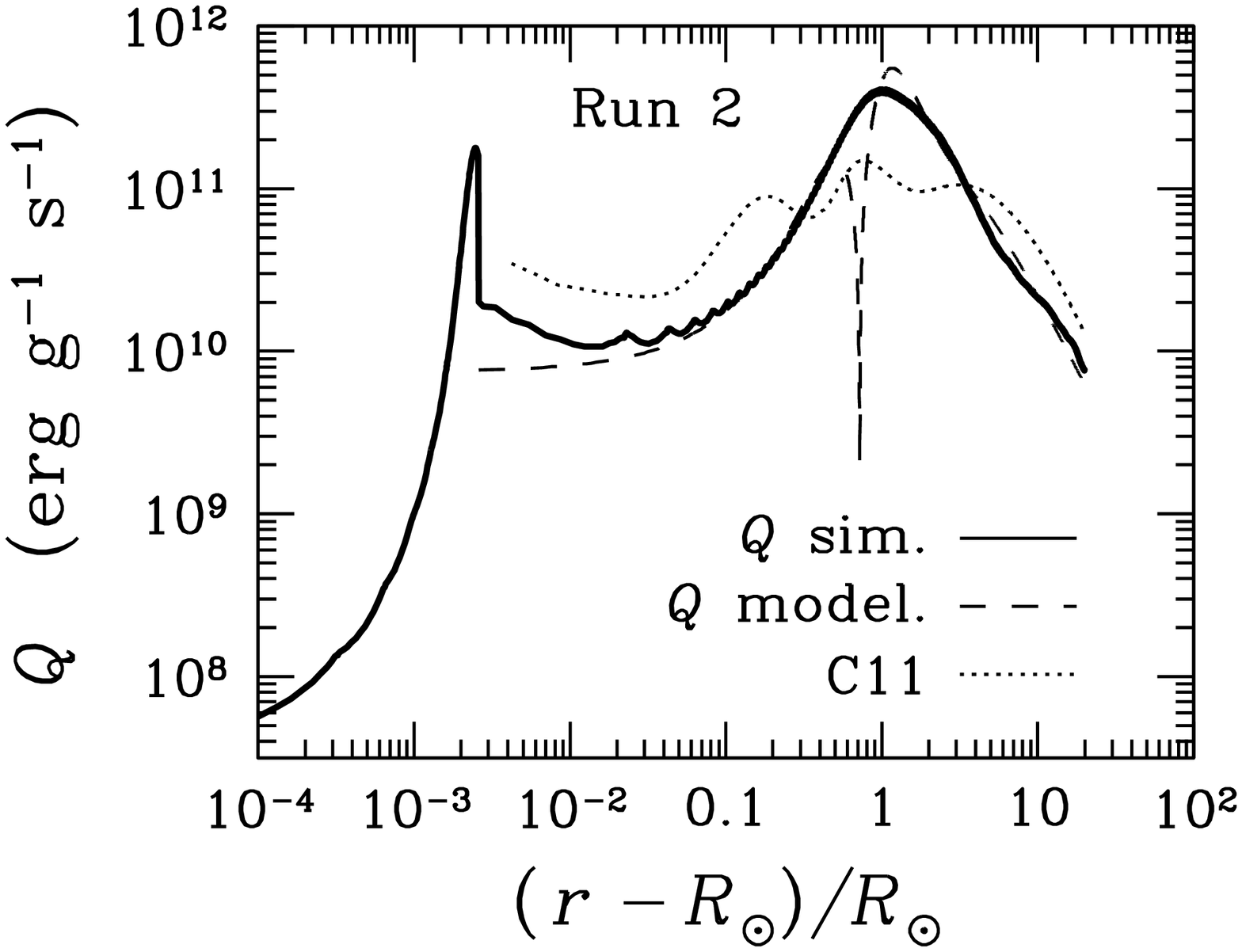}
}
\centerline{
\includegraphics[trim = 0cm 4cm 0cm 0cm, width=6cm]{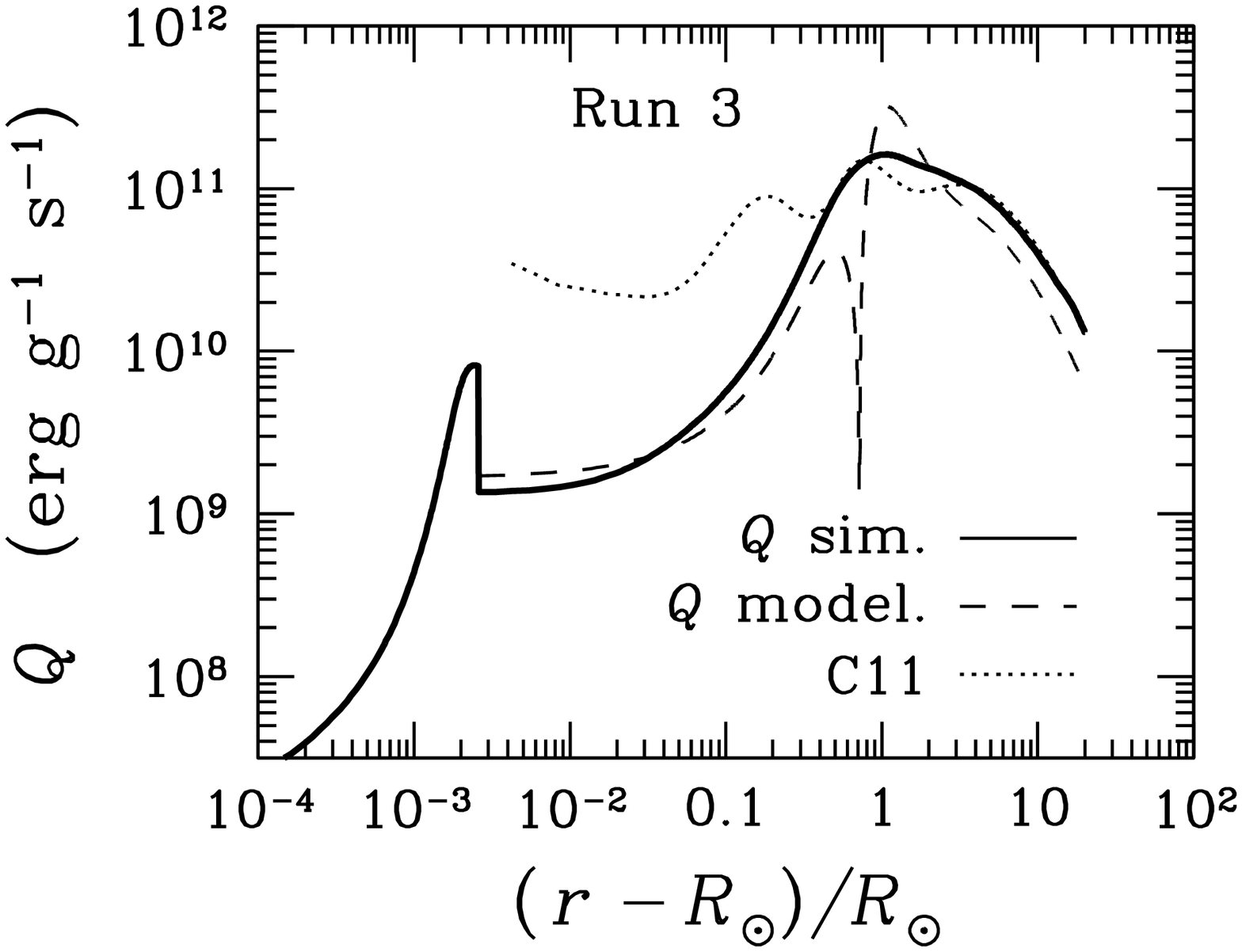}
}
\caption{The turbulent-heating rate per unit mass~$Q$ in Runs 1
through~3 and in the analytic model of Section~\ref{sec:model}. The
dotted line labeled C11 is the turbulent-heating rate in the
solar-wind model of \cite{chandran11}, which approximates the heating
needed to power the fast solar wind.
\label{fig:LAQtot} 
}
\end{figure}

In figure~\ref{fig:LAQtot}  we plot the
rate~$Q$ at which energy is dissipated per unit mass by hyper-dissipation
in our simulations
\cite[see][]{perez13} as a function of~$r$, as well as the turbulent
heating rate in the analytic model described in Section~\ref{sec:model}.
The amplitudes of the turbulent fluctuations in our simulations are
consistent with the results of several observational studies that were
summarized in figure~9 of \cite{cranmer05}, including non-thermal line widths in coronal holes inferred from
{\em SUMER} (Solar Ultraviolet Measurements of Emitted Radiation) and
{\em UVCS} (Ultraviolet Coronagraph Spectrometer) measurements
\citep{banerjee98,esser99}.  For comparison, the r.m.s. amplitudes of the
fluctuating velocity~$\delta v_{\rm rms}$ at $r=r_{\rm tr}$ in Runs~1,
2, and~3 are, respectively, $30.4$~km/s, $30.0$~km/s, and
$26.7$~km/s.\footnote{In contrast to $\bm{z}^\pm$, $\delta \bm{v}$ is
  continuous across the transition region, and it makes no difference
  whether we evaluate $\delta v_{\rm rms}$
  at, just above,
  or just below the transition region.}
The values of $\delta v_{\rm rms}$ at $r=2R_{\odot}$ in Runs~1, 2,
and~3 are, respectively, $170$~km/s, $157$~km/s, and $146$~km/s.
Because the turbulence amplitudes in our simulations are consistent
with the aforementioned observations, the turbulent-heating rate in
each of our simulations can be used to estimate the rate at which
transverse, non-compressive MHD
turbulence would heat the solar wind as a function of~$r$ if the
correlation lengths and correlation time at $r=r_{\rm b}$ in the
simulation were realistic.\footnote{An important caveat to this
  statement is that we have neglected the interaction between non-compressive
  fluctuations and compressive fluctuations, including phase mixing,
  which we discuss in Section~\ref{sec:phase}.}

To estimate the amount of turbulent heating that would be needed to
power the solar wind, we also plot in figure~\ref{fig:LAQtot}  
the turbulent-heating rate in the one-dimensional (flux-tube) solar-wind
model of \cite{chandran11}. This
model included Coulomb collisions,
super-radial expansion of the magnetic field,
separate energy equations for the protons and electrons,
proton temperature anisotropy, a transition between Spitzer
conductivity near the Sun and a Hollweg collisionless heat flux at
larger~$r$, and enhanced pitch-angle scattering by
temperature-anisotropy instabilities in regions in which the plasma is
either mirror or firehose unstable. The model agreed with a number of
remote observations of coronal holes and in-situ measurements of fast-solar-wind streams.

The turbulent-heating rate in the \cite{chandran11} model, which we
denote~$Q_{\rm C11}$, is for the most part comparable to (i.e., within a factor of 3 of)
the heating rate in our numerical simulations. The simulated
heating rates in Runs~1 and~3 are in fact strikingly close to~$Q_{\rm C11}$ at
$r\gtrsim 4R_{\odot} $.  However, in all three runs, $Q > Q_{\rm C11}$ at $r = 2
- 3 R_{\odot}$. This latter discrepancy is largest in the case of Run~2,
in which $Q \simeq 3 Q_{\rm C11}$ at $r=2 R_{\odot}$. Although Run~2
has the largest heating rate of all three simulations at $r=2 R_{\odot}$,
the simulated heating rate in Run~2 is smaller than~$Q_{\rm C11}$ at
$r\gtrsim 5 R_{\odot}$ by a factor of~$\sim 2$. 

The only region in which the simulated heating rate differs from
$Q_{\rm C11}$ by a factor $\gtrsim 4$ is at
$r< 1.3 R_{\odot}$ in Run~3, where $Q/Q_{\rm C11}$ falls below~0.1. Even in Runs~1 and~2, the simulated
heating rate at $r< 1.3R_{\odot}$
is smaller than~$Q_{\rm C11}$ by a factor of~$\sim 2$. The finding
that $Q\lesssim 0.5 Q_{\rm C11}$ at $r< 1.3 R_{\odot}$ in all three runs
may indicate the presence
of additional heating mechanisms in the actual low corona, such as
 compressive fluctuations, a possibility previously considered by 
\cite{cranmer07} and \cite{verdini10}.

Recently, \cite{vanballegooijen16,vanballegooijen17} carried out a
series of direct numerical simulations of reflection-driven MHD
turbulence and concluded that such turbulence is unable to provide
enough heating to power the solar wind.  The reason we reach a
  different conclusion is likely that we use the two-fluid solar-wind model of
\cite{chandran11} to estimate the amount of heating required, whereas
\cite{vanballegooijen16,vanballegooijen17} used a one-fluid solar-wind model (A. van Ballegooijen, private
communication). In the \cite{chandran11} two-fluid model, the electron
temperature is lower than the proton temperature, and thus
less heat is conducted back to the chromosphere than in a one-fluid
solar-wind model.

\subsection{Simulation results: Elsasser power spectra}
\label{sec:powerspectra} 

\begin{figure}
\centerline{
\includegraphics[trim = 0cm 0cm 0cm 0cm, width=6cm]{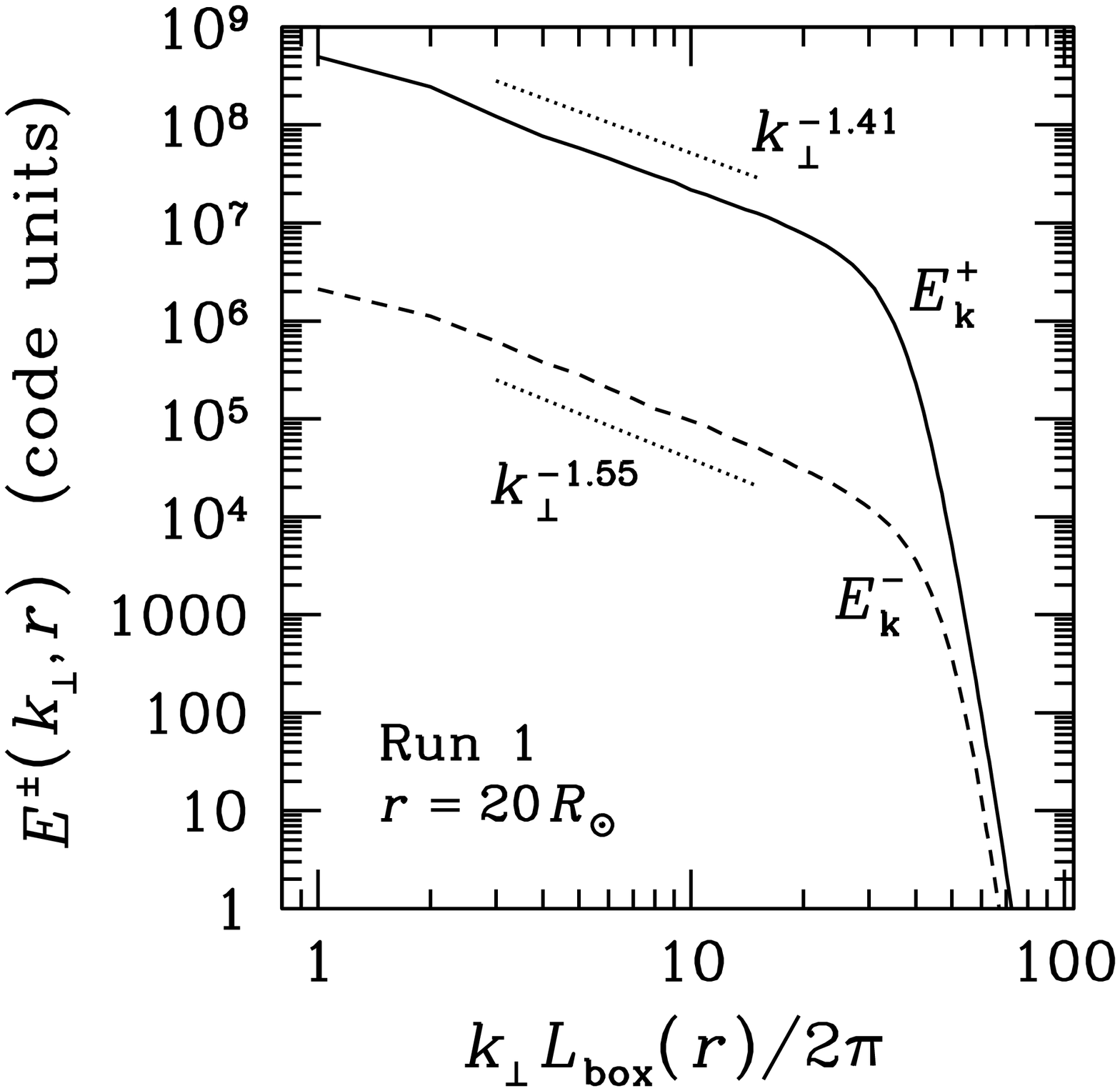}
\hspace{0.5cm} 
\includegraphics[trim = 0cm 4cm 0cm 0cm, width=6cm]{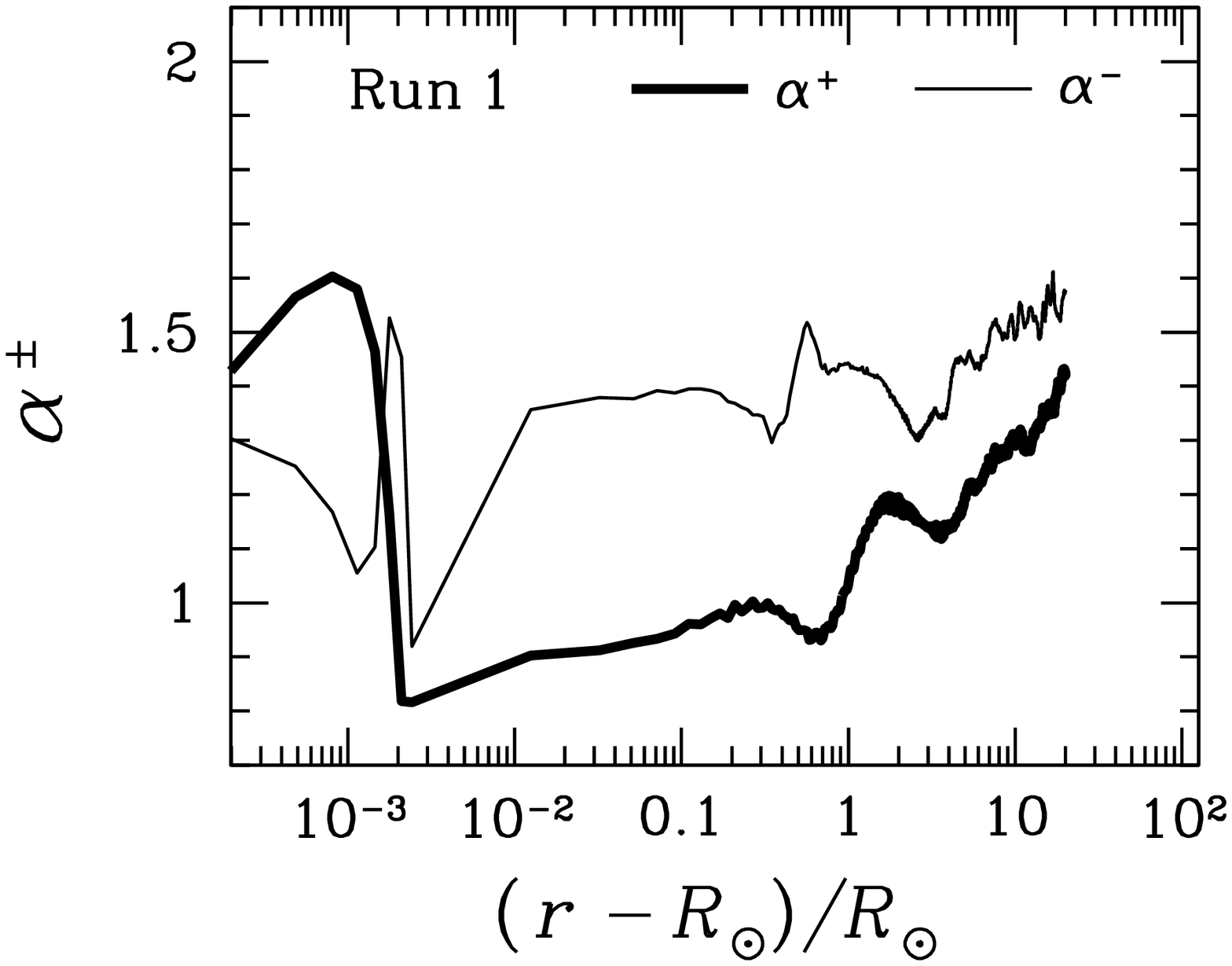}
}
\centerline{
\includegraphics[trim = 0cm 4cm 0cm 0cm, width=6cm]{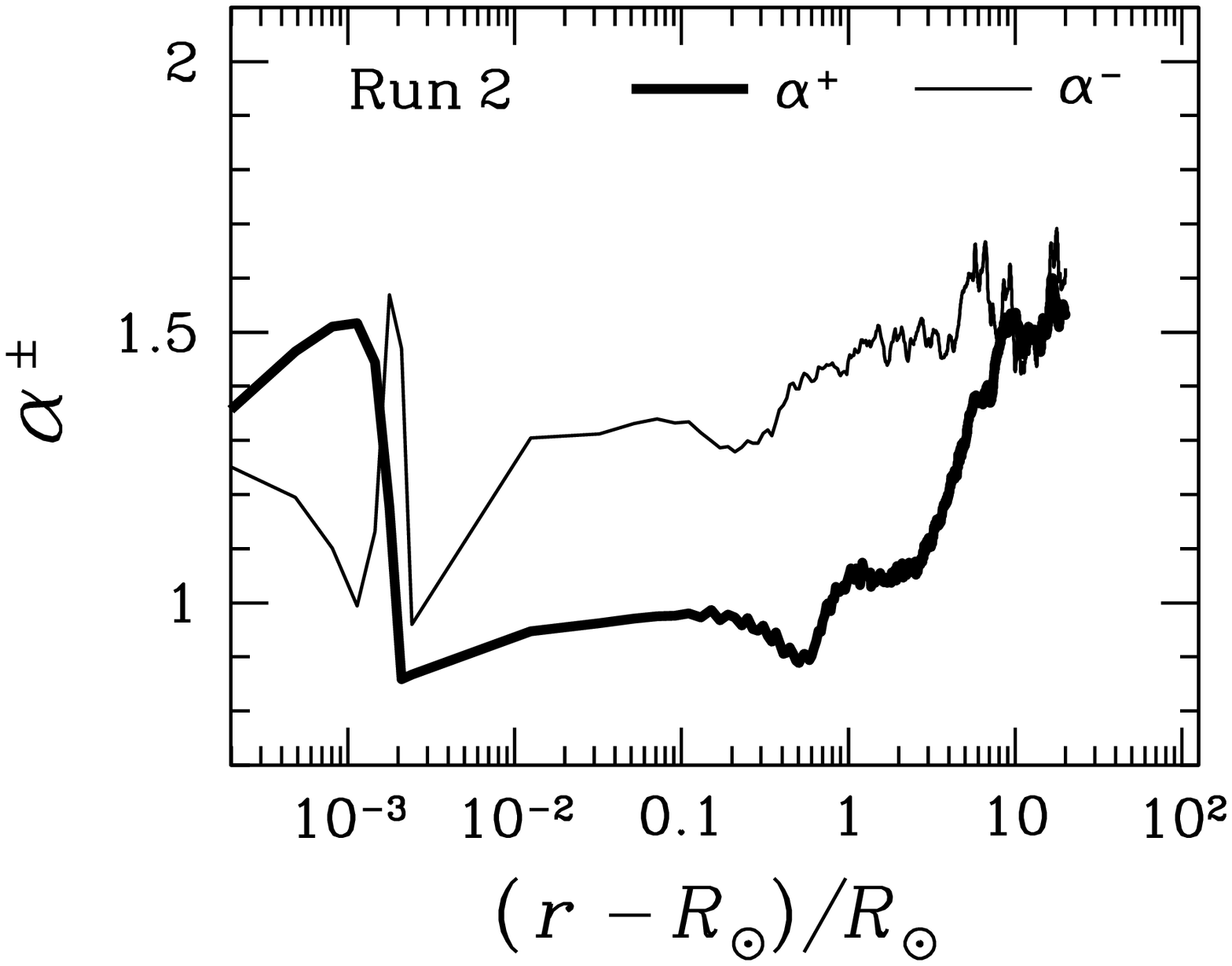}
\hspace{0.5cm} 
\includegraphics[trim = 0cm 4cm 0cm 0cm, width=6cm]{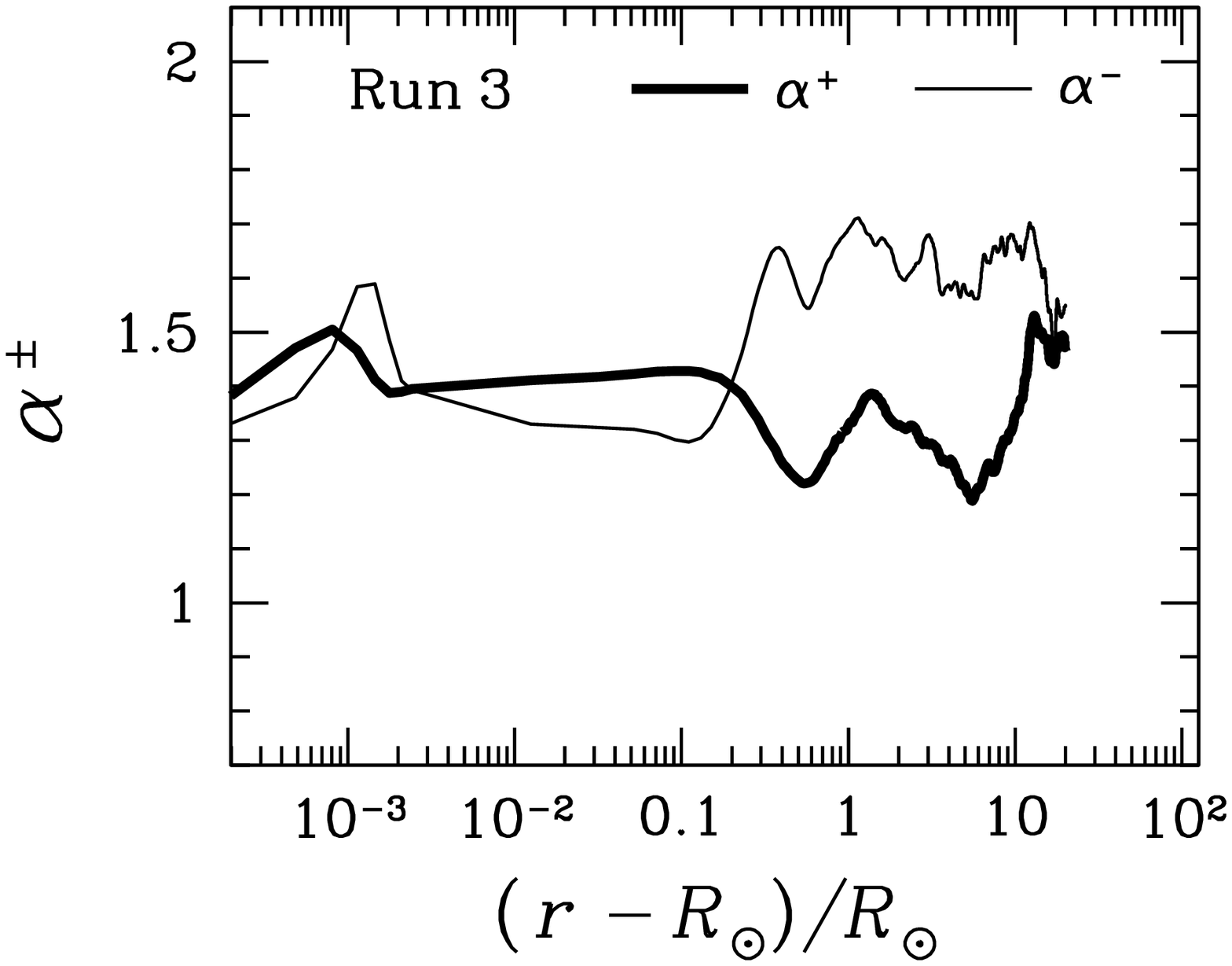}
}
\caption{\label{fig:alpha_pm}
Top-left panel: the Elsasser power spectra~$ E^\pm(k_\perp,r)$ defined in
 ~(\ref{eq:defEpm}) as 
  functions of perpendicular wavenumber~$k_\perp$ at $r=20 R_{\odot}$
  in Run~1. Top-right and bottom panels:
the spectral indices $\alpha^+(r)$ and~$\alpha^-(r)$ defined in (\ref{eq:defalpha}) in our three numerical simulations.
}
\end{figure}

We define the perpendicular Elsasser power spectra 
\begin{equation}
E^\pm(k_\perp,r) = k_\perp \int_{0}^{2\pi} \d\phi \overline{
\left|\bm{\tilde{z}}^\pm(k_\perp,\phi,r,t)\right|^2 },
\label{eq:defEpm} 
\end{equation} 
where $\bm{\tilde{z}}^\pm(k_\perp, \phi)$ is the Fourier transform
of~$\bm{z}^\pm$ in~$x$ and~$y$ (see figure~\ref{fig:clebsch}), $\phi$
is the polar angle in the $(k_x, k_y)$ plane, and 
$\overline{(\dots)}$ indicates a time average. 
As illustrated in the top-left panel of figure~\ref{fig:alpha_pm}, we find that $E^\pm(k_\perp)$ exhibits an approximate
power-law scaling of the form
\begin{equation}
E^\pm (k_\perp,r) \propto k_\perp^{-\alpha^\pm(r)}
\label{eq:defalpha} 
\end{equation}
from $k_\perp \simeq 3 \times 2\pi/L_{\rm box}$ to 
$k_\perp \simeq 15\times 2\pi/L_{\rm box}$ at all~$r$ in all three of our
simulations. We evaluate $\alpha^\pm(r)$ by fitting $E^\pm(k_\perp,r)$ to
a power law within this range of wave numbers, and plot the resulting
values of $\alpha^\pm(r)$ in figure~\ref{fig:alpha_pm}.  

Although we drive only large-scale
($k_\perp \leq 3 \times 2\pi/L_{\rm box}$) velocity fluctuations at
the photosphere, figure~\ref{fig:alpha_pm} shows that there is
broad-spectrum turbulence throughout the chromosphere. This is because
of the strong reflection of $\bm{z}^+$ fluctuations at the transition
region and the strong reflection of~$\bm{z}^-$ fluctuations (at
all~$k_\perp$) at the photosphere (enforced by the fixed-velocity
boundary condition at $r=R_{\odot}$), which together lead to
comparatively ``balanced'' turbulence (meaning that
$z^+_{\rm rms} \simeq z^-_{\rm rms}$) in the chromosphere, as shown
previously by \cite{vanballegooijen11}.  In the low chromosphere,
$\alpha^\pm \simeq 1.3 - 1.5$ in all four simulations, which is
similar to the value $\alpha^\pm \simeq 3/2$ that arises in numerical
simulations of homogeneous, balanced, RMHD turbulence
\citep{mason08,perez12,beresnyak12}. On the other hand, $\alpha^+$
decreases from $\simeq 1.5$ to $\simeq 0.8$ as $r$ increases from
$1.001 R_{\odot}$ to $r_{\rm tr} = 1.0026 R_{\odot}$ in Runs~1
and~2. This spectral flattening arises because the Alfv\'en-speed
gradient in the upper chromosphere acts as a high-pass filter on
outward-propagating~AWs in Runs~1 and~2, causing lower-$k_\perp$ (and
hence lower-frequency -- see \cite{goldreich95}) $\bm{z}^+$
fluctuations to undergo non-WKB reflection, and allowing
higher-$k_\perp$ (and hence higher-frequency) $\bm{z}^+$ fluctuations
to propagate unhindered to the transition region
\citep{velli93,reville18}.  The difference in Run~3 is that
$L_{\rm box}$ is larger, and thus $\bm{z}^+$ fluctuations do not reach
sufficiently large $k_\perp$ values that they can avoid non-WKB
reflection in the upper chromosphere.  \footnote{The transition
  region in our simulations is a density discontinuity, which reflects
  $\bm{z}^+$ fluctuations with an efficiency that is independent
  of~$k_\perp$ \citep{vanballegooijen11}; reflection at the transition
  region helps explain why $\alpha^-$ is comparatively small in the
  upper chromosphere in Runs~1 and~2, but it does not act like a
  high-pass filter in our simulations.}  The idea that $\bm{z}^+$
fluctuations at the high-$k_\perp$ end of the inertial range propagate
through the chromosphere more easily in Runs~1 and~2 than in Run~3 is
consistent with the fact that $z^+_{\rm rms}(r_{\rm b})$ and
$\delta B_{\rm rms}$ are somewhat larger in Runs~1 and~2 than in Run~3
(see table~\ref{tab:parameters} and figure~\ref{fig:zpm_profiles}).

As $r$ increases from $r_{\rm tr}$ to $r_{\rm A}$ and beyond,
$\alpha^\pm$ approaches $\simeq 3/2$ in all three runs. In Runs~1
and~2, the increase in $\alpha^+$ as $r$ increases from $r_{\rm tr}$
to $r_{\rm A}$ is not steady. In Run~1, $\alpha^+$ decreases as $r$
increases from $2.8 R_{\odot}$ to $4.2 R_{\odot}$, and in Run~2,
$\alpha^+$ plateaus around a value of~1 between $r=2 R_{\odot}$ and
$r=3R_{\odot}$. This behavior suggests that, in these two simulations,
the turbulent dynamics at
$2 R_{\odot} \lesssim r \lesssim 4 R_{\odot}$ drives $\alpha^+$ towards
a value close to~1, and the tendency for $\alpha^+$ to evolve
towards~3/2 sets in at $r\gtrsim 4 R_{\odot}$. We discuss these trends
further in Section~\ref{sec:124}.

\section{Two-component analytic model of reflection-driven MHD turbulence}
\label{sec:model} 

\cite{chandran09c} (hereafter CH09) developed an analytic model of
reflection-driven MHD turbulence in the solar corona and solar
wind. This model can reproduce the radial profile of
$z^+_{\rm rms}$ in our numerical simulations fairly accurately,
provided the constant~$\chi$ introduced in Equation~(34) of CH09 is
treated as an adjustable free parameter that is allowed to take on
different values in different simulations. With the best-fit values
of~$\chi$, the CH09 model also reproduces the turbulent-heating
profiles in Runs~1 and~2 reasonably well. However, the model is
significantly less accurate at reproducing~$Q(r)$ in Run~3 and
deviates markedly from the $z^-_{\rm rms}$ profiles in all three runs.
Moreover, the CH09 model does not explain the differences between the
best-fit values of~$\chi$ for Runs~1, 2, and~3 (which are,
respectively, 0.55, 0.72, and 0.36), or explain how these values can
be determined from the perpendicular correlation length and
correlation time of the AWs launched by the Sun. These shortcomings
indicate that there are important physical processes operating in our numerical
simulations that were not accounted for by CH09.

In order to elucidate these processes, we develop a new
analytic model of reflection-driven MHD turbulence at
\begin{equation}
r \geq r_{\rm b},
\label{eq:rlimit} 
\end{equation} 
where $r_{\rm b}$ is the radius of the coronal base defined
in~(\ref{eq:defrb}).  The reader who is not interested in the
technical details may wish to skip to Section~\ref{sec:comparison},
which summarizes the free parameters and boundary conditions of the
model and compares the model with our simulation results. 

We begin by dividing the $\bm{z}^+$ fluctuations into two components
as described in Section~\ref{sec:twocomp}:
\begin{equation}
\bm{z}^+ = \bm{z}^+_{\rm HF} + \bm{z}^+_{\rm LF} \qquad
\bm{g} = \bm{g}_{\rm HF} + \bm{g}_{\rm LF}.
\label{eq:2compg} 
\end{equation} 
The quantities $\bm{z}^+_{\rm HF}$ and
$\bm{z}^+_{\rm LF}$ have different radial correlation lengths
(see~(\ref{eq:defgLF})), but we take them to have the same
perpendicular outer scale~$L_\perp$.\footnote{This is an
  over-simplification for Runs~1 and~2, because $\alpha^+(r_{\rm b})
  \simeq 0.8$ in these runs, indicating that much of the $z^+_{\rm HF}$ energy is
  concentrated at high~$k_\perp$. We neglect this spectral flattening in our
  analytic model, however, because there is minimal flattening
  in Run~3 and because we wish to keep the model as simple as possible.}
  We make the simplifying approximation that the HF and LF fluctuations are uncorrelated; i.e.,
$\overline{\bm{g}_{\rm HF} \cdot \bm{g}_{\rm LF}} = 0$, where
$\overline{(\dots)}$ indicates a time average.
Non-WKB reflection is more efficient for low-frequency AWs than for
high-frequency AWs \citep{heinemann80,velli93}. We thus take
$\bm{f}$ to be a ``low-frequency quantity'' that is correlated with $\bm{g}_{\rm LF}$ but not~$\bm{g}_{\rm HF}$.
Upon taking the dot product of~(\ref{eq:Hg})  with $2\bm{g}_{\rm HF}$, averaging, and
assuming a statistical steady state, we obtain
\begin{equation} 
(U+v_{\rm A}) \diff{}{r} g_{\mathrm{HF, rms}}^2
= 2 \overline{ \bm{R} \cdot \bm{g}_{\rm HF}},
\label{eq:gHf1} 
\end{equation} 
where
$g_{\mathrm{HF, rms}} = \overline{|\bm{g}_{\mathrm{HF}}|^2}^{1/2}$
(with analogous definitions for $g_{\mathrm{LF,  rms}}$,
$z^+_{\rm HF, rms}$, $z^+_{\rm LF,  rms}$, $f_{\mathrm{rms}}$,  and
$z^-_{\rm rms}$), and $\bm{R}$ represents
the right-hand side of~(\ref{eq:Hg}). 
The nonlinear
term on the right-hand side of~(\ref{eq:Hg}) acts to cascade $\bm{g}_{\rm HF}$ fluctuations to small scales
at which the fluctuations dissipate. We set
$\overline{ \bm{R} \cdot \bm{g}_{\rm HF} } = - \gamma^+_{\rm HF} g_{\rm HF, rms}^2$,
where $\gamma^+_{\rm HF}$
is the cascade rate of the outer-scale $\bm{g}_{\rm HF}$ fluctuations.
Equation~(\ref{eq:gHf1}) then becomes
\begin{equation}
(U+v_{\rm A}) \diff{}{r} g_{\mathrm{HF, rms}}^2 = -
2\gamma_{\rm HF}^+ g_{\rm HF, rms}^2.
\label{eq:gHf1a} 
\end{equation} 
We follow \cite{velli89} and \cite{verdini09a} in taking the 
outer-scale $\bm{z}^-$ fluctuations 
to be anomalously coherent in a reference
frame that propagates outward with the $\bm{z}^+$ fluctuations, because the
$\bm{z}^-$ fluctuations are produced by sources that propagate
outward at speed $U+v_{\rm A}$. We thus estimate $\gamma^+_{\rm HF}$
using a strong-turbulence scaling regardless of the value of
$z^-_{\rm HF, rms}$, setting 
\begin{equation}
\gamma_{\rm HF}^+ = \frac{c_{\rm diss} z^-_{\rm rms}}{L_\perp},
\label{eq:gammaHF} 
\end{equation} 
where
$c_{\rm diss}$ is a dimensionless free parameter.

Using a similar procedure, but this time for~$g_{\rm LF}$, we find that
\begin{equation}
(U+v_{\rm A}) \diff{}{r} g_{\mathrm{LF,  rms}}^2 = -
2\gamma_{\rm LF}^+ g_{\rm LF,  rms}^2,
\label{eq:gLf1} 
\end{equation} 
where $\gamma_{\rm LF}^+$ is the rate at which outer-scale $\bm{g}_{\rm
  LF}$ fluctuations
cascade to small scales and dissipate. In writing~(\ref{eq:gLf1}), we have dropped a term containing $\overline{
\bm{f} \cdot \bm{g}_{\rm LF}}$ on the assumption that $f \ll g_{\rm LF}$.
We set
\begin{equation}
\gamma_{\rm LF}^+ = \frac{c_{\rm diss}  z^-_{\rm rms}A}{L_\perp},
\label{eq:gammaLF} 
\end{equation} 
where the dimensionless coefficient $A$ models the weakening of
nonlinear interactions between $\bm{g}_{\rm LF}$ and $\bm{f}$ as these
two fluctuation types become increasingly aligned with each other. 
We discuss how we determine~$A$ in Section~\ref{sec:A} below. In
order to compare our model with our simulation results, we take~$A$ to
be related to $\sin\theta$ in~(\ref{eq:defsinthetarms}) via
the equation
\begin{equation}
\sin\theta = \frac{0.55 (Ag_{\rm LF, rms}^2 + g_{\rm HF, rms}^2)}{g_{\rm
    LF, rms}^2 + g_{\rm HF, rms}^2},
\label{eq:sinthetaA} 
\end{equation} 
which expresses the idea that only low-frequency $\bm{g}$ fluctuations
align with $\bm{f}$ fluctuations, while both low-frequency and
high-frequency $\bm{g}$ fluctuations contribute to the average that is
used to compute~$\sin\theta$ in~(\ref{eq:defsinthetarms}). The
factor of 0.55 in~(\ref{eq:sinthetaA}) is included because
this is the typical value of the right-hand side of~(\ref{eq:defsinthetarms}) for outer-scale fluctuations in
homogeneous RMHD turbulence \citep{chandran15}.

\subsection{Amplitude of the inward-propagating fluctuations}
\label{sec:zrms} 

To determine $z^-_{\rm rms}$, we assume that $\bm{z}^-$ cascades
primarily via interactions with $\bm{z}^+_{\rm LF}$ (or, equivalently,
$\bm{g}_{\rm LF}$). The outer-scale $\bm{z}^-$ cascade 
rate then depends upon the critical-balance parameter \citep{goldreich95,boldyrev06}
\begin{equation}
\chi^-_{\rm LF} = \frac{z^+_{\rm LF,  rms} L^+_{r, \rm LF} A}{L_\perp v_{\rm A}},
\label{eq:chiMinusLf} 
\end{equation} 
where $L^+_{r, \rm LF}$ is the radial correlation length of the
$\bm{g}_{\rm LF}$ fluctuations. The critical-balance
parameter $\chi^-_{\rm LF}$ is an estimate of
the fractional change in an outer-scale $\bm{z}^-$ fluctuation that results
from a single ``collision'' with an outer-scale $\bm{g}_{\rm LF}$ fluctuation
lasting a time $\Delta t \sim L_{r, \rm LF}^+/v_{\rm A}$ \citep{lithwick07}.

If $\chi^-_{\rm LF} \ll 1$, then each such collision causes only a
small perturbation to the outer-scale $\bm{z}^-$ fluctuation, and the
turbulence is weak. In this limit, the effects
of successive collisions add like a random walk, and roughly
\begin{equation}
N = (\chi^-_{\rm LF})^{-2}
\label{eq:defN} 
\end{equation} 
collisions are needed for nonlinear interactions to
cause an order-unity change in the outer-scale $\bm{z}^-$ fluctuation. 
The outer-scale $\bm{z}^-$ cascade time scale~$t^-_{\rm NL}$  is then~$\sim N L_{r,\rm  LF}^+/v_{\rm A}$. 
The generation of outer-scale $\bm{z}^-$ (or $\bm{f}$) fluctuations by non-WKB reflection
in this weak-turbulence regime can also be viewed as a random-walk-like process.
Equation~(\ref{eq:Hf})  implies that, in a reference frame~$S^-$ that
propagates with radial velocity~$U-v_{\rm A}$, the increment to $\bm{f}$ from non-WKB reflection
during a time $\Delta t = L_{r, \rm LF}^+/v_{\rm A}$ is  of order
\begin{equation}
\Delta f \sim \left(\frac{U-v_{\rm A}}{2v_{\rm A}}\right) \left|
  \diff{v_{\rm A}}{r}\right| g_{\rm LF,  rms} \Delta t.
\label{eq:Deltaf} 
\end{equation} 
It follows from (\ref{eq:deffg}) 
that the corresponding increment to $\bm{z}^-$ is of order
\begin{equation}
\Delta z^- \sim \left(\frac{U+v_{\rm A}}{2v_{\rm A}}\right) \left|
  \diff{v_{\rm A}}{r}\right| z^+_{\rm LF,  rms} \Delta t.
\label{eq:Deltazm} 
\end{equation} 
The r.m.s. value of $\bm{z}^-$ is approximately the ``amount'' of $\bm{z}^-$ that
``builds up'' in frame~$S^-$ by non-WKB reflection during the
cascade/damping time scale
$ \sim N \Delta t$. The resulting value of~$z^-_{\rm rms}$ is $\sim N^{1/2} \Delta
z^-$, or, equivalently,
\begin{equation}
z^-_{\rm rms} \sim  \frac{L_\perp}{A} \left(\frac{U+v_{\rm A}}{2v_{\rm A}}\right) \left|\diff{v_{\rm A}}{r}\right|.
\label{eq:zminus} 
\end{equation} 

If $\chi^-_{\rm LF} \gtrsim 1$, then  the outer-scale $\bm{z}^-$
fluctuations are sheared coherently throughout their lifetimes, 
the turbulence is strong,
and $t_{\rm NL}^- \sim
L_\perp/(z^+_{\rm LF,  rms} A)$. In this case, $z^-_{\rm rms}$ is
approximately the rate at which $\bm{z}^-$ fluctuations are produced by non-WKB
reflection multiplied by $t^-_{\rm NL}$, which again leads
to~(\ref{eq:zminus}). This estimate, with $A\rightarrow 1$, is the same as that obtained
by \cite{chandran09c} for the strong-turbulence limit. In
the limits $U\rightarrow 0$ and $A\rightarrow 1$,~(\ref{eq:zminus}) is also the same as
the estimate by \cite{dmitruk02} for the strong-turbulence limit. 

Since~(\ref{eq:zminus}) holds in both the weak and
strong-turbulence regimes, we set
\begin{equation}
z^-_{\rm rms} =  \frac{c^- L_\perp}{A} \left(\frac{U+v_{\rm
      A}}{2v_{\rm A}}\right) \left|\diff{v_{\rm A}}{r}\right|
\label{eq:zminus2} 
\end{equation} 
regardless of the value of~$\chi^-_{\rm LF}$, where $c^-$ is a
dimensionless piecewise constant function that has one value at
$r>r_{\rm m}$ and a smaller value at $r\leq r_{\rm m}$, where
$r_{\rm m}$ is defined in~(\ref{eq:rm}). Before discussing $c^-$ further, we
note an immediate consequence of~(\ref{eq:zminus2}), that
$z^-_{\rm rms}$ increases as~$A$ (or equivalently~$\sin \theta$)
decreases. This is because reducing~$\sin\theta$ decreases the rate at
which outer-scale $\bm{z}^-$ fluctuations cascade without decreasing
the rate at which they are produced by non-WKB reflection.

\subsection{Suppression of inward-propagating fluctuations at $r<r_{\rm m}$}
\label{sec:suppression}

The reason we take $c^-$ to have a smaller value at $r< r_{\rm m}$
than at $r>r_{\rm m}$ is that the non-WKB-reflection source term for $\bm{z}^-$ 
fluctuations reverses direction at $r=r_{\rm m}$, since $\d v_{\rm
  A}/\d r$ changes sign. Since $\bm{g}_{\rm
  LF}$ has a large radial correlation length, when
$\bm{z}^-$ fluctuations produced via non-WKB reflection at $r>r_{\rm m}$ propagate to $r<r_{\rm
  m}$, they tend to cancel out the $\bm{z}^-$ fluctuations that are
produced via non-WKB reflection at $r<r_{\rm m}$, reducing~$z^-_{\rm rms}$. If the
$\bm{z}^-$ fluctuations at $r=r_{\rm m}$ can propagate a radial distance
$\sim (r_{\rm m} -R_{\odot})$ before cascading and dissipating, then this cancellation
effect is large. On the other hand, if the $\bm{z}^-$ fluctuations at
$r=r_{\rm m}$ can only propagate a radial distance~$\ll (r_{\rm m} -
R_{\odot})$ before cascading and dissipating, then little cancellation
occurs. To account for this phenomenology, we set 
\begin{equation}
c^- = \left\{\begin{array}{ll}
c^-_{\rm I} & \mbox{ \hspace{0.3cm} if $r\leq r_{\rm m}$}\vspace{0.2cm}  \\
c^-_{\rm O} & \mbox{ \hspace{0.3cm} if $r> r_{\rm m}$} 
\end{array}
\right.,
\label{eq:cminus} 
\end{equation} 
where $c^-_{\rm O}$ is a dimensionless free parameter,
\begin{equation}
c^-_{\rm I} = \frac{c^-_{\rm O}}{1+M},
\label{eq:cminusLG} 
\end{equation} 
\begin{equation}
M= \frac{v_{\rm A m}L_{\perp m}}{z^+_{\rm LF  m} L_\nabla},
\label{eq:defS} 
\end{equation} 
$L_\nabla$ is a free parameter with dimensions of length,
and $v_{\rm A m}$, $L_{\perp m}$, and $z^+_{\rm LF  m}$ are the
values of $v_{\rm A}$, $L_\perp$, and $z^+_{\rm LF,  rms}$ at $r=r_{\rm m}$.
As we argue below (see~(\ref{eq:sintheta})), $A$ is
of order unity at $r< r_{\rm m }$, which means that
$M$ is the approximate radial distance an outer-scale $\bm{z}^-$ fluctuation at
$r=r_{\rm m}$ propagates before cascading to smaller scales, divided by~$L_\nabla$.
We can rewrite~$M$ in terms of quantities evaluated at 
$r=r_{\rm b}$ by making the approximations that 
$v_{\rm A} \gg U$ at $r\leq r_{\rm
  m}$ and $g_{\rm LF, rms}(r_{\rm m}) \simeq g_{\rm LF, rms}(r_{\rm b})$ and by using~(\ref{eq:cons}). This yields
\begin{equation}
M = \frac{v_{\rm A b}L_{\perp \rm b} }{z^+_{\rm LF  b} L_\nabla}\left( \frac{v_{\rm A m}}{v_{\rm
      A b}}\right)^{1/2} ,
\label{eq:S2} 
\end{equation} 
where $v_{\rm A b}$, $L_{\perp b}$, and $z^+_{\rm LF  b}$ are the
values of $v_{\rm A}$, $L_\perp$, and $z^+_{\rm LF,  rms}$ at $r=r_{\rm b}$.

\subsection{Alignment factor and critical-balance parameter}
\label{sec:A} 

To estimate the alignment factor~$A$ introduced in~(\ref{eq:gammaLF}), we first note that
nonlinear interactions between counter-propagating AWs produce
negative residual energy, with $\bm{z}^-$ anti-parallel to~$\bm{z}^+$
(i.e., an excess of magnetic energy over kinetic energy) \citep{muller05,boldyrev11a}.
At $r> r_{\rm m}$, $\d v_{\rm A}/\d r< 0$, and it follows
from~(\ref{eq:deffg}) and~(\ref{eq:Hf}) that non-WKB reflection also
acts to produce negative residual energy. On the other hand, at $r<
r_{\rm m}$, $\d v_{\rm A}/\d r>0$, and non-WKB reflection acts to produce
positive residual energy. In other words, at $r< r_{\rm m}$, linear
processes (non-WKB reflection) and non-linear processes have competing
effects on the alignment of~$\bm{z}^-$. Based on these arguments,
we conjecture that
at $r< r_{\rm m}$ the outer-scale fluctuations do not develop
significant alignment, and that at $r> r_{\rm
  m}$ the outer-scale $\bm{z}^+_{\rm LF}$ and $\bm{z}^-$ fluctuations
become increasingly aligned as the $\bm{z}^+_{\rm LF}$ fluctuations
``decay'' via nonlinear interactions.
We also conjecture that $A$ is a decreasing function
of $\tau_{v}^{\rm (ph)}$, because a larger $\tau_{v}^{\rm
  (ph)}$ increases the efficiency of non-WKB
reflection, which produces $\bm{z}^-$ fluctuations that are aligned
with~$\bm{z}^+_{\rm LF}$. In addition, we conjecture that $A$ is a decreasing function of
\begin{equation}
\Gamma = \frac{z^+_{\rm LF,rms} L^+_{r,\rm LF}}{L_\perp v_{\rm A}},
\label{eq:defGamma} 
\end{equation} 
which is the critical-balance parameter~$\chi^-_{\rm LF}$ in~(\ref{eq:chiMinusLf}) without the factor of $A$.
There are two reasons for taking $A$ to decrease with increasing~$\Gamma$. The first is that
when $\Gamma \ll 1$, outer-scale $\bm{z}^-$ fluctuations can propagate
through many different outer-scale $\bm{z}^+_{\rm LF}$ fluctuations before
cascading to smaller scales. The $\bm{z}^-$ fluctuations  that are
co-located with a particular outer-scale $\bm{z}^+_{\rm LF}$ ``eddy'' of radial
extent~$\sim L_{r,\rm LF}^+$ are
thus a mixture of the $\bm{z}^-$ fluctuations produced by the non-WKB
reflection of that $\bm{z}^+_{\rm LF}$ eddy and $\bm{z}^-$
fluctuations that were initially produced by the non-WKB reflection
of $\bm{z}^+_{\rm LF}$ eddies located farther from the Sun. 
The greater the number of distinct
outer-scale $\bm{z}^+_{\rm
  LF}$ eddies whose reflections contribute to the value of $\bm{z}^-$ at any single
point, the less aligned the $\bm{z}^-$ field will be with any
individual $\bm{z}^+_{\rm LF}$ eddy. Moreover, when $\Gamma>1 $,
shearing of the $\bm{z}^-$ fluctuations by $\bm{z}^+_{\rm
  LF}$ rotates the $\bm{z}^-$ fluctuations into alignment
with $\bm{z}^+_{\rm LF}$, and the resulting value of~$A$ is a
decreasing function of~$\Gamma$ \citep{chandran15}.
We quantify the foregoing conjectures by setting
\begin{equation}
A = \left\{\begin{array}{ll}
A_0 & \mbox{ \hspace{0.3cm} if $r<r_{\rm m}$} \vspace{0.4cm} \\
\displaystyle A_0\left[1 + \frac{\tau_{v}^{\rm (ph)}\Gamma}{\tau_\theta} \ln\left(\frac{g^2_{\rm
        LF m}}{g^2_{\rm LF,rms}}\right)\right]^{-1} & \mbox{ \hspace{0.3cm} if $r>r_{\rm m}$} 
\end{array}
\right.,
\label{eq:sintheta} 
\end{equation} 
where the dimensionless constant~$A_0$ and the time constant $\tau_\theta$ are free parameters.

In the linear, short-wavelength, AW propagation problem, if an AW is
launched into a coronal hole by a boundary condition imposed at the
transition region and photosphere, and if the AW period is~$P$, then
the radial wavelength of the AW at radius~$r$ is $(U +v_{\rm
  A})P$.
That is, the wave period remains constant as the wave propagates away
from the Sun, and the radial wavelength varies in proportion to the
wave phase velocity. We take nonlinear, non-WKB $\bm{z}^+$ fluctuations to
behave in the same way, setting
\begin{equation}
\frac{L_{r,\rm LF}^+}{L_{r, \rm LF b}^+} = \frac{U + v_{\rm A}}{U_{\rm b} + v_{\rm
    Ab}}
\label{eq:Lrplus0},  
\end{equation} 
where $L_{r,  \rm LFb}^+$, $U_{\rm b}$, and $v_{\rm Ab}$ are the values,
respectively, of $L_{r, \rm LF}^+$, $U$, and~$v_{\rm A}$ evaluated at
$r=r_{\rm b}$, 
and likewise for $L_{r, \rm HF}^+$. It then follows from
(\ref{eq:cons}), (\ref{eq:Lbox}), 
(\ref{eq:defLperp}), and~(\ref{eq:Lrplus0})
that
\begin{equation}
\Gamma = \Gamma_{\rm b} \frac{g_{\rm LF,  rms}}{g_{\rm LF  b}}\sqrt{\frac{v_{\rm A}}{v_{\rm A b}}},
\label{eq:Gammaval} 
\end{equation} 
where $\Gamma_{\rm b}$ is the value of~$\Gamma$ at $r=r_{\rm b}$.

\subsection{Solving for the fluctuation-amplitude profiles}
\label{sec:solving}

Upon combining~(\ref{eq:gLf1}), (\ref{eq:gammaLF}),
(\ref{eq:zminus2}), and~(\ref{eq:cminus}), we obtain
\begin{equation}
\diff{}{r} \ln g_{\rm LF,rms}^2 = \left\{\begin{array}{ll}
                                             \displaystyle - c_{\rm I}
                                              \diff{}{r} \ln
                                             v_{\rm A} & \mbox{
                                                                \hspace{0.3cm}
                                                                if
                                                                $r\leq
                                                                r_{\rm
                                                                m}$}
                                                                \vspace{0.2cm}
                                             \\
                                            \displaystyle c_{\rm O}  \diff{}{r} \ln v_{\rm A} & \mbox{ \hspace{0.3cm} if $r>r_{\rm m}$} 
\end{array}\right.,
\label{eq:gLf2} 
\end{equation} 
where
\begin{equation}
c_{\rm I} \equiv c_{\rm diss} c^-_{\rm I} \qquad c_{\rm O} \equiv
c_{\rm diss} c^-_{\rm O} .
\label{eq:defc} 
\end{equation} 
After integrating~(\ref{eq:gLf2}), we find that
\begin{equation}
\frac{g_{\rm LF,  rms}^2}{g_{\rm LF b}^2} =
\left\{ \begin{array}{ll}
\displaystyle \left(\frac{v_{\rm A b}}{v_{\rm A}}\right)^{c_{\rm I}} & \mbox{
                                              \hspace{0.3cm} if
                                                  $r_{\rm b} < r <
                                                  r_{\rm m}$}
                                                               \vspace{0.2cm} \\
\displaystyle \left(\frac{v_{\rm A b} }{v_{\rm A m}}\right)^{c_{\rm I}} 
\left(\frac{v_{\rm A} }{v_{\rm A m}}\right)^{c_{\rm O}}  & \mbox{                                       \hspace{0.3cm} if
                                                  $r > r_{\rm m}$} 
\end{array}
\right.,
\label{eq:gLFsolve} 
\end{equation} 
where $g_{\rm LF  b}$ is the value of $g_{\rm
  LF,  rms}$ at $r=r_{\rm b}$. Upon combining~(\ref{eq:gHf1a}), (\ref{eq:gammaHF}),
(\ref{eq:zminus2}), and (\ref{eq:cminus}), we obtain
\begin{equation}
\diff{}{r} \ln g_{\rm HF,rms}^2 = \left\{\begin{array}{ll} \displaystyle
- \frac{c_{\rm I}}{A}  \diff{}{r} \ln v_{\rm A} &
                                                                    \mbox{
                                                                    \hspace{0.3cm}
                                                                    if
                                                                    $r\leq
                                                                    r_{\rm
                                                                    m}$}
                                                                          \vspace{0.2cm} 
                                             \\
 \displaystyle
 \frac{c_{\rm O}}{A}  \diff{}{r} \ln v_{\rm A} &
                                                                    \mbox{
                                                                    \hspace{0.3cm}
                                                                    if
                                                                    $r>
                                                                    r_{\rm
                                                                    m}$}
\end{array} \right. .
\label{eq:gHf2} 
\end{equation} 
With the aid of~(\ref{eq:sintheta}) and~(\ref{eq:Gammaval}),
we integrate~(\ref{eq:gHf2}) to obtain
\begin{equation}  
\frac{g_{\rm HF,rms}^2}{g_{\rm HFb}^2}= \left\{ \begin{array}{ll}
\displaystyle  \left(\frac{v_{\rm A b}}{v_{\rm
                             A}}\right)^{c_{\rm I}/A_0} & \mbox{ \hspace{0.cm} if
  $r< r_{\rm m}$} \vspace{0.4cm} \\
\displaystyle  \left(\frac{v_{\rm A b}}{v_{\rm
                             Am}}\right)^{c_{\rm I}/A_0}
 w^{c_{\rm O}/A_0} e^{-H}
                                           & \mbox{ \hspace{0.0cm} if $r>r_{\rm m}$}
\end{array}
\right.,
\label{eq:gHfsolve2} 
\end{equation} 
where
\begin{equation}
H = \frac{c_\theta c_{\rm O}^2 \Gamma_{\rm b}}{\sigma^2 A_0} \left(\frac{v_{\rm
      A m}}{v_{\rm A b}}\right)^{(1-c_{\rm I})/2} (\sigma w^\sigma \ln
w - w^\sigma + 1),
\label{eq:defA} 
\end{equation} 
\begin{equation}
w = \frac{v_{\rm A}}{v_{\rm A m}}
\qquad
\sigma = \frac{1+c_{\rm O}}{2},
\label{eq:defsigma} 
\end{equation} 
$g_{\rm HF b}$ is the value of $g_{\rm
  HF, rms}$ at $r=r_{\rm b}$, and $c_\theta = \tau_{v}^{\rm (ph)}/\tau_\theta$.

\subsection{Turbulent-heating rate}
\label{sec:Qanalytic} 

The turbulent-heating rate in our model is 
\begin{equation}
Q = \frac{\rho}{2} \left[\gamma^+_{\rm HF} (z^+_{\rm HF,rms})^2 +
  \gamma^+_{\rm LF} (z^+_{\rm LF,rms})^2
+  \gamma^- (z^-_{\rm rms})^2\right],
\label{eq:defQ2} 
\end{equation} 
where 
\begin{equation}
\gamma^- = \gamma^-_{\rm LF} + \gamma^-_{\rm HF}
\label{eq:defgammaminus} 
\end{equation} 
is the cascade
rate of the outer-scale $\bm{z}^-$ fluctuations, and $\gamma^-_{\rm
  LF}$ ($\gamma^-_{\rm HF}$) is the contribution to $\gamma^-$ from 
interactions between $\bm{z^}-$ fluctuations and LF (HF) $\bm{z}^+$ fluctuations. To allow for either
weakly turbulent ($\chi^-_{\rm LF} < 1$) or strongly turbulent
($\chi^-_{\rm LF} \geq 1$) shearing of $\bm{z}^-$
fluctuations by $\bm{z}^+_{\rm LF}$ fluctuations, we set
\begin{equation}
\gamma^-_{\rm LF} = \frac{ c_{\rm diss} z^+_{\rm LF,rms}
  A}{L_\perp} \times \left\{\begin{array}{ll}
\chi^-_{\rm LF} & \mbox{ \hspace{0.3cm} if $\chi^-_{\rm LF} \leq 1$}
                  \vspace{0.2cm} 
                                       \\
1 & \mbox{ \hspace{0.3cm} if $\chi^-_{\rm LF} > 1$}
\end{array}
\right. .
\label{eq:gammaminusLF} 
\end{equation} 
In analogy to~(\ref{eq:chiMinusLf}), we define the
critical-balance parameter for the shearing of $\bm{z}^-$ fluctuations
by $\bm{z}^+_{\rm HF}$ fluctuations to be
\begin{equation}
\chi^-_{\rm HF} = \frac{z^+_{\rm HF,  rms} L^+_{r, \rm HF} }{L_\perp v_{\rm A}},
\label{eq:chiMinusHf} 
\end{equation} 
where we have omitted the factor of~$A$, because we take 
$\bm{z}^-$ to be aligned with $\bm{z}^+_{\rm LF}$ but not with
$\bm{z}^+_{\rm HF}$. We then set
\begin{equation}
\gamma^-_{\rm HF} =  \frac{c_{\rm diss} z^+_{\rm HF,rms}}{L_\perp} \times \left\{\begin{array}{ll}
\chi^-_{\rm HF} & \mbox{ \hspace{0.3cm} if $\chi^-_{\rm HF} \leq 1$}
                  \vspace{0.2cm} 
                                       \\
1 & \mbox{ \hspace{0.3cm} if $\chi^-_{\rm HF} > 1$}
\end{array}
\right. .
\label{eq:gammaminusHF} 
\end{equation} 

\subsection{Comparison with simulation results}
\label{sec:comparison} 

To compare our model with one of our numerical simulations, 
we treat $L_\perp(r_{\rm  b}) = L_{\rm box}(r_{\rm b})/3$, $L^+_{r, \rm
  LF}(r_{\rm b})$, $L^+_{r, \rm HF}(r_{\rm b})$, and $z^+_{\rm
  rms}(r_{\rm b})$
as boundary conditions in our model,
which we determine using the measured values of these quantities in that
particular simulation.
Also, motivated by figure~\ref{fig:HFLF}, we
set
\begin{equation}
\frac{g_{\rm HFb}^2}{g_{\rm LF b}^2}  = 1
\label{eq:HFLFratio} 
\end{equation} 
in all our model solutions. We take $L^+_{r, \rm HF}(r_{\rm b})$  in our simulations to be
the radial separation~$\Delta r$ at which $C(r_{\rm b},\Delta r) =
1/2$, where
\begin{equation}
C(r, \Delta r) = \frac{\langle \bm{g}(\tilde{x},\tilde{y},r,t) \cdot \bm{g}(\tilde{x},\tilde{y},
  r+\Delta r, t)\rangle}{\langle |\bm{g}(\tilde{x},\tilde{y},r,t)|^2\rangle}
\label{eq:defC} 
\end{equation} 
is the radial autocorrelation function of the~$\bm{g}$
fluctuations, and
$\tilde{x}$ and $\tilde{y}$ are defined
following~(\ref{eq:defgHFrms}). 
On the other hand, because figure~\ref{fig:HFLF}  shows that
the LF fluctuations are
energetically dominant at $r=r_{\rm max}$, we
define $L^+_{r, \rm LF}(r_{\rm max})$ to be the value of~$\Delta r$ at
which $C(r_{\rm max}, -\Delta r) = 1/2$. Applying~(\ref{eq:Lrplus0}), we then set
$L^+_{r, \rm LF}(r_{\rm b}) = L^+_{r, \rm LF}(r_{\rm max})(U_{\rm b} +
v_{\rm A b})/[U(r_{\rm max}) + v_{\rm A}(r_{\rm max})] = 0.886
L^+_{\rm LF}(r_{\rm max})$.
The values of $L_{\rm box}(r_{\rm b})$, $L^+_{r, \rm LF}(r_{\rm b})$, $L^+_{r, \rm HF}(r_{\rm b})$, and
$z^+_{\rm rms}(r_{\rm b})$ in our three simulations are listed
in table~\ref{tab:BCs}.

\begin{table}
\caption{\vspace{0.1cm} Boundary Conditions in Our Analytic Model for Matching
  Runs~1 through~3
\label{tab:BCs} }
\begin{center}
\begin{tabular}{lcccc}
  \vspace{-0.4cm} 
  &&  \\ 
  \hline \hline 
  \vspace{-0.25cm} 
  & & \\
  Quantity & & Run 1 & Run 2 & Run 3 \\
  \vspace{-0.25cm} 
  && \\
  \hline
  \vspace{-0.25cm} && \\
  $z^+_{\rm rms}(r_{\rm b})$ & $\dots\dots\dots\dots$ & $61 \mbox{ km/s}$
    & $55 \mbox{ km/s}$ & $41 \mbox{ km/s}$\\
  $L^+_{r, \rm  HF}(r_{\rm b})$ & $\dots\dots\dots\dots$ & $0.015 R_{\odot}$ &
                                                                              $0.015
                                                                              R_{\odot}$
           & $0.10 R_{\odot}$\\
  $L^+_{r, \rm  LF}(r_{\rm b})$ & $\dots\dots\dots\dots$ & $0.071
                                                           R_{\odot}$ &
                                                                       $0.27
                                                                       R_{\odot}$
           & $0.35 R_{\odot}$ \\
  $L_\perp(r_{\rm b})$& $\dots\dots\dots\dots$ & $ \hspace{0.1cm} 1.4 \times
                                                           10^3 \mbox{
                                                           km}$
                                                 \hspace{0.1cm}  &
                                                                   \hspace{0.1cm}  $ 1.4
                                                                   \times
                                                                   10^3
                                                                   \mbox{
                                                                   km}$
                                                                   \hspace{0.1cm}
           & \hspace{0.1cm}  $ 5.3 \times 10^3 \mbox{ km}$
             \hspace{0.1cm} \\
  \vspace{-0.2cm} 
  \\
  \hline
\end{tabular}
\end{center}
\medskip
{\footnotesize $z^+_{\rm rms}$ is the r.m.s. amplitude of the
  outward-propagating Elsasser variable, $L^+_{r, \rm LF}$ is the
  radial correlation length of the low-frequency outward-propagating
  Heinemann-Olbert variable $\bm{g}_{\rm LF}$,
  $L^+_{r, \rm HF}$ is the
  radial correlation length of the high-frequency outward-propagating
  Heinemann-Olbert variable $\bm{g}_{\rm HF}$, $L_\perp$ is the
  perpendicular outer scale (see~(\ref{eq:defLperp})),
  and $r_{\rm b}$ is the radius of the coronal base defined in (\ref{eq:defrb}).
}
\vspace{0.3cm} 
\end{table}

We take the free parameters $c_{\rm diss}$, $c^-_{\rm O}$,
$\tau_\theta$, $A_0$, and~$L_\nabla$ to be the same regardless of the simulation with which 
we are comparing our model. We then vary these free parameters to
optimize the agreement between our model and all three
simulations. We list the resulting parameters in
table~\ref{tab:bestfit}. 

Figures~\ref{fig:zpm_profiles} through~\ref{fig:LAQtot} show the radial
profiles of $z^+_{\rm rms}$, $z^-_{\rm rms}$, $z^+_{\rm HF, rms}$,
$z^+_{\rm LF, rms}$, $\sin\theta$, and~$Q$ that result from our model
using the best-fit parameters in table~\ref{tab:bestfit} and the
boundary conditions in table~\ref{tab:BCs}.  As these figures show,
our model reproduces a number of trends seen in the simulations. For
example, in both the model and simulations,
$z^+_{\rm HF, rms}/z^+_{\rm LF, rms}$ and $\sin\theta$ decrease with
increasing~$r$, particularly in Run~2. The radial decrease in
$z^+_{\rm HF}/z^+_{\rm LF}$ is consistent with our expectation that
high-frequency $\bm{z}^+$ fluctuations cascade and dissipate more
rapidly than low-frequency $\bm{z}^+$ fluctuations, because 
high-frequency $\bm{z}^+$ fluctuations are not aligned with $\bm{z}^-$. In our
model, the radial decrease in $\sin \theta$ is related both to the
comparatively rapid cascade of the unaligned high-frequency $\bm{z}^+$
fluctuations and the fact that the low-frequency $\bm{z}^+$ fluctuations
become increasingly aligned with~$\bm{z}^-$ as they interact nonlinearly
with~$\bm{z}^-$. We note that the decrease in
$\sin\theta$ coincides with an increase in $z^-_{\rm rms}$ for the
reasons described following~(\ref{eq:zminus2}).  The model reproduces
the $z^\pm_{\rm rms}$ profiles in the simulations fairly accurately.  The turbulent-heating rates in the
model and simulations also agree quite well, but the heating rate in the
model is somewhat smaller than in Run~3 at $r > 3 R_{\odot}$. The most
notable failing of the model is that $z^-_{\rm rms} = Q = 0$ at
$r=r_{\rm m}$, because our estimate of $z^-_{\rm rms}$ is proportional
to the local value of $\d v_{\rm A}/\d r$, which vanishes at
$r=r_{\rm m}$.  A more realistic model would account for the fact that
$\bm{z}^-$ fluctuations propagate a finite distance before cascading
and dissipating, which would smooth out the profiles of
$z^-_{\rm rms}$ and~$Q$ in the vicinity of $r=r_{\rm m}$. Importantly,
despite the aforementioned differences between the model and our
numerical results, varying the boundary conditions in the model to
match the measured conditions in the simulations largely accounts for
the differences between the $z^\pm_{\rm rms}$ and $Q$ profiles in
Runs~1, 2, and~3 without any modification to the free parameters in
table~\ref{tab:bestfit}. This suggests that the model provides a
  reasonably accurate representation of the dominant physical
  processes that control these radial profiles.

\begin{table}
\caption{\vspace{0.1cm} Best-Fit Free Parameters in our Analytic Model
\label{tab:bestfit} }
\begin{center}
\begin{tabular}{lcc}
  \vspace{-0.4cm} 
  &&  \\ 
  \hline \hline 
  \vspace{-0.25cm} 
  & & \\
  Parameter & & Value \\
  \vspace{-0.25cm} 
  && \\
  \hline
  \vspace{-0.25cm} && \\
  $c_{\rm diss}$ & $\dots\dots\dots\dots$ & $0.2$ \\
  $c^-_{\rm O}$ & $\dots\dots\dots\dots$ & $1.8$ \\
  $\tau_\theta$ & $\dots\dots\dots\dots$ & $3.2$~min \\
  $A_0$ & $\dots\dots\dots\dots$ & 0.6 \\
  $L_\nabla$ & $\dots\dots\dots\dots$& $0.15 R_{\odot}$ \\
  \vspace{-0.2cm} 
  \\
  \hline
\end{tabular}
\end{center}
\medskip
{\footnotesize The quantity $c_{\rm diss}$ is a 
coefficient appearing in the cascade/damping rates~$\gamma^+_{\rm HF}$
and $\gamma^+_{\rm LF}$ ((\ref{eq:gammaHF}) and
(\ref{eq:gammaLF})), $c^-_{\rm O}$ is a coefficient
in our estimate of~$z^-_{\rm rms}$ (see (\ref{eq:zminus2})
through~(\ref{eq:cminusLG})),
$\tau_\theta$ and $A_0$ are constants appearing in our estimate of the alignment
angle ((\ref{eq:sintheta})), and $L_\nabla$ is a length scale
that affects the degree to which $\bm{z}^-$ fluctuations produced by
non-WKB reflection at $r>r_{\rm m} \simeq 1.7 R_{\odot}$ cancel out the
$\bm{z}^-$ fluctuations produced by non-WKB reflection at $r<r_{\rm
  m}$ (see~(\ref{eq:zminus2}) through~(\ref{eq:defS})).}
\vspace{0.3cm} 
\end{table}

\section{Previous studies of the Elsasser power spectra in MHD turbulence}
\label{sec:homogeneous} 

In this section, we review previous studies of the Elsasser power
spectra in homogeneous RMHD turbulence and reflection-driven MHD
turbulence. The reader already familiar with this
literature may wish to skip directly to Section~\ref{sec:124}. We
follow the convention of describing the turbulent $\bm{z}^\pm$
fluctuations as collections of AW packets, using $\lambda$ to denote
the length scale of a wave packet measured perpendicular to the
magnetic field, $l^\pm_\lambda$ to denote the correlation length
measured along the magnetic field of $\bm{z}^\pm$ wave packets with
perpendicular scale~$\lambda$, and $ \delta z^\pm_\lambda$ to denote
the amplitude of wave packets at scale~$\lambda$ -- i.e., the r.m.s.
increment in~$\bm{z}^\pm$ across a distance~$\lambda$ perpendicular to
the magnetic field.

\subsection{Balanced, homogeneous RMHD turbulence}
\label{sec:balanced} 

In ``balanced turbulence,''  the statistical
properties of $\bm{z}^+$ and $\bm{z}^-$ fluctuations are identical. In particular,
\begin{equation}
\delta z^+_\lambda = \delta z^-_\lambda \qquad l^+_\lambda = l^-_\lambda,
\label{eq:balanced}
\end{equation} 
and the cross helicity (the difference
between the energies per unit mass of $\bm{z}^+$ and $\bm{z}^-$ fluctuations) is
zero.  In homogeneous RMHD turbulence, 
the strongest nonlinear interactions are local in scale, meaning that
$\delta z^\pm_\lambda$ fluctuations are cascaded primarily by 
$\bm{z}^\mp$ fluctuations at perpendicular scales comparable to~$\lambda$.
To understand how a $\delta z^\pm_\lambda$ wave packet cascades, it is
helpful to consider a propagating ``slice'' of the wave packet --
i.e., a single cross section of the wave packet in the plane
perpendicular to the background magnetic field \citep[see,
e.g.,][]{lithwick07}. This slice ``collides'' with a series of
counter-propagating $\delta z^\mp_\lambda$ wave packets. Each
collision has a duration of \begin{equation} t_\lambda^\pm \sim
  \frac{l^\mp_\lambda}{v_{\rm A}} .
\label{eq:dt} 
\end{equation} 
The instantaneous rate at which $\delta z^\mp_\lambda$ wave packets shear $\delta z^\pm_\lambda$ wave packets is~$\sim \delta z^\mp_\lambda/\lambda$. During a single collision, the aforementioned ``slice'' of the $\delta z^\pm_\lambda$ fluctuation undergoes a fractional distortion of order~\citep{goldreich95,goldreich97,lithwick07}
\begin{equation}
\chi^\pm_\lambda = \frac{\delta z^\mp_\lambda l^\mp_\lambda}{\lambda v_{\rm A}}.
\label{eq:chilambda} 
\end{equation} 

\subsubsection{Weak balanced turbulence}
\label{sec:weak} 

If $\chi^\pm_\lambda \ll 1$, a $\delta z^\pm_\lambda$ wave packet
undergoes only a small fractional change during each collision,
and the turbulence is weak. \cite{ng96,ng97} and \cite{goldreich97} advanced
a phenomenological model of weak, incompressible,
MHD turbulence in which
the effects of consecutive collisions are uncorrelated and add like a
random walk. After $N$ collisions, the r.m.s. fractional change in a
$\delta z^\pm_\lambda$ wave packet is $\sim N^{1/2}
\chi^\pm_\lambda$. After $N \sim
(\chi^\pm_\lambda)^{-2}$ collisions, the r.m.s. fractional distortion of
the wave packet grows to a value of order unity, and the energy
contained within the wave packet cascades to smaller scales. The
cascade time scale is thus
\begin{equation}
\tau_\lambda^\pm \sim \left(\chi^\pm_\lambda\right)^{-2} t^\pm_\lambda \sim \frac{\lambda^2 v_{\rm A}}{(\delta z^\mp_\lambda)^2 l^\mp_\lambda}.
\label{eq:tcascwk} 
\end{equation} 
Because neither the $\delta z^+_\lambda$ nor $\delta z^-_\lambda$ wave
packet is altered significantly during any single collision, the leading and trailing edges of a $\delta z^\pm_\lambda$ wave packet are sheared in virtually the same way during each collision, and the parallel length scale of the wave packets does not change as the fluctuation energy cascades to smaller~$\lambda$ \citep{shebalin83}; i.e.,
\begin{equation}
l^\pm_\lambda  \propto \lambda^0.
\label{eq:lconst} 
\end{equation} 
In the inertial range, 
the $\bm{z}^\pm$ energy-cascade rate (per unit mass), $\epsilon^\pm$, is
independent of scale:
\begin{equation}
\epsilon^\pm  \sim \frac{(\delta z^\pm_\lambda)^2}{\tau_\lambda^\pm}
\propto \lambda^0.
\label{eq:defepspm} 
\end{equation} 
Equations~(\ref{eq:balanced}), (\ref{eq:tcascwk}), (\ref{eq:lconst}),
and~(\ref{eq:defepspm}) imply that
\begin{equation}
\delta z^\pm_\lambda \propto \lambda^{1/2}.
\label{eq:dzpmwk} 
\end{equation} 
The scaling of the one-dimensional power spectrum of the $\bm{z}^\pm$
fluctuations, denoted $E^\pm(k_\perp)$, follows from the relation
\begin{equation}
k_\perp E^\pm(k_\perp) \sim \left(\delta z^\pm_\lambda\right)^2 \Big|_{\lambda = k_\perp^{-1}},
\label{eq:dzEk} 
\end{equation} 
where $k_\perp$ is the component of the wave vector perpendicular to
the background magnetic field.
Equations~(\ref{eq:dzpmwk}) and (\ref{eq:dzEk})  imply that 
\begin{equation}
E^\pm(k_\perp) \propto k_\perp^{-2}.
\label{eq:Epmwk} 
\end{equation}

The scaling in~(\ref{eq:Epmwk}) has been found in 
direct numerical simulations \citep{perez08a} as well as 
in exact solutions to the weak-turbulence 
wave kinetic equations for incompressible MHD turbulence
\citep{galtier00}. It is worth noting, however, that in weak-turbulence theory all AWs are cascaded by $k_\parallel =
0$ modes, where $k_\parallel$ is the wave vector component
along~$\bm{B}_0$, and these zero-frequency modes violate 
the assumptions of weak-turbulence theory. Several studies have addressed this issue,
as well as its consequences for imbalanced turbulence \citep{boldyrev09a,schekochihin12,meyrand15},
as discussed further in Section~\ref{sec:wkI}.

\subsubsection{Strong balanced turbulence}
\label{sec:strong} 

If $\chi^\pm_\lambda \gtrsim 1$, then each slice of a $\delta
z^\pm_\lambda$ wave packet is strongly distorted during a single
collision, the turbulence is strong, and $\bm{z}^\pm$ energy at
scale~$\lambda$ cascades to smaller scales on the time scale
\begin{equation}
\tau^\pm_\lambda \sim \frac{\lambda}{\delta z^\mp_\lambda},
\label{eq:tcascstr} 
\end{equation} 
leading to a scale-independent energy-cascade rate 
\begin{equation}
\epsilon^\pm  \sim \frac{(\delta z^\pm_\lambda)^2 \delta
  z^\mp_\lambda}{\lambda} \propto \lambda^0.
\label{eq:epsstrong} 
\end{equation} 
Equations~(\ref{eq:balanced}) and (\ref{eq:epsstrong})  imply that
\begin{equation}
\delta z^\pm_\lambda \propto \lambda^{1/3},
\label{eq:zpmstr} 
\end{equation} 
which implies via~(\ref{eq:dzEk}) that
\begin{equation}
E^\pm(k_\perp) \propto k_\perp^{-5/3}.
\label{eq:Epmscal} 
\end{equation}

\cite{goldreich95} conjectured that in strong, balanced, RMHD turbulence (and also in anisotropic, incompressible, MHD turbulence), the linear and nonlinear
time scales of each wave packet are comparable, i.e.,
\begin{equation}
\chi^\pm_\lambda \sim 1.
\label{eq:CB0} 
\end{equation} 
Numerical simulations confirm that this ``critical-balance'' conjecture
describes
strong RMHD turbulence not only on average \citep{cho00}, but
structure by structure~\citep{mallet15}.
Together, (\ref{eq:zpmstr}) and~(\ref{eq:CB0})  imply  that
\begin{equation}
l^\pm_\lambda \propto \lambda^{2/3}.
\label{eq:lscalCB} 
\end{equation}

Several studies have argued, on the basis of numerical simulations and
theoretical arguments, that the inertial-range power spectrum in
strong, balanced, RMHD turbulence is flatter than in the
Goldreich-Sridhar model and closer to $k_\perp^{-3/2}$, because of
scale-dependent dynamic alignment
\citep{boldyrev05,boldyrev06,mason08,perez12} and/or intermittency
\citep{maron01,chandran15,mallet17a}. On the other hand,
\cite{beresnyak12,beresnyak14} argued for a scaling closer
to~$k_\perp^{-5/3}$ based on the Reynolds-number scaling of the amplitude of
dissipation-scale structures. A possible
resolution of the disagreement between these two sets of studies was
provided by \cite{mallet17b,mallet17c} and \cite{loureiro17,loureiro17b}, who investigated the disruption of
sheet-like structures in RMHD turbulence by the tearing instability
and magnetic reconnection \citep[see also][]{pucci14,pucci18,vech18}.

\subsection{Imbalanced RMHD turbulence in homogeneous plasmas}
\label{sec:imbalanced} 

In ``imbalanced turbulence,'' one of the Elsasser variables,
say~$\bm{z}^+$, has a substantially higher r.m.s. amplitude than the other:
\begin{equation}
z^+_{\rm rms} > z^-_{\rm rms}.
\label{eq:imbalanced} 
\end{equation} 
Equation~(\ref{eq:imbalanced}) includes the highly imbalanced case, in
which $z^+_{\rm rms} \gg z^-_{\rm rms}$, as well as moderately
imbalanced turbulence, in which, e.g., $z^+_{\rm rms} \simeq 2 z^-_{\rm rms}$.

\subsubsection{Weak imbalanced turbulence}
\label{sec:wkI} 

When~(\ref{eq:imbalanced}) is satisfied and
\begin{equation}
\chi^+_\lambda \ll 1 \qquad \chi^-_\lambda \ll 1,
\end{equation} 
the turbulence is both imbalanced and weak. \cite{galtier00} showed that in the weak-turbulence theory of
imbalanced incompressible MHD turbulence, 
\begin{equation}
\alpha^+ + \alpha^- = 4,
\label{eq:G00} 
\end{equation} 
where 
\begin{equation}
E^\pm(k_\perp) \propto
k_\perp^{-\alpha^\pm},
\label{eq:defalpha2} 
\end{equation} 
the homogeneous-turbulence version of~(\ref{eq:defalpha}).
\cite{lithwick03} argued that in weak incompressible MHD turbulence,
the spectra are ``pinned'' at the dissipation
wavenumber~$k_{\perp \rm d}$, with
$E^+(k_{\perp \rm d}) = E^-(k_{\perp \rm d})$, and that the more
energetic Elsasser variable has the steeper inertial-range power
spectrum. \cite{boldyrev09a} espoused a different picture, in which
a ``condensate'' of magnetic fluctuations at
$k_\parallel = 0$ 
dominates the energy cascade, leading to a state in
which~$\alpha^+ = \alpha^- = 2$. \cite{schekochihin12} developed a
theory accounting for both weakly turbulent AWs with
nonzero~$k_\parallel$ and 2D modes with $k_\parallel= 0$, and found
that $\alpha^+=\alpha^-=2$ for the
weakly turbulent modes and~$\alpha^+=\alpha^-=1$ for the 2D modes in the imbalanced
case.

\subsubsection{Strong imbalanced turbulence}
\label{sec:LGS07} 

When~(\ref{eq:imbalanced}) is satisfied and $\chi^+_\lambda$
or $\chi^-_\lambda$ is $\gtrsim 1$, the turbulence is considered
strong. A number of authors have developed models of strong
imbalanced MHD turbulence
\citep[e.g.,][]{beresnyak08,chandran08a,beresnyak09,perez09a,perez10b,podesta10}. Here
we focus on the study by \cite{lithwick07} (hereafter LGS), who
explored an assumption about the forcing of outer-scale $\bm{z}^-$
fluctuations that turns out to be particularly relevant to
inhomogeneous reflection-driven MHD turbulence in the solar wind.

LGS assumed, in addition to~(\ref{eq:imbalanced}), that
\begin{equation}
\chi^-_\lambda \gtrsim 1.
\label{eq:chiplimb} 
\end{equation} 
Equation~(\ref{eq:chiplimb}) implies that
$\delta z^-_\lambda$~fluctuations are sheared on a
time scale~$\lambda/\delta z^+_\lambda$ that is comparable to or less
than the time~$l^+_\lambda/v_{\rm A}$ for a slice of a $\delta z^-_\lambda$ wave packet
to pass through a counter-propagating
$\delta z^+_\lambda$ wave packet. The cascade time scale for
$\delta z^-_\lambda$ wave packets is therefore
\begin{equation}
\tau^-_\lambda \sim \frac{\lambda}{\delta z^+_\lambda}.
\label{eq:tauminus} 
\end{equation} 
LGS argued that, since a $\delta z^-_\lambda$ wave packet cascades
after it has propagated along the background magnetic field for a
distance~$\sim v_{\rm A} \tau^-_\lambda$, the parallel correlation
length of the $\delta z^-_\lambda$ wave packet is
\begin{equation}
l^-_\lambda \sim v_{\rm A} \tau^-_\lambda \sim \frac {v_{\rm A}
  \lambda }{\delta z^+_\lambda}.
\label{eq:lminuslambda} 
\end{equation} 
LGS further argued that, since $\delta z^+_\lambda$ wave packets
separated by a distance $l^-_\lambda$ along the magnetic field are
sheared by uncorrelated $\delta z^-_\lambda$ fluctuations, 
\begin{equation}
l^+_\lambda \sim l^-_\lambda.
\label{eq:equal} 
\end{equation} 
It follows from~(\ref{eq:chilambda}), (\ref{eq:lminuslambda}) and~(\ref{eq:equal})
that 
\begin{equation}
\chi^-_\lambda \sim 1 \qquad \chi^+_\lambda \sim \frac{\delta z^-_\lambda}{\delta
  z^+_\lambda} < 1.
\label{eq:chipl1} 
\end{equation} 

The apparent implication of the second half of (\ref{eq:chipl1}),
  particularly when $\delta z^-_\lambda/\delta z^+_\lambda \ll 1$, is
that $\delta z^+_\lambda$ wave packets cascade in a weakly turbulent
manner, through multiple, uncorrelated collisions with
$\delta z^-_\lambda$ wave packets, each of which leads to a small
fractional change in the $\delta z^+_\lambda$ wave packet of order
$\chi^+_\lambda$ (see Section~\ref{sec:weak}).  LGS argued, however,
that each $\delta z^+_\lambda$ wave packet is in fact sheared
coherently throughout its lifetime,  even when $\chi^+_\lambda \sim
\delta z^-_\lambda/\delta z^+_\lambda  \ll 1$. To establish this conclusion, LGS
considered the ``$\bm{z}^+$ frame,'' which moves with $\bm{z}^+$~fluctuations at
speed~$v_{\rm A}$ along~$\bm{B}_0$ relative to the background
plasma. They then proposed a thought experiment in which the amplitude
of $\bm{z}^-$ is infinitesimal, so that $\bm{z}^-$ has negligible effect upon~$\bm{z}^+$. The $\bm{z}^+$
vector field is then time-independent in the $\bm{z}^+$ frame.  If the
$\bm{z}^+$ fluctuations are initialized with a power-law spectrum spanning
the entire inertial range, and if $\bm{z}^-$ fluctuations are continuously
injected at the outer scale with an arbitrarily long coherence
time~$T$ in the $\bm{z}^+$ frame, then the $\bm{z}^-$ fluctuations will
cascade to small scales and set up not just a statistical steady
state, but an actual steady state in the $\bm{z}^+$ frame in which the
$\bm{z}^-$ vector field is independent of time. This latter conclusion
follows because $\bm{z}^-$ is nonlinearly distorted by $\bm{z}^+$,
which is constant in time in the $\bm{z}^+$ frame. The $\delta z^-_\lambda$
fluctuations encountered by a $\delta z^+_\lambda$ wave packet are
therefore coherent for an arbitrarily long time, and in particular for
a time much longer than the crossing time
\begin{equation}
t_{\rm cross, \lambda}^+ \sim \frac{l^-_\lambda}{v_{\rm A}}
\label{eq:tcross} 
\end{equation} 
required for a slice of the $\delta
z^+_\lambda$ wave packet to propagate through a $\delta z^-_\lambda$
wave packet.

Building upon this thought experiment, LGS proceeded to consider the more
realistic case in which $\delta z^-_\lambda$ is finite, but still
small compared to $\delta z^+_\lambda$ at all~$\lambda$. They made a key assumption, which we call the
``coherence assumption,'' that the coherence time~$T$ (at a fixed
position in the
$\bm{z}^+$ frame) of the forcing of outer-scale $\bm{z}^-$ fluctuations is at
least as long as the $\bm{z}^+$ cascade time at the outer scale, as
was the case in
the thought experiment above. When the coherence assumption holds,
the dominant mechanism for decorrelating the $\delta z^-_\lambda$
fluctuations encountered by a $\delta z^+_\lambda$ wave packet is the
variation of the $\bm{z}^+$ vector field, not the crossing of
counter-propagating wave packets, and the
$\delta z^+_\lambda$ wave packet is sheared coherently throughout its
lifetime. The~$\bm{z}^+$ cascade time scale at scale~$\lambda$ then becomes
\begin{equation}
\tau_\lambda^+ \sim \frac{\lambda}{\delta z^-_\lambda},
\label{eq:tauplus} 
\end{equation} 
and
\begin{equation}
\epsilon^+ \sim \frac{(\delta z^+_{\lambda})^2}{\tau_\lambda^+} \sim \frac{(\delta z^+_\lambda)^2 \delta z^-_\lambda}{\lambda}.
\label{eq:LGS07epsp} 
\end{equation} 
Because of~(\ref{eq:chipl1}), $\tau_\lambda^- \sim \lambda/\delta z^+_\lambda$,
and
\begin{equation}
\epsilon^- \sim \frac{(\delta z^-_{\lambda})^2}{\tau_\lambda^-} \sim \frac{(\delta z^-_\lambda)^2 \delta z^+_\lambda}{\lambda}.
\label{eq:LGS07epsm} 
\end{equation} 
Setting $\epsilon^\pm\propto \lambda^0$, LGS combined
Equations~(\ref{eq:LGS07epsp}) and (\ref{eq:LGS07epsm}) to obtain 
\begin{equation}
\delta z^+_\lambda \propto \delta z^-_\lambda \propto  \lambda^{1/3},
\label{eq:zpzmLGS07} 
\end{equation} 
which, via~(\ref{eq:dzEk}), implies that
\begin{equation}
E^\pm(k_\perp) \propto k_\perp^{-5/3}.
\label{eq:LGSEkp} 
\end{equation} 

\subsection{Anomalous coherence in reflection-driven MHD turbulence}
\label{eq:vetal} 

\cite{velli89} (hereafter VGM) proposed a model of reflection-driven MHD turbulence in which the Elsasser power spectra were isotropic functions of the wavenumber~$k$, denoted $E^\pm(k)$. They divided the $\bm{z}^\pm$ fluctuations into ``primary'' and ``secondary'' components, where the primary components of $\bm{z}^\pm$ had the usual phase velocities of $\bm{U} \pm \bm{v}_{\rm A}$. The secondary components of $\bm{z}^\pm$ were driven modes produced by the reflection of $\bm{z}^\mp$ fluctuations and as a consequence had phase velocities of $\bm{U} \mp \bm{v}_{\rm A}$. VGM considered the super-Alfv\'enic solar wind at $r>r_{\rm A}$ 
and took $\bm{z}^-$ to be dominated by secondary fluctuations. VGM estimated the r.m.s. amplitude of the secondary component of $\bm{z}^-$ at scale~$k^{-1}$, which we denote $z_{k,\rm s}^-$, to be 
\begin{equation}
z_{k,\rm s}^- \sim \frac{z_{k,\rm p}^+}{k v_{\rm A}\tau_{\rm r}},
\label{eq:zkms} 
\end{equation} 
where $z_{k,\rm p}^+$ is the r.m.s. amplitude of the primary component of
$\bm{z}^+$ at scale~$1/k$, $\tau_{\rm r}$ is the reflection time scale (which depends only on the radial profile of the background flow),  $z^+_{k, \rm p}/\tau_{\rm r}$ is the rate at which $z^-_{k,\rm s}$ fluctuations are produced by the reflection of $z^+_{k,\rm p}$ fluctuations, and $1/(kv_{\rm A})$ is the time it takes for the secondary $\bm{z}^-$ fluctuations at scale~$1/k$ to propagate out of the primary $\bm{z}^+$ fluctuations that produced them. VGM argued that the secondary $\bm{z}^-$ fluctuations shear the $\bm{z}^+$ fluctuations coherently in time, since both fluctuation types have phase velocities of $\bm{U} + \bm{v}_{\rm A}$.
They then set the $\bm{z}^+$ cascade power to be
\begin{equation}
\epsilon^+ \sim k z^-_{k,\rm s} (z^+_{k,\rm p})^2 \sim \frac{ (z^+_{k,\rm p})^3}{v_{\rm A} \tau_{\rm r}}
\label{eq:epspV89} 
\end{equation} 
and took $\epsilon^+$ to be independent of~$k$, obtaining $z^+_{k,\rm p} \propto k^0$.
Equations~(\ref{eq:dzEk}) and (\ref{eq:zkms}) then yield
\begin{equation}
E^+(k) \propto k^{-1} \qquad E^-(k) \propto k^{-3}.
\label{eq:V89km1} 
\end{equation} 

It is useful to compare the VGM model with the LGS model discussed in
Section~\ref{sec:LGS07}. In both models, the $\bm{z}^-$ fluctuations are
anomalously coherent in the reference frame of the $\bm{z}^+$ fluctuations.
In the LGS model, this coherence is introduced via the ``coherence
assumption'' discussed in Section~\ref{sec:LGS07}. VGM argued that
this coherence arises because of the physics of AW reflection.  A key
difference between the models is that VGM neglected the
``tertiary'' small-scale $\bm{z}^-$ fluctuations that are produced as
secondary $\bm{z}^-$ fluctuations cascade to small scales. In the LGS
model, these tertiary $\bm{z}^-$ fluctuations are anomalously coherent in
the $\bm{z}^+$ reference frame and drive the
Elsasser spectra towards a $k_\perp^{-5/3}$ scaling rather than a
$k^{-1}$~scaling.

\subsection{Inverse cascade in reflection-driven MHD turbulence}
\label{sec:vB17} 

\cite{vanballegooijen17} carried out direct
numerical simulations of reflection-driven MHD turbulence in the solar
corona and solar wind using a methodology similar to the one we
have employed. Using their simulation data, they computed the rate~$\epsilon^\pm(k_\perp,r,t)$ at which
nonlinear interactions transfer $\bm{z}^\pm$ energy from perpendicular wavenumbers less
than~$k_\perp$ to perpendicular wavenumbers greater than~$k_\perp$
(their Equation~(17) divided by~$\rho$, with $R\rightarrow L_{\rm
  box}/2$, $f_{\pm, k} \rightarrow \phi_{\pm, k}$, and $a \rightarrow
k_\perp L_{\rm box}/2$),
\begin{equation}
\epsilon^\pm(k_\perp, r,t) = \frac{1}{[L_{\rm
    box}(r)]^2}\sum_{k_{\perp l} > k_\perp}\sum_{k_{\perp j} <
  k_\perp} \sum_{k_{\perp i}} M_{lji}
\left(k_{\perp i}^2 - k_{\perp j}^2 - k_{\perp l}^2 \right) \phi_{\pm, l} \phi_{\pm, j} \phi_{\mp,i},
\label{eq:vBeps} 
\end{equation} 
where $\phi_{\pm k}$ is the Fourier transform (in $x$ and~$y$) of the Elsasser stream
function~$\phi_\pm$ (defined such that $\bm{z}^\pm = \nabla \phi_\pm \times \bm{B}_0/B_0$),
and $M_{lji}$ is a dimensionless mode-coupling coefficient that depends upon
$k_{\perp l}$, $k_{\perp j}$, and $k_{\perp i}$, but not upon the mode
amplitudes. They found that $\epsilon^+(k_\perp, r, t)$ became negative across a
broad range of~$k_\perp$ within a
modest range of radii 
just larger than the radius (or radii) at which $\diffIL{v_{\rm A}}{r}$
changes signs -- e.g., just beyond the Alfv\'en-speed maximum at
$r=r_{\rm m} $ in the subset of their simulations in which the background
density was smooth.

To explain their findings, they considered two locations, one just
inside the $r=r_{\rm m}$ surface at $r=r_1$ and one just outside
the $r=r_{\rm m}$ surface at $r=r_2$, such that $|\diffIL{v_{\rm A}}{r}|$ was
the same at the two radii. They noted that, because $z^+_{\rm rms} \gg
z^-_{\rm rms}$, $\bm{z}^-$ fluctuations cascade much
more rapidly than $\bm{z}^+$ fluctuations. There thus exists a range of
values of $r_2 - r_1$ for which the time $\Delta t_{12}$ required for
$\bm{z}^+$ fluctuations to propagate from $r_1$ to $r_2$ is  small compared
to the outer-scale $\bm{z}^+$ cascade time scale 
but large compared to the outer-scale $\bm{z}^-$ cascade time scale. For values
of $r_2 - r_1$ in this range,
\begin{eqnarray} 
\phi_{+, k}(r_2, t) & \simeq & \phi_{+, k}(r_1, t-\Delta t_{12})
\label{eq:fplus} 
\\
\phi_{-, k}(r_2, t) & \simeq & - \phi_{-, k}(r_1, t-\Delta t_{12}),
\label{eq:fminus} 
\end{eqnarray} 
where $\Delta t_{12}$ is the $\bm{z}^+$ propagation time between $r_1$
and~$r_2$. Equation~(\ref{eq:fplus})  holds because nonlinear
interactions do not have enough time to substantially alter the $\bm{z}^+$
fluctuations during their transit from~$r_1$ to
$r_2$. Equation~(\ref{eq:fminus}) follows from 
the change in sign of $\diffIL{v_{\rm A}}{r}$ at $r=r_{\rm
  m}$.\footnote{Because the $\bm{z}^-$ fluctuations cascade very rapidly
near $r=r_{\rm m}$, 
(\ref{eq:Hf}) can be solved approximately by balancing
the last term on the left-hand side and the
first term on the right-hand side. In this approximation, 
changing the sign of $\diffIL{v_{\rm A}}{r}$ changes the sign of $\bm{f}$.}
Because $\phi_{-,k}$ changes sign and
$\phi_{+, k}$ remains almost unchanged, $\epsilon^+$ in~(\ref{eq:vBeps}) changes sign between $r_1$ and
$r_2$. Between the coronal base and the Alfv\'en-speed maximum,
nonlinear interactions set up the usual direct cascade of energy from
large scales to small scales, causing $\epsilon^+$ to be positive at
$r_1$. At $r_2$, $\epsilon^+$ thus becomes negative, indicating an
inverse cascade.

\cite{vanballegooijen17} found that as the $\bm{z}^+$ fluctuations
propagate farther beyond $r=r_{\rm m}$, they gradually adjust to the
new value of $\bm{z}^-$, and the direct cascade of energy from large
scales to small scales resumes. This transition back to a direct
cascade occurs first at large~$k_\perp$ (at which the nonlinear time
is short) and later at small~$k_\perp$.  In one of their simulations,
there is an inverse cascade of $\bm{z}^+$ energy throughout the
  region between the Alfv\'en-speed maximum at $1.4 R_{\odot}$ and an
  outer radius of~$r=2.5 R_{\odot}$. In a second simulation, there is
an inverse cascade between the Alfv\'en-speed maximum at
  $1.6 R_{\odot}$ and an outer radius of~$4 R_{\odot}$. Since the
outer-scale $\bm{z}^+$ cascade time is comparable to the time required
for $v_{\rm A}$ to change by a factor of~2 in the $\bm{z}^+$ reference
frame \citep{dmitruk02,chandran09c}, the above results indicate that
the inverse cascade persists (in the $\bm{z}^+$ reference frame) for a
time comparable to the outer-scale $\bm{z}^+$ energy-cascade time
scale.

Although $\epsilon^+$ became negative between $r\simeq r_{\rm m}$ and
$r\simeq 2 r_{\rm m}$ in the numerical simulations of
\cite{vanballegooijen17},
the energy-dissipation rate (computed from the dissipation terms added
to the governing equations) decreased by only a factor of~$\simeq 2$ within the inverse-cascade region. The
reason for this is that the direct-cascade region at $r<r_{\rm m}$ had
already ``done the work'' of transporting $\bm{z}^+$ energy to
large~$k_\perp$, and the inverse cascade between $r \simeq r_{\rm m}$
and $r\simeq 2 r_{\rm m}$ was unable to completely evacuate the high-$k_\perp$
part of the spectrum. 

\section{The Elsasser power spectra in our numerical simulations}
\label{sec:124} 

In our numerical simulations, $\alpha^+$ and $\alpha^-$
approach~$\simeq 3/2$ as $r$ increases to values $\gtrsim r_{\rm A}$,
as illustrated in figure~\ref{fig:alpha_pm}. These spectral indices
are broadly consistent with the LGS model of strong imbalanced
turbulence.  As discussed in Section~\ref{sec:LGS07}, the central
assumption of the LGS model is the ``coherence assumption'' -- that
outer-scale $\bm{z}^-$ fluctuations are injected in a manner that remains
coherent over the lifetime of the outer-scale $\bm{z}^+$ fluctuations when
viewed in the ``$\bm{z}^+$ reference frame,'' which moves along~$\bm{B}_0$
at the same velocity ($\bm{U} + \bm{v}_{\rm A}$) as the $\bm{z}^+$
fluctuations. It is difficult, at least for us, to justify this
assumption with any generality for homogeneous RMHD
turbulence. However, the coherence assumption is often satisfied in
reflection-driven MHD turbulence, because the outer-scale $\bm{z}^-$
fluctuations are produced by the reflection of outer-scale $\bm{z}^+$
fluctuations, and by definition these $\bm{z}^+$ fluctuations remain
coherent in the $\bm{z}^+$ reference frame throughout their lifetimes.  A
second requirement of the LGS model is that
$\chi^-_\lambda \gtrsim 1$. This requirement is
marginally satisfied at $r\gtrsim r_{\rm A}$ in all three simulations, as we will document in
greater detail in a separate publication. The LGS model thus provides
a credible explanation for the Elsasser power spectra at $r\gtrsim
r_{\rm A}$ in Runs~1 through~3.
The discrepancy between the predicted $\alpha^\pm = 5/3$ scaling and the measured
$\alpha^\pm \simeq 3/2$ scaling may result from some combination of
intermittency and scale-dependent dynamic alignment, as in homogeneous
RMHD turbulence (see Section~\ref{sec:strong}).

As discussed in Section~\ref{sec:powerspectra} \citep[see
also][]{velli93,reville18}, the steep Alfv\'en-speed gradient in the
upper chromosphere acts as a high-pass
filter. High-$k_\perp$ $\bm{z}^+$ fluctuations, which have
large nonlinear frequencies and hence large linear frequencies
\citep{goldreich95}, can propagate through this region with minimal
reflection. In contrast, low-$k_\perp$ $\bm{z}^+$ fluctuations undergo
strong non-WKB reflection as they propagate from the lower
chromosphere to the transition region. This selective transmission
accounts for the very small value of~$\alpha^+$ just above the
transition region in Runs~1 and~2.  The $\bm{z}^+$ spectrum in Run~3
does not flatten in the same way, presumably because the nonlinear time scale 
is larger than in Runs~1 and~2 because of the larger
value of~$L_\perp$, causing all the $\bm{z}^+$ fluctuations in Run~3 
to undergo significant reflection in the upper chromosphere.

As discussed in Section~\ref{sec:vB17}, \cite{vanballegooijen17}
showed that the $\bm{z}^+$ fluctuations undergo a transient inverse
cascade at $r_{\rm m} \lesssim r \lesssim 2 r_{\rm m}$, where
$r_{\rm m}$ is the location of the Alfv\'en-speed maximum
($1.71 R_{\odot}$ in our simulations). This inverse cascade results
from the change in sign of $\diffIL{v_{\rm A}}{r}$ at $r=r_{\rm m}$,
which reverses the direction of the fast-cascading $\bm{z}^-$
fluctuations, which in turn reverses the sign of~$\epsilon^+$
in~(\ref{eq:vBeps}).  The tendency for $\bm{z}^-$ fluctuations to
reverse direction at $r=r_{\rm m}$ destroys the anomalous coherence of
the $\bm{z}^-$ fluctuations in the $\bm{z}^+$ reference frame near
$r=r_{\rm m}$, making the LGS model inapplicable. We do not have a
detailed theory for how the spectra should scale between $r=r_{\rm m}$
and $r=2 r_{\rm m}$ in the presence of this inverse cascade, but the
simulation results indicates that the $\bm{z}^+$ spectrum flattens
significantly in this region relative to the LGS prediction.

\section{Other parameter regimes and lack of universality}
\label{sec:universality} 

One of the principal sources of uncertainty in modeling MHD turbulence
in the solar-wind acceleration region concerns the dominant length scales
and time scales of the AWs launched by the Sun. For the
correlation lengths and correlation times that we have considered in
this work, the two-component analytic model developed in
Section~\ref{sec:model} reasonably approximates our simulation
results, and the Elsasser power spectra in our simulations evolve, at least
approximately, towards the scalings of the LGS model at $r\gtrsim
r_{\rm A}$. However, we have
also carried out another simulation with higher-frequency photospheric
forcing and the same perpendicular correlation length as in Run~3. This
additional simulation is not well described by either our
two-component model or the LGS model. For example, at very large~$r$,
the $z^+$
power spectrum evolves towards a $k_\perp^{-1}$ scaling, albeit at radii
for which $\delta B \simeq B_0$. We will describe this simulation in
more detail in a future publication, but we mention it now to caution
the reader that the picture we have developed in this paper does not
apply universally for all combinations of correlation times and
correlation lengths at the photosphere.

\section{Phase Mixing}
\label{sec:phase} 

By focusing on transverse, non-compressive fluctuations and neglecting
density fluctuations, we neglect ``phase mixing'' \citep{heyvaerts83},
by which we mean the process in which an initially planar AW phase
front becomes corrugated as it propagates through a medium in which
$v_{\rm A}$  (or $U$) varies across the magnetic field. This corrugation
corresponds to a transfer of fluctuation energy to
larger~$k_\perp$. Phase mixing could provide the additional heating
that seems to be needed (see figure~\ref{fig:LAQtot}) to power the
fast solar wind at $r\lesssim 1.3 R_{\odot}$ over and above the
heating provided by reflection-driven MHD turbulence. Observations of
comet Lovejoy show that the density varies by a factor of~$\sim 6$
over distances of a few thousand~km perpendicular to~$\bm{B}_0$ at
$r= 1.3 R_{\odot}$ in both closed-field regions and open-field regions
\citep{raymond14}. On the other hand, {\em Helios} radio occultation
data indicate that the fractional density variations are
$\simeq 0.1 - 0.2$ at $5 R_{\odot} < r < 20 R_{\odot}$
\citep{hollweg10}. We conjecture that the transition from large
$\delta n/n_0$ at $r\simeq 1.3 R_{\odot}$ to small $\delta n/n_0$ at
$r\gtrsim 5 R_{\odot}$ results from mixing of density fluctuations by
the non-compressive component of the turbulence, which acts to reduce $\delta n/n_0$ as plasma flows
away from the Sun. The limited radial extent of the
large-$\delta n/n_0$ region suggests that most of the phase mixing
occurs close to the Sun. Moreover, since phase mixing is more
effective for AWs with larger parallel wavenumbers and frequencies,
phase mixing at $r\lesssim 5 R_{\odot}$ may act as a low-pass filter,
by preferentially removing high-frequency AW fluctuation energy.
Future investigations of reflection-driven MHD turbulence that account
for phase mixing will be important for developing a more complete
understanding of solar-wind turbulence and its role in the origin of
the solar wind.

\section{Conclusion}
\label{sec:conclusion} 

We have carried out three direct numerical simulations of
reflection-driven MHD turbulence within a narrow magnetic flux tube
that extends from the photosphere, through the chromosphere, through a
coronal hole, and out to a maximum heliocentric distance of
$21 R_{\odot}$. Our simulations assume fixed, observationally
motivated profiles for $\rho$, $U$, and $B_0$ and solve only for the
non-compressive, transverse components of the fluctuating magnetic
field and velocity. In each simulation, the turbulence is driven by an
imposed, randomly evolving, photospheric velocity field that has a
single characteristic time scale and length scale.  Because
outward-propagating AWs undergo strong reflection at the transition
region, there is an approximately equal mix of $\bm{z}^+$ and
$\bm{z}^-$ fluctuations in the chromosphere, and vigorous turbulence
develops within the chromosphere \citep{vanballegooijen11}. As a
result, the waves that escape into the corona have a broad spectrum of
wavenumbers and frequencies. In the corona and solar wind,
outward-propagating $\bm{z}^+$ fluctuations undergo partial non-WKB
reflection, thereby generating inward-propagating $\bm{z}^-$
fluctuations, but $z^+_{\rm rms} \gg z^-_{\rm rms}$.

In order to explain the radial profiles of $z^\pm_{\rm rms}$ and the
turbulent-heating rate in our simulations, we have developed an
analytic model of reflection-driven MHD turbulence that relies on the following
conjectures: (i) the Sun launches two populations of $\bm{z}^+$
fluctuations into the corona, a short-radial-correlation-length (HF) population and a
long-radial-correlation-length (LF) population; (ii) non-WKB
reflection of LF $\bm{z}^+$ fluctuations is the dominant source of $\bm{z}^-$
fluctuations; (iii) LF $\bm{z}^+$ fluctuations become aligned with
$\bm{z}^-$ at $r> r_{\rm m}$, where $r_{\rm m}$
is defined in (\ref{eq:rm}), causing LF $\bm{z}^+$ fluctuations to
cascade and dissipate more slowly
than HF $\bm{z}^+$ fluctuations; (iv) the change in sign of
$\d v_{\rm A}/\d r$ at $r=r_{\rm m} $ leads to a
reduction in $z^-_{\rm rms}$ at $r<r_{\rm m}$; and 
(v) $\bm{z}^-$ fluctuations are anomalously
coherent in a reference frame that moves outward with the $\bm{z}^+$
fluctuations, because the $\bm{z}^-$ fluctuations are produced by 
the outward-propagating $\bm{z}^+$ fluctuations via non-WKB reflection
\citep{velli89,verdini09a}.

To compare our analytic model and numerical results, we determine the
inner boundary conditions in our model by setting the quantities
listed in the left column of table~\ref{tab:BCs} equal to their
measured or inferred values at the coronal base in our simulations. We
then vary the five free parameters in our model (see
table~\ref{tab:bestfit}) to maximize the agreement between the model
and simulations, using a single set of free-parameter values to match
all three simulations. The resulting best-fit profiles of
$z^\pm_{\rm rms}$ and $Q$ in our model agree reasonably well with our
numerical results. The turbulent-heating rate in our simulations is
also comparable to the turbulent-heating rate in the solar-wind model
of \cite{chandran11} at $r\gtrsim 1.3 R_{\odot}$, which agreed with a
number of observational constraints. This suggests that MHD turbulence
can account for much of the heating that occurs in the fast solar
wind.

The inertial-range Elsasser power spectra in our simulations vary with
radius.  In the lower chromosphere, the spectral indices $\alpha^+$
and $\alpha^-$ (defined in~(\ref{eq:defalpha})) are $\simeq 3/2$,
consistent with theories of balanced RMHD turbulence
(Section~\ref{sec:balanced}).  In Runs~1 and~2, $\alpha^+$ drops with
increasing~$r$ in the upper chromosphere, reaching values less than~1
just above the transition region.  We attribute this spectral
flattening to the steep Alfv\'en-speed gradient in the upper
chromosphere, which acts as a high-pass filter
\citep{velli93,reville18}, as discussed in
Section~\ref{sec:powerspectra}. Much farther from the Sun, at
$r\gtrsim 10 R_{\odot}$, $\alpha^+$ and $\alpha^-$ are reasonably
close to~3/2 in all three runs, in approximate agreement with the
LGS model of strong imbalanced turbulence, which is
reviewed in Section~\ref{sec:LGS07}. However, at smaller radii,
between $r\simeq r_{\rm m} = 1.7 R_{\odot}$ and $r\simeq2 r_{\rm m}$, $\alpha^+$ 
hovers near unity in Runs~1 and~2. We attribute this latter
behavior to a disruption of the anomalous coherence of inertial-range
$\bm{z}^-$ fluctuations in the $\bm{z}^+$ reference frame. This disruption is
caused by the sign change in $\diffIL{v_{\rm A}}{r}$ at $r=r_{\rm m}$,
which, as shown by \cite{vanballegooijen17}, leads to an inverse
cascade of $\bm{z}^+$ energy in this same region (Section~\ref{sec:vB17}).

As mentioned in Section~\ref{sec:universality}, we have carried out
additional, as-yet-unpublished, numerical simulations similar to the
ones we report here, but with different photospheric boundary
conditions.  For some values of the correlation length and correlation
time of the photospheric velocity field, the fluctuations at
$r\gtrsim 10 R_{\odot}$ conform to neither the analytic
model of Section~\ref{sec:model} nor the LGS model described in
Section~\ref{sec:LGS07}. Determining how the properties of
non-compressive turbulence at $r\gtrsim 10 R_{\odot}$ depend upon the
photospheric boundary conditions remains an open
problem. Further work is also needed to determine how compressive and
non-compressive fluctuations interact and evolve as they propagate
away from the Sun and also to investigate the role of non-transverse
(e.g., spherically polarized)
fluctuations \citep[see, e.g.,][]{vasquez96,horbury18,squire19}.

Observations have led to a detailed picture of solar-wind turbulence at
$r\simeq 1 \mbox{
  au}$~\citep[e.g.,][]{belcher71,matthaeus82,bruno05,podesta07,horbury08,chen12,wicks13a}.
With the recent launch of NASA's Parker Solar Probe \citep{fox16}, it will
soon become possible to measure velocity and density fluctuations
\citep{kasper16} as well as
electric-field and
magnetic-field fluctuations \citep{bale16} at heliocentric distances
as small as~$9.8 R_{\odot}$. Such measurements will provide critical
tests for numerical and theoretical models such as the ones we have
presented here.

\acknowledgements We thank M. Asgari-Targhi, 
A. Schekochihin, and A. van Ballegooijen for helpful discussions. We also thank the three reviewers for their comments and suggestions,
which helped improve the manuscript. This work
was supported in part by  NASA grants  NNX11AJ37G, NNX15AI80,
NNX16AG81G, NNX16AH92G, NNX17AI18G, and 80NSSC19K0829, NASA
grant NNN06AA01C to the Parker Solar Probe FIELDS Experiment, and NSF
grant PHY-1500041. High-performance-computing resources were provided by the Argonne
Leadership Computing Facility (ALCF) at Argonne National Laboratory,
which is supported by the Office of Science of the U.S. Department of
Energy under contract DE-AC02-06CH11357. The ALCF resources were
granted under INCITE projects from 2012 to 2014. High-performance
computing resources were also provided by the Texas Advanced Computing
Center under the NSF-XSEDE Project TG-ATM100031.

\bibliography{articles}

\begin{thebibliography}{102}
\expandafter\ifx\csname natexlab\endcsname\relax\def\natexlab#1{#1}\fi
\def\au#1{#1} \def\ed#1{#1} \def\yr#1{#1}\def\at#1{#1}\def\jt#1{\textit{#1}}
  \def\bt#1{#1}\def\bvol#1{\textbf{#1}} \def\vol#1{#1} \def\pg#1{#1}
  \def\publ#1{#1}\def\arxiv#1{#1}\def\org#1{#1}\def\st#1{\textit{#1}}

\bibitem[{Bale} {\em et~al.\/}(2016){Bale}, {Goetz}, {Harvey}, {Turin},
  {Bonnell}, {Dudok de Wit}, {Ergun}, {MacDowall}, {Pulupa}, {Andre}, {Bolton},
  {Bougeret}, {Bowen}, {Burgess}, {Cattell}, {Chandran}, {Chaston}, {Chen},
  {Choi}, {Connerney}, {Cranmer}, {Diaz-Aguado}, {Donakowski}, {Drake},
  {Farrell}, {Fergeau}, {Fermin}, {Fischer}, {Fox}, {Glaser}, {Goldstein},
  {Gordon}, {Hanson}, {Harris}, {Hayes}, {Hinze}, {Hollweg}, {Horbury},
  {Howard}, {Hoxie}, {Jannet}, {Karlsson}, {Kasper}, {Kellogg}, {Kien},
  {Klimchuk}, {Krasnoselskikh}, {Krucker}, {Lynch}, {Maksimovic}, {Malaspina},
  {Marker}, {Martin}, {Martinez-Oliveros}, {McCauley}, {McComas}, {McDonald},
  {Meyer-Vernet}, {Moncuquet}, {Monson}, {Mozer}, {Murphy}, {Odom},
  {Oliverson}, {Olson}, {Parker}, {Pankow}, {Phan}, {Quataert}, {Quinn},
  {Ruplin}, {Salem}, {Seitz}, {Sheppard}, {Siy}, {Stevens}, {Summers}, {Szabo},
  {Timofeeva}, {Vaivads}, {Velli}, {Yehle}, {Werthimer} \& {Wygant}]{bale16}
{\sc \au{{Bale}, S.~D.}, \au{{Goetz}, K.}, \au{{Harvey}, P.~R.}, \au{{Turin},
  P.}, \au{{Bonnell}, J.~W.}, \au{{Dudok de Wit}, T.}, \au{{Ergun}, R.~E.},
  \au{{MacDowall}, R.~J.}, \au{{Pulupa}, M.}, \au{{Andre}, M.}, \au{{Bolton},
  M.}, \au{{Bougeret}, J.-L.}, \au{{Bowen}, T.~A.}, \au{{Burgess}, D.},
  \au{{Cattell}, C.~A.}, \au{{Chandran}, B.~D.~G.}, \au{{Chaston}, C.~C.},
  \au{{Chen}, C.~H.~K.}, \au{{Choi}, M.~K.}, \au{{Connerney}, J.~E.},
  \au{{Cranmer}, S.}, \au{{Diaz-Aguado}, M.}, \au{{Donakowski}, W.},
  \au{{Drake}, J.~F.}, \au{{Farrell}, W.~M.}, \au{{Fergeau}, P.}, \au{{Fermin},
  J.}, \au{{Fischer}, J.}, \au{{Fox}, N.}, \au{{Glaser}, D.}, \au{{Goldstein},
  M.}, \au{{Gordon}, D.}, \au{{Hanson}, E.}, \au{{Harris}, S.~E.}, \au{{Hayes},
  L.~M.}, \au{{Hinze}, J.~J.}, \au{{Hollweg}, J.~V.}, \au{{Horbury}, T.~S.},
  \au{{Howard}, R.~A.}, \au{{Hoxie}, V.}, \au{{Jannet}, G.}, \au{{Karlsson},
  M.}, \au{{Kasper}, J.~C.}, \au{{Kellogg}, P.~J.}, \au{{Kien}, M.},
  \au{{Klimchuk}, J.~A.}, \au{{Krasnoselskikh}, V.~V.}, \au{{Krucker}, S.},
  \au{{Lynch}, J.~J.}, \au{{Maksimovic}, M.}, \au{{Malaspina}, D.~M.},
  \au{{Marker}, S.}, \au{{Martin}, P.}, \au{{Martinez-Oliveros}, J.},
  \au{{McCauley}, J.}, \au{{McComas}, D.~J.}, \au{{McDonald}, T.},
  \au{{Meyer-Vernet}, N.}, \au{{Moncuquet}, M.}, \au{{Monson}, S.~J.},
  \au{{Mozer}, F.~S.}, \au{{Murphy}, S.~D.}, \au{{Odom}, J.}, \au{{Oliverson},
  R.}, \au{{Olson}, J.}, \au{{Parker}, E.~N.}, \au{{Pankow}, D.}, \au{{Phan},
  T.}, \au{{Quataert}, E.}, \au{{Quinn}, T.}, \au{{Ruplin}, S.~W.},
  \au{{Salem}, C.}, \au{{Seitz}, D.}, \au{{Sheppard}, D.~A.}, \au{{Siy}, A.},
  \au{{Stevens}, K.}, \au{{Summers}, D.}, \au{{Szabo}, A.}, \au{{Timofeeva},
  M.}, \au{{Vaivads}, A.}, \au{{Velli}, M.}, \au{{Yehle}, A.}, \au{{Werthimer},
  D.} \& \au{{Wygant}, J.~R.}} \yr{2016}  \at{{The FIELDS Instrument Suite for
  Solar Probe Plus. Measuring the Coronal Plasma and Magnetic Field, Plasma
  Waves and Turbulence, and Radio Signatures of Solar Transients}}.  \jt{\ssr}
  \bvol{204},  \pg{49--82}.

\bibitem[{Banerjee} {\em et~al.\/}(1998){Banerjee}, {Teriaca}, {Doyle} \&
  {Wilhelm}]{banerjee98}
{\sc \au{{Banerjee}, D.}, \au{{Teriaca}, L.}, \au{{Doyle}, J.~G.} \&
  \au{{Wilhelm}, K.}} \yr{1998}  \at{{Broadening of SI VIII lines observed in
  the solar polar coronal holes}}.  \jt{\aap}  \bvol{339},  \pg{208--214}.

\bibitem[Barnes(1966)]{barnes66}
{\sc \au{Barnes, A.}} \yr{1966}  \at{Collisionless damping of hydromagnetic
  waves}.  \jt{Physics of Fluids}  \bvol{9},  \pg{1483--1495}.

\bibitem[{Belcher} \& {Davis}(1971)]{belcher71}
{\sc \au{{Belcher}, J.~W.} \& \au{{Davis}, Jr., L.}} \yr{1971}
  \at{{Large-amplitude Alfv{\'e}n waves in the interplanetary medium, 2.}}
  \jt{\jgr}  \bvol{76},  \pg{3534--3563}.

\bibitem[{Beresnyak}(2012)]{beresnyak12}
{\sc \au{{Beresnyak}, A.}} \yr{2012}  \at{{Basic properties of
  magnetohydrodynamic turbulence in the inertial range}}.  \jt{\mnras}
  \bvol{422},  \pg{3495--3502},  \arxiv{arXiv: 1111.5329}.

\bibitem[{Beresnyak}(2014)]{beresnyak14}
{\sc \au{{Beresnyak}, A.}} \yr{2014}  \at{{Spectra of Strong
  Magnetohydrodynamic Turbulence from High-resolution Simulations}}.
  \jt{\apjl}  \bvol{784},  \pg{L20},  \arxiv{arXiv: 1401.4177}.

\bibitem[{Beresnyak} \& {Lazarian}(2008)]{beresnyak08}
{\sc \au{{Beresnyak}, A.} \& \au{{Lazarian}, A.}} \yr{2008}  \at{{Strong
  Imbalanced Turbulence}}.  \jt{\apj}  \bvol{682},  \pg{1070--1075},
  \arxiv{arXiv: 0709.0554}.

\bibitem[{Beresnyak} \& {Lazarian}(2009)]{beresnyak09}
{\sc \au{{Beresnyak}, A.} \& \au{{Lazarian}, A.}} \yr{2009}  \at{{Structure of
  Stationary Strong Imbalanced Turbulence}}.  \jt{\apj}  \bvol{702},
  \pg{460--471},  \arxiv{arXiv: 0904.2574}.

\bibitem[{Boldyrev}(2005)]{boldyrev05}
{\sc \au{{Boldyrev}, S.}} \yr{2005}  \at{{On the Spectrum of
  Magnetohydrodynamic Turbulence}}.  \jt{\apjl}  \bvol{626},  \pg{L37--L40},
  \arxiv{arXiv: arXiv:astro-ph/0503053}.

\bibitem[{Boldyrev}(2006)]{boldyrev06}
{\sc \au{{Boldyrev}, S.}} \yr{2006}  \at{{Spectrum of Magnetohydrodynamic
  Turbulence}}.  \jt{Physical Review Letters}  \bvol{96}~(11),  \pg{115002--+},
   \arxiv{arXiv: arXiv:astro-ph/0511290}.

\bibitem[{Boldyrev} \& {Perez}(2009)]{boldyrev09a}
{\sc \au{{Boldyrev}, S.} \& \au{{Perez}, J.~C.}} \yr{2009}  \at{{Spectrum of
  Weak Magnetohydrodynamic Turbulence}}.  \jt{Physical Review Letters}
  \bvol{103}~(22),  \pg{225001},  \arxiv{arXiv: 0907.4475}.

\bibitem[{Boldyrev} {\em et~al.\/}(2011){Boldyrev}, {Perez}, {Borovsky} \&
  {Podesta}]{boldyrev11a}
{\sc \au{{Boldyrev}, S.}, \au{{Perez}, J.~C.}, \au{{Borovsky}, J.~E.} \&
  \au{{Podesta}, J.~J.}} \yr{2011}  \at{{Spectral Scaling Laws in
  Magnetohydrodynamic Turbulence Simulations and in the Solar Wind}}.
  \jt{\apjl}  \bvol{741},  \pg{L19},  \arxiv{arXiv: 1106.0700}.

\bibitem[{Bruno} \& {Carbone}(2005)]{bruno05}
{\sc \au{{Bruno}, R.} \& \au{{Carbone}, V.}} \yr{2005}  \at{{The Solar Wind as
  a Turbulence Laboratory}}.  \jt{Living Reviews in Solar Physics}  \bvol{2},
  \pg{4--+}.

\bibitem[Canuto {\em et~al.\/}(1988)Canuto, Hussaini, Quarteroni \&
  Zang]{CanHus88}
{\sc \au{Canuto, C.}, \au{Hussaini, M.}, \au{Quarteroni, A.} \& \au{Zang, T.}}
  \yr{1988} {\em Spectral Methods in Fluid Dynamics\/}.
  \publ{Springer-Verlag}.

\bibitem[{Chandran}(2005)]{chandran05a}
{\sc \au{{Chandran}, B.~D.~G.}} \yr{2005}  \at{{Weak Compressible
  Magnetohydrodynamic Turbulence in the Solar Corona}}.  \jt{Phys. Rev. Lett.}
  \bvol{95}~(26),  \pg{265004--+},  \arxiv{arXiv: arXiv:astro-ph/0511586}.

\bibitem[{Chandran}(2008)]{chandran08b}
{\sc \au{{Chandran}, B.~D.~G.}} \yr{2008}  \at{{Weakly Turbulent
  Magnetohydrodynamic Waves in Compressible Low-{$\beta$} Plasmas}}.  \jt{Phys.
  Rev. Lett.}  \bvol{101}~(23),  \pg{235004--+},  \arxiv{arXiv: 0810.5360}.

\bibitem[{Chandran}(2008a)]{chandran08a}
{\sc \au{{Chandran}, B.~D.~G.}} \yr{2008a}  \at{{Strong Anisotropic MHD
  Turbulence with Cross Helicity}}.  \jt{\apj}  \bvol{685},  \pg{646--658},
  \arxiv{arXiv: 0801.4903}.

\bibitem[{Chandran}(2018)]{chandran18a}
{\sc \au{{Chandran}, B. D.~G.}} \yr{2018}  \at{{Parametric instability, inverse
  cascade and the range of solar-wind turbulence}}.  \jt{Journal of Plasma
  Physics}  \bvol{84},  \pg{905840106}.

\bibitem[{Chandran} {\em et~al.\/}(2011){Chandran}, {Dennis}, {Quataert} \&
  {Bale}]{chandran11}
{\sc \au{{Chandran}, B.~D.~G.}, \au{{Dennis}, T.~J.}, \au{{Quataert}, E.} \&
  \au{{Bale}, S.~D.}} \yr{2011}  \at{{Incorporating Kinetic Physics into a
  Two-fluid Solar-wind Model with Temperature Anisotropy and Low-frequency
  Alfv{\'e}n-wave Turbulence}}.  \jt{\apj}  \bvol{743},  \pg{197},
  \arxiv{arXiv: 1110.3029}.

\bibitem[{Chandran} \& {Hollweg}(2009)]{chandran09c}
{\sc \au{{Chandran}, B.~D.~G.} \& \au{{Hollweg}, J.~V.}} \yr{2009}
  \at{{Alfv{\'e}n Wave Reflection and Turbulent Heating in the Solar Wind from
  1 Solar Radius to 1 AU: An Analytical Treatment}}.  \jt{\apj}  \bvol{707},
  \pg{1659--1667},  \arxiv{arXiv: 0911.1068}.

\bibitem[{Chandran} {\em et~al.\/}(2015{\natexlab{{\em a\/}}}){Chandran},
  {Perez}, {Verscharen}, {Klein} \& {Mallet}]{chandran15b}
{\sc \au{{Chandran}, B.~D.~G.}, \au{{Perez}, J.~C.}, \au{{Verscharen}, D.},
  \au{{Klein}, K.~G.} \& \au{{Mallet}, A.}} \yr{2015{\natexlab{{\em a\/}}}}
  \at{{On the Conservation of Cross Helicity and Wave Action in Solar-wind
  Models with Non-WKB Alfv{\'e}n Wave Reflection}}.  \jt{\apj}  \bvol{811},
  \pg{50},  \arxiv{arXiv: 1509.01135}.

\bibitem[{Chandran} {\em et~al.\/}(2015{\natexlab{{\em b\/}}}){Chandran},
  {Schekochihin} \& {Mallet}]{chandran15}
{\sc \au{{Chandran}, B.~D.~G.}, \au{{Schekochihin}, A.~A.} \& \au{{Mallet},
  A.}} \yr{2015{\natexlab{{\em b\/}}}}  \at{{Intermittency and Alignment in
  Strong RMHD Turbulence}}.  \jt{\apj}  \bvol{807},  \pg{39},  \arxiv{arXiv:
  1403.6354}.

\bibitem[{Chen} {\em et~al.\/}(2012){Chen}, {Mallet}, {Schekochihin},
  {Horbury}, {Wicks} \& {Bale}]{chen12}
{\sc \au{{Chen}, C.~H.~K.}, \au{{Mallet}, A.}, \au{{Schekochihin}, A.~A.},
  \au{{Horbury}, T.~S.}, \au{{Wicks}, R.~T.} \& \au{{Bale}, S.~D.}} \yr{2012}
  \at{{Three-dimensional Structure of Solar Wind Turbulence}}.  \jt{\apj}
  \bvol{758},  \pg{120},  \arxiv{arXiv: 1109.2558}.

\bibitem[{Cho} \& {Lazarian}(2003)]{cho03a}
{\sc \au{{Cho}, J.} \& \au{{Lazarian}, A.}} \yr{2003}  \at{Compressible
  magnetohydrodynamic turbulence: Mode coupling, scaling relations, anisotropy,
  viscosity-damped regime and astrophysical implications}.  \jt{\mnras}
  \bvol{345},  \pg{325--339}.

\bibitem[{Cho} \& {Vishniac}(2000)]{cho00}
{\sc \au{{Cho}, J.} \& \au{{Vishniac}, E.~T.}} \yr{2000}  \at{{The Anisotropy
  of Magnetohydrodynamic Alfv{\'e}nic Turbulence}}.  \jt{\apj}  \bvol{539},
  \pg{273--282},  \arxiv{arXiv: arXiv:astro-ph/0003403}.

\bibitem[{Cohen} \& {Dewar}(1974)]{cohen74}
{\sc \au{{Cohen}, R.~H.} \& \au{{Dewar}, R.~L.}} \yr{1974}  \at{{On the
  backscatter instability of solar wind Alfven waves}}.  \jt{\jgr}  \bvol{79},
  \pg{4174--4178}.

\bibitem[{Cranmer} \& {van Ballegooijen}(2005)]{cranmer05}
{\sc \au{{Cranmer}, S.~R.} \& \au{{van Ballegooijen}, A.~A.}} \yr{2005}  \at{On
  the generation, propagation, and reflection of {A}lfv{\' e}n waves from the
  solar photosphere to the distant heliosphere}.  \jt{\apjs}  \bvol{156},
  \pg{265--293}.

\bibitem[{Cranmer} {\em et~al.\/}(2007){Cranmer}, {van Ballegooijen} \&
  {Edgar}]{cranmer07}
{\sc \au{{Cranmer}, S.~R.}, \au{{van Ballegooijen}, A.~A.} \& \au{{Edgar},
  R.~J.}} \yr{2007}  \at{{Self-consistent Coronal Heating and Solar Wind
  Acceleration from Anisotropic Magnetohydrodynamic Turbulence}}.  \jt{\apjs}
  \bvol{171},  \pg{520--551},  \arxiv{arXiv: arXiv:astro-ph/0703333}.

\bibitem[{De Pontieu} {\em et~al.\/}(2007){De Pontieu}, {McIntosh}, {Carlsson},
  {Hansteen}, {Tarbell}, {Schrijver}, {Title}, {Shine}, {Tsuneta}, {Katsukawa},
  {Ichimoto}, {Suematsu}, {Shimizu} \& {Nagata}]{depontieu07}
{\sc \au{{De Pontieu}, B.}, \au{{McIntosh}, S.~W.}, \au{{Carlsson}, M.},
  \au{{Hansteen}, V.~H.}, \au{{Tarbell}, T.~D.}, \au{{Schrijver}, C.~J.},
  \au{{Title}, A.~M.}, \au{{Shine}, R.~A.}, \au{{Tsuneta}, S.},
  \au{{Katsukawa}, Y.}, \au{{Ichimoto}, K.}, \au{{Suematsu}, Y.},
  \au{{Shimizu}, T.} \& \au{{Nagata}, S.}} \yr{2007}  \at{{Chromospheric
  Alfv{\'e}nic Waves Strong Enough to Power the Solar Wind}}.  \jt{Science}
  \bvol{318},  \pg{1574--7}.

\bibitem[{Dmitruk} \& {Matthaeus}(2003)]{dmitruk03}
{\sc \au{{Dmitruk}, P.} \& \au{{Matthaeus}, W.~H.}} \yr{2003}
  \at{{Low-Frequency Waves and Turbulence in an Open Magnetic Region:
  Timescales and Heating Efficiency}}.  \jt{\apj}  \bvol{597},
  \pg{1097--1105}.

\bibitem[{Dmitruk} {\em et~al.\/}(2002){Dmitruk}, {Matthaeus}, {Milano},
  {Oughton}, {Zank} \& {Mullan}]{dmitruk02}
{\sc \au{{Dmitruk}, P.}, \au{{Matthaeus}, W.~H.}, \au{{Milano}, L.~J.},
  \au{{Oughton}, S.}, \au{{Zank}, G.~P.} \& \au{{Mullan}, D.~J.}} \yr{2002}
  \at{Coronal heating distribution due to low-frequency, wave-driven
  turbulence}.  \jt{\apj}  \bvol{575},  \pg{571--577}.

\bibitem[{Esser} {\em et~al.\/}(1999){Esser}, {Fineschi}, {Dobrzycka},
  {Habbal}, {Edgar}, {Raymond}, {Kohl} \& {Guhathakurta}]{esser99}
{\sc \au{{Esser}, R.}, \au{{Fineschi}, S.}, \au{{Dobrzycka}, D.}, \au{{Habbal},
  S.~R.}, \au{{Edgar}, R.~J.}, \au{{Raymond}, J.~C.}, \au{{Kohl}, J.~L.} \&
  \au{{Guhathakurta}, M.}} \yr{1999}  \at{Plasma properties in coronal holes
  derived from measurements of minor ion spectral lines and polarized white
  light intensity}.  \jt{\apjl}  \bvol{510},  \pg{L63--L67}.

\bibitem[{Feldman} {\em et~al.\/}(1997){Feldman}, {Habbal}, {Hoogeveen} \&
  {Wang}]{feldman97}
{\sc \au{{Feldman}, W.~C.}, \au{{Habbal}, S.~R.}, \au{{Hoogeveen}, G.} \&
  \au{{Wang}, Y.}} \yr{1997}  \at{{Experimental constraints on pulsed and
  steady state models of the solar wind near the Sun}}.  \jt{\jgr}  \bvol{102},
   \pg{26905--26918}.

\bibitem[{Fox} {\em et~al.\/}(2016){Fox}, {Velli}, {Bale}, {Decker},
  {Driesman}, {Howard}, {Kasper}, {Kinnison}, {Kusterer}, {Lario}, {Lockwood},
  {McComas}, {Raouafi} \& {Szabo}]{fox16}
{\sc \au{{Fox}, N.~J.}, \au{{Velli}, M.~C.}, \au{{Bale}, S.~D.}, \au{{Decker},
  R.}, \au{{Driesman}, A.}, \au{{Howard}, R.~A.}, \au{{Kasper}, J.~C.},
  \au{{Kinnison}, J.}, \au{{Kusterer}, M.}, \au{{Lario}, D.}, \au{{Lockwood},
  M.~K.}, \au{{McComas}, D.~J.}, \au{{Raouafi}, N.~E.} \& \au{{Szabo}, A.}}
  \yr{2016}  \at{{The Solar Probe Plus Mission: Humanity's First Visit to Our
  Star}}.  \jt{\ssr}  \bvol{204},  \pg{7--48}.

\bibitem[{Galeev} \& {Oraevskii}(1963)]{galeev63}
{\sc \au{{Galeev}, A.~A.} \& \au{{Oraevskii}, V.~N.}} \yr{1963}  \at{{The
  Stability of Alfv{\'e}n Waves}}.  \jt{Soviet Physics Doklady}  \bvol{7},
  \pg{988}.

\bibitem[{Galtier} {\em et~al.\/}(2000){Galtier}, {Nazarenko}, {Newell} \&
  {Pouquet}]{galtier00}
{\sc \au{{Galtier}, S.}, \au{{Nazarenko}, S.~V.}, \au{{Newell}, A.~C.} \&
  \au{{Pouquet}, A.}} \yr{2000}  \at{A weak turbulence theory for
  incompressible magnetohydrodynamics}.  \jt{Journal of Plasma Physics}
  \bvol{63},  \pg{447--488}.

\bibitem[{Goldreich} \& {Sridhar}(1995)]{goldreich95}
{\sc \au{{Goldreich}, P.} \& \au{{Sridhar}, S.}} \yr{1995}  \at{{Toward a
  theory of interstellar turbulence. 2: Strong alfvenic turbulence}}.
  \jt{\apj}  \bvol{438},  \pg{763--775}.

\bibitem[Goldreich \& Sridhar(1997)]{goldreich97}
{\sc \au{Goldreich, P.} \& \au{Sridhar, S.}} \yr{1997}  \at{Magnetohydrodynamic
  turbulence revisited}.  \jt{\apj}  \bvol{485},  \pg{680--688}.

\bibitem[{Heinemann} \& {Olbert}(1980)]{heinemann80}
{\sc \au{{Heinemann}, M.} \& \au{{Olbert}, S.}} \yr{1980}  \at{{Non-WKB Alfven
  waves in the solar wind}}.  \jt{\jgr}  \bvol{85},  \pg{1311--1327}.

\bibitem[{Heyvaerts} \& {Priest}(1983)]{heyvaerts83}
{\sc \au{{Heyvaerts}, J.} \& \au{{Priest}, E.~R.}} \yr{1983}  \at{{Coronal
  heating by phase-mixed shear Alfven waves}}.  \jt{\aap}  \bvol{117},
  \pg{220--234}.

\bibitem[{Hollweg} {\em et~al.\/}(2010){Hollweg}, {Cranmer} \&
  {Chandran}]{hollweg10}
{\sc \au{{Hollweg}, J.~V.}, \au{{Cranmer}, S.~R.} \& \au{{Chandran}, B.~D.~G.}}
  \yr{2010}  \at{{Coronal Faraday Rotation Fluctuations and a
  Wave/Turbulence-driven Model of the Solar Wind}}.  \jt{\apj}  \bvol{722},
  \pg{1495--1503}.

\bibitem[{Hollweg} \& {Isenberg}(2002)]{hollweg02}
{\sc \au{{Hollweg}, J.~V.} \& \au{{Isenberg}, P.~A.}} \yr{2002}
  \at{{Generation of the fast solar wind: A review with emphasis on the
  resonant cyclotron interaction}}.  \jt{Journal of Geophysical Research (Space
  Physics)}  \bvol{107},  \pg{1147--+}.

\bibitem[{Hollweg} \& {Isenberg}(2007)]{hollweg07}
{\sc \au{{Hollweg}, J.~V.} \& \au{{Isenberg}, P.~A.}} \yr{2007}
  \at{{Reflection of Alfv{\'e}n waves in the corona and solar wind: An impulse
  function approach}}.  \jt{Journal of Geophysical Research (Space Physics)}
  \bvol{112},  \pg{8102--+}.

\bibitem[{Horbury} {\em et~al.\/}(2008){Horbury}, {Forman} \&
  {Oughton}]{horbury08}
{\sc \au{{Horbury}, T.~S.}, \au{{Forman}, M.} \& \au{{Oughton}, S.}} \yr{2008}
  \at{{Anisotropic Scaling of Magnetohydrodynamic Turbulence}}.  \jt{Physical
  Review Letters}  \bvol{101}~(17),  \pg{175005--+},  \arxiv{arXiv: 0807.3713}.

\bibitem[{Horbury} {\em et~al.\/}(2018){Horbury}, {Matteini} \&
  {Stansby}]{horbury18}
{\sc \au{{Horbury}, T.~S.}, \au{{Matteini}, L.} \& \au{{Stansby}, D.}}
  \yr{2018}  \at{{Short, large-amplitude speed enhancements in the near-Sun
  fast solar wind}}.  \jt{\mnras}  \bvol{478},  \pg{1980--1986}.

\bibitem[{Kasper} {\em et~al.\/}(2016){Kasper}, {Abiad}, {Austin},
  {Balat-Pichelin}, {Bale}, {Belcher}, {Berg}, {Bergner}, {Berthomier},
  {Bookbinder}, {Brodu}, {Caldwell}, {Case}, {Chandran}, {Cheimets}, {Cirtain},
  {Cranmer}, {Curtis}, {Daigneau}, {Dalton}, {Dasgupta}, {DeTomaso},
  {Diaz-Aguado}, {Djordjevic}, {Donaskowski}, {Effinger}, {Florinski}, {Fox},
  {Freeman}, {Gallagher}, {Gary}, {Gauron}, {Gates}, {Goldstein}, {Golub},
  {Gordon}, {Gurnee}, {Guth}, {Halekas}, {Hatch}, {Heerikuisen}, {Ho}, {Hu},
  {Johnson}, {Jordan}, {Korreck}, {Larson}, {Lazarus}, {Li}, {Livi}, {Ludlam},
  {Maksimovic}, {McFadden}, {Marchant}, {Maruca}, {McComas}, {Messina},
  {Mercer}, {Park}, {Peddie}, {Pogorelov}, {Reinhart}, {Richardson},
  {Robinson}, {Rosen}, {Skoug}, {Slagle}, {Steinberg}, {Stevens}, {Szabo},
  {Taylor}, {Tiu}, {Turin}, {Velli}, {Webb}, {Whittlesey}, {Wright}, {Wu} \&
  {Zank}]{kasper16}
{\sc \au{{Kasper}, J.~C.}, \au{{Abiad}, R.}, \au{{Austin}, G.},
  \au{{Balat-Pichelin}, M.}, \au{{Bale}, S.~D.}, \au{{Belcher}, J.~W.},
  \au{{Berg}, P.}, \au{{Bergner}, H.}, \au{{Berthomier}, M.}, \au{{Bookbinder},
  J.}, \au{{Brodu}, E.}, \au{{Caldwell}, D.}, \au{{Case}, A.~W.},
  \au{{Chandran}, B.~D.~G.}, \au{{Cheimets}, P.}, \au{{Cirtain}, J.~W.},
  \au{{Cranmer}, S.~R.}, \au{{Curtis}, D.~W.}, \au{{Daigneau}, P.},
  \au{{Dalton}, G.}, \au{{Dasgupta}, B.}, \au{{DeTomaso}, D.},
  \au{{Diaz-Aguado}, M.}, \au{{Djordjevic}, B.}, \au{{Donaskowski}, B.},
  \au{{Effinger}, M.}, \au{{Florinski}, V.}, \au{{Fox}, N.}, \au{{Freeman},
  M.}, \au{{Gallagher}, D.}, \au{{Gary}, S.~P.}, \au{{Gauron}, T.},
  \au{{Gates}, R.}, \au{{Goldstein}, M.}, \au{{Golub}, L.}, \au{{Gordon},
  D.~A.}, \au{{Gurnee}, R.}, \au{{Guth}, G.}, \au{{Halekas}, J.}, \au{{Hatch},
  K.}, \au{{Heerikuisen}, J.}, \au{{Ho}, G.}, \au{{Hu}, Q.}, \au{{Johnson},
  G.}, \au{{Jordan}, S.~P.}, \au{{Korreck}, K.~E.}, \au{{Larson}, D.},
  \au{{Lazarus}, A.~J.}, \au{{Li}, G.}, \au{{Livi}, R.}, \au{{Ludlam}, M.},
  \au{{Maksimovic}, M.}, \au{{McFadden}, J.~P.}, \au{{Marchant}, W.},
  \au{{Maruca}, B.~A.}, \au{{McComas}, D.~J.}, \au{{Messina}, L.},
  \au{{Mercer}, T.}, \au{{Park}, S.}, \au{{Peddie}, A.~M.}, \au{{Pogorelov},
  N.}, \au{{Reinhart}, M.~J.}, \au{{Richardson}, J.~D.}, \au{{Robinson}, M.},
  \au{{Rosen}, I.}, \au{{Skoug}, R.~M.}, \au{{Slagle}, A.}, \au{{Steinberg},
  J.~T.}, \au{{Stevens}, M.~L.}, \au{{Szabo}, A.}, \au{{Taylor}, E.~R.},
  \au{{Tiu}, C.}, \au{{Turin}, P.}, \au{{Velli}, M.}, \au{{Webb}, G.},
  \au{{Whittlesey}, P.}, \au{{Wright}, K.}, \au{{Wu}, S.~T.} \& \au{{Zank},
  G.}} \yr{2016}  \at{{Solar Wind Electrons Alphas and Protons (SWEAP)
  Investigation: Design of the Solar Wind and Coronal Plasma Instrument Suite
  for Solar Probe Plus}}.  \jt{\ssr}  \bvol{204},  \pg{131--186}.

\bibitem[{Lithwick} \& {Goldreich}(2003)]{lithwick03}
{\sc \au{{Lithwick}, Y.} \& \au{{Goldreich}, P.}} \yr{2003}  \at{Imbalanced
  weak magnetohydrodynamic turbulence}.  \jt{\apj}  \bvol{582},
  \pg{1220--1240},  \arxiv{arXiv: astro-ph/0208046}.

\bibitem[{Lithwick} {\em et~al.\/}(2007){Lithwick}, {Goldreich} \&
  {Sridhar}]{lithwick07}
{\sc \au{{Lithwick}, Y.}, \au{{Goldreich}, P.} \& \au{{Sridhar}, S.}} \yr{2007}
   \at{Imbalanced strong {MHD} turbulence}.  \jt{\apj}  \bvol{655},
  \pg{269--274},  \arxiv{arXiv: astro-ph/0607243}.

\bibitem[{Loureiro} \& {Boldyrev}(2017{\natexlab{{\em a\/}}})]{loureiro17b}
{\sc \au{{Loureiro}, N.~F.} \& \au{{Boldyrev}, S.}} \yr{2017{\natexlab{{\em
  a\/}}}}  \at{{Collisionless Reconnection in Magnetohydrodynamic and Kinetic
  Turbulence}}.  \jt{\apj}  \bvol{850},  \pg{182}.

\bibitem[{Loureiro} \& {Boldyrev}(2017{\natexlab{{\em b\/}}})]{loureiro17}
{\sc \au{{Loureiro}, N.~F.} \& \au{{Boldyrev}, S.}} \yr{2017{\natexlab{{\em
  b\/}}}}  \at{{Role of Magnetic Reconnection in Magnetohydrodynamic
  Turbulence}}.  \jt{Physical Review Letters}  \bvol{118}~(24),  \pg{245101}.

\bibitem[{Luo} \& {Melrose}(2006)]{luo06}
{\sc \au{{Luo}, Q.} \& \au{{Melrose}, D.}} \yr{2006}  \at{Anisotropic weak
  turbulence of {A}lfv{\'e}n waves in collisionless astrophysical plasmas}.
  \jt{\mnras}  \bvol{368},  \pg{1151--1158},  \arxiv{arXiv:
  arXiv:astro-ph/0602295}.

\bibitem[{Mallet} \& {Schekochihin}(2017)]{mallet17a}
{\sc \au{{Mallet}, A.} \& \au{{Schekochihin}, A.~A.}} \yr{2017}  \at{{A
  statistical model of three-dimensional anisotropy and intermittency in strong
  Alfv{\'e}nic turbulence}}.  \jt{\mnras}  \bvol{466},  \pg{3918--3927},
  \arxiv{arXiv: 1606.00466}.

\bibitem[{Mallet} {\em et~al.\/}(2015){Mallet}, {Schekochihin} \&
  {Chandran}]{mallet15}
{\sc \au{{Mallet}, A.}, \au{{Schekochihin}, A.~A.} \& \au{{Chandran},
  B.~D.~G.}} \yr{2015}  \at{{Refined critical balance in strong Alfv{\'e}nic
  turbulence}}.  \jt{\mnras}  \bvol{449},  \pg{L77--L81},  \arxiv{arXiv:
  1406.5658}.

\bibitem[{Mallet} {\em et~al.\/}(2017{\natexlab{{\em a\/}}}){Mallet},
  {Schekochihin} \& {Chandran}]{mallet17c}
{\sc \au{{Mallet}, A.}, \au{{Schekochihin}, A.~A.} \& \au{{Chandran}, B.
  D.~G.}} \yr{2017{\natexlab{{\em a\/}}}}  \at{{Disruption of Alfv{\'e}nic
  turbulence by magnetic reconnection in a collisionless plasma}}.  \jt{Journal
  of Plasma Physics}  \bvol{83},  \pg{905830609}.

\bibitem[{Mallet} {\em et~al.\/}(2017{\natexlab{{\em b\/}}}){Mallet},
  {Schekochihin} \& {Chandran}]{mallet17b}
{\sc \au{{Mallet}, A.}, \au{{Schekochihin}, A.~A.} \& \au{{Chandran},
  B.~D.~G.}} \yr{2017{\natexlab{{\em b\/}}}}  \at{{Disruption of sheet-like
  structures in Alfv{\'e}nic turbulence by magnetic reconnection}}.
  \jt{\mnras}  \bvol{468},  \pg{4862--4871},  \arxiv{arXiv: 1612.07604}.

\bibitem[{Maron} \& {Goldreich}(2001)]{maron01}
{\sc \au{{Maron}, J.} \& \au{{Goldreich}, P.}} \yr{2001}  \at{{Simulations of
  Incompressible Magnetohydrodynamic Turbulence}}.  \jt{\apj}  \bvol{554},
  \pg{1175--1196},  \arxiv{arXiv: arXiv:astro-ph/0012491}.

\bibitem[{Mason} {\em et~al.\/}(2008){Mason}, {Cattaneo} \&
  {Boldyrev}]{mason08}
{\sc \au{{Mason}, J.}, \au{{Cattaneo}, F.} \& \au{{Boldyrev}, S.}} \yr{2008}
  \at{{Numerical measurements of the spectrum in magnetohydrodynamic
  turbulence}}.  \jt{\pre}  \bvol{77}~(3),  \pg{036403--+},  \arxiv{arXiv:
  0706.2003}.

\bibitem[{Matthaeus} \& {Goldstein}(1982)]{matthaeus82}
{\sc \au{{Matthaeus}, W.~H.} \& \au{{Goldstein}, M.~L.}} \yr{1982}
  \at{Measurement of the rugged invariants of magnetohydrodynamic turbulence in
  the solar wind}.  \jt{\jgr}  \bvol{87},  \pg{6011--6028}.

\bibitem[{Meyrand} {\em et~al.\/}(2019){Meyrand}, {Kanekar}, {Dorland} \&
  {Schekochihin}]{meyrand19}
{\sc \au{{Meyrand}, R.}, \au{{Kanekar}, A.}, \au{{Dorland}, W.} \&
  \au{{Schekochihin}, A.~A.}} \yr{2019}  \at{{Fluidization of collisionless
  plasma turbulence}}.  \jt{Proceedings of the National Academy of Science}
  \bvol{116},  \pg{1185--1194},  \arxiv{arXiv: 1808.04284}.

\bibitem[{Meyrand} {\em et~al.\/}(2015){Meyrand}, {Kiyani} \&
  {Galtier}]{meyrand15}
{\sc \au{{Meyrand}, R.}, \au{{Kiyani}, K.~H.} \& \au{{Galtier}, S.}} \yr{2015}
  \at{{Weak magnetohydrodynamic turbulence and intermittency}}.  \jt{Journal of
  Fluid Mechanics}  \bvol{770},  \pg{R1}.

\bibitem[{M{\"u}ller} \& {Grappin}(2005)]{muller05}
{\sc \au{{M{\"u}ller}, W.} \& \au{{Grappin}, R.}} \yr{2005}  \at{{Spectral
  Energy Dynamics in Magnetohydrodynamic Turbulence}}.  \jt{Physical Review
  Letters}  \bvol{95}~(11),  \pg{114502--+},  \arxiv{arXiv:
  arXiv:physics/0509019}.

\bibitem[{Ng} \& {Bhattacharjee}(1996)]{ng96}
{\sc \au{{Ng}, C.~S.} \& \au{{Bhattacharjee}, A.}} \yr{1996}  \at{{Interaction
  of Shear-Alfven Wave Packets: Implication for Weak Magnetohydrodynamic
  Turbulence in Astrophysical Plasmas}}.  \jt{\apj}  \bvol{465},  \pg{845--+}.

\bibitem[{Ng} \& {Bhattacharjee}(1997)]{ng97}
{\sc \au{{Ng}, C.~S.} \& \au{{Bhattacharjee}, A.}} \yr{1997}  \at{{Scaling of
  anisotropic spectra due to the weak interaction of shear-Alfv{\'e}n wave
  packets}}.  \jt{Physics of Plasmas}  \bvol{4},  \pg{605--610}.

\bibitem[{Perez} \& {Boldyrev}(2008)]{perez08a}
{\sc \au{{Perez}, J.~C.} \& \au{{Boldyrev}, S.}} \yr{2008}  \at{{On Weak and
  Strong Magnetohydrodynamic Turbulence}}.  \jt{\apjl}  \bvol{672},
  \pg{L61--L64},  \arxiv{arXiv: arXiv:0712.2086}.

\bibitem[{Perez} \& {Boldyrev}(2009)]{perez09a}
{\sc \au{{Perez}, J.~C.} \& \au{{Boldyrev}, S.}} \yr{2009}  \at{{Role of
  Cross-Helicity in Magnetohydrodynamic Turbulence}}.  \jt{Physical Review
  Letters}  \bvol{102}~(2),  \pg{025003--+},  \arxiv{arXiv: 0807.2635}.

\bibitem[{Perez} \& {Boldyrev}(2010)]{perez10b}
{\sc \au{{Perez}, J.~C.} \& \au{{Boldyrev}, S.}} \yr{2010}  \at{{Strong
  magnetohydrodynamic turbulence with cross helicity}}.  \jt{Physics of
  Plasmas}  \bvol{17}~(5),  \pg{055903},  \arxiv{arXiv: 1004.3798}.

\bibitem[{Perez} \& {Chandran}(2013)]{perez13}
{\sc \au{{Perez}, J.~C.} \& \au{{Chandran}, B.~D.~G.}} \yr{2013}  \at{{Direct
  Numerical Simulations of Reflection-Driven, Reduced MHD Turbulence from the
  Sun to the Alfven Critical Point}}.  \jt{\apj}  \bvol{776},  \pg{124},
  \arxiv{arXiv: 1308.4046}.

\bibitem[{Perez} {\em et~al.\/}(2012){Perez}, {Mason}, {Boldyrev} \&
  {Cattaneo}]{perez12}
{\sc \au{{Perez}, J.~C.}, \au{{Mason}, J.}, \au{{Boldyrev}, S.} \&
  \au{{Cattaneo}, F.}} \yr{2012}  \at{{On the Energy Spectrum of Strong
  Magnetohydrodynamic Turbulence}}.  \jt{Physical Review X}  \bvol{2}~(4),
  \pg{041005},  \arxiv{arXiv: 1209.2011}.

\bibitem[{Podesta} \& {Bhattacharjee}(2010)]{podesta10}
{\sc \au{{Podesta}, J.~J.} \& \au{{Bhattacharjee}, A.}} \yr{2010}  \at{{Theory
  of Incompressible Magnetohydrodynamic Turbulence with Scale-dependent
  Alignment and Cross-helicity}}.  \jt{\apj}  \bvol{718},  \pg{1151--1157},
  \arxiv{arXiv: 0903.5041}.

\bibitem[{Podesta} {\em et~al.\/}(2007){Podesta}, {Roberts} \&
  {Goldstein}]{podesta07}
{\sc \au{{Podesta}, J.~J.}, \au{{Roberts}, D.~A.} \& \au{{Goldstein}, M.~L.}}
  \yr{2007}  \at{{Spectral Exponents of Kinetic and Magnetic Energy Spectra in
  Solar Wind Turbulence}}.  \jt{\apj}  \bvol{664},  \pg{543--548}.

\bibitem[{Pucci} \& {Velli}(2014)]{pucci14}
{\sc \au{{Pucci}, F.} \& \au{{Velli}, M.}} \yr{2014}  \at{{Reconnection of
  Quasi-singular Current Sheets: The ``Ideal'' Tearing Mode}}.  \jt{\apjl}
  \bvol{780},  \pg{L19}.

\bibitem[{Pucci} {\em et~al.\/}(2018){Pucci}, {Velli}, {Tenerani} \& {Del
  Sarto}]{pucci18}
{\sc \au{{Pucci}, F.}, \au{{Velli}, M.}, \au{{Tenerani}, A.} \& \au{{Del
  Sarto}, D.}} \yr{2018}  \at{{Onset of fast ``ideal'' tearing in thin current
  sheets: Dependence on the equilibrium current profile}}.  \jt{Physics of
  Plasmas}  \bvol{25}~(3),  \pg{032113},  \arxiv{arXiv: 1801.08412}.

\bibitem[{Raymond} {\em et~al.\/}(2014){Raymond}, {McCauley}, {Cranmer} \&
  {Downs}]{raymond14}
{\sc \au{{Raymond}, J.~C.}, \au{{McCauley}, P.~I.}, \au{{Cranmer}, S.~R.} \&
  \au{{Downs}, C.}} \yr{2014}  \at{{The Solar Corona as Probed by Comet Lovejoy
  (C/2011 W3)}}.  \jt{\apj}  \bvol{788},  \pg{152},  \arxiv{arXiv: 1405.1639}.

\bibitem[{R{\'e}ville} {\em et~al.\/}(2018){R{\'e}ville}, {Tenerani} \&
  {Velli}]{reville18}
{\sc \au{{R{\'e}ville}, V.}, \au{{Tenerani}, A.} \& \au{{Velli}, M.}} \yr{2018}
   \at{{Parametric Decay and the Origin of the Low-frequency Alfv{\'e}nic
  Spectrum of the Solar Wind}}.  \jt{\apj}  \bvol{866},  \pg{38}.

\bibitem[{Richardson} \& {Schwarzschild}(1950)]{richardson50}
{\sc \au{{Richardson}, R.~S.} \& \au{{Schwarzschild}, M.}} \yr{1950}  \at{{On
  the Turbulent Velocities of Solar Granules.}}  \jt{\apj}  \bvol{111},
  \pg{351}.

\bibitem[{Sagdeev} \& {Galeev}(1969)]{sagdeev69}
{\sc \au{{Sagdeev}, R.~Z.} \& \au{{Galeev}, A.~A.}} \yr{1969} {\em {Nonlinear
  Plasma Theory}\/}.

\bibitem[{Schekochihin} {\em et~al.\/}(2009){Schekochihin}, {Cowley},
  {Dorland}, {Hammett}, {Howes}, {Quataert} \& {Tatsuno}]{schekochihin09}
{\sc \au{{Schekochihin}, A.~A.}, \au{{Cowley}, S.~C.}, \au{{Dorland}, W.},
  \au{{Hammett}, G.~W.}, \au{{Howes}, G.~G.}, \au{{Quataert}, E.} \&
  \au{{Tatsuno}, T.}} \yr{2009}  \at{{Astrophysical Gyrokinetics: Kinetic and
  Fluid Turbulent Cascades in Magnetized Weakly Collisional Plasmas}}.
  \jt{\apjs}  \bvol{182},  \pg{310--377},  \arxiv{arXiv: 0704.0044}.

\bibitem[{Schekochihin} {\em et~al.\/}(2012){Schekochihin}, {Nazarenko} \&
  {Yousef}]{schekochihin12}
{\sc \au{{Schekochihin}, A.~A.}, \au{{Nazarenko}, S.~V.} \& \au{{Yousef},
  T.~A.}} \yr{2012}  \at{{Weak Alfv{\'e}n-wave turbulence revisited}}.
  \jt{\pre}  \bvol{85}~(3),  \pg{036406},  \arxiv{arXiv: 1110.6682}.

\bibitem[{Schekochihin} {\em et~al.\/}(2016){Schekochihin}, {Parker},
  {Highcock}, {Dellar}, {Dorland} \& {Hammett}]{schekochihin16}
{\sc \au{{Schekochihin}, A.~A.}, \au{{Parker}, J.~T.}, \au{{Highcock}, E.~G.},
  \au{{Dellar}, P.~J.}, \au{{Dorland}, W.} \& \au{{Hammett}, G.~W.}} \yr{2016}
  \at{{Phase mixing versus nonlinear advection in drift-kinetic plasma
  turbulence}}.  \jt{Journal of Plasma Physics}  \bvol{82}~(2),
  \pg{905820212},  \arxiv{arXiv: 1508.05988}.

\bibitem[Shebalin {\em et~al.\/}(1983)Shebalin, Matthaeus \&
  Montgomery]{shebalin83}
{\sc \au{Shebalin, J.~V.}, \au{Matthaeus, W.} \& \au{Montgomery, D.}} \yr{1983}
   \at{Anisotropy in {MHD} turbulence due to a mean magnetic field}.
  \jt{Journal of Plasma Physics}  \bvol{29},  \pg{525}.

\bibitem[{Shoda} {\em et~al.\/}(2019){Shoda}, {Suzuki}, {Asgari-Targhi} \&
  {Yokoyama}]{shoda19}
{\sc \au{{Shoda}, M.}, \au{{Suzuki}, T.~K.}, \au{{Asgari-Targhi}, M.} \&
  \au{{Yokoyama}, T.}} \yr{2019}  \at{{Three-dimensional Simulation of the Fast
  Solar Wind Driven by Compressible Magnetohydrodynamic Turbulence}}.  \jt{The
  Astrophysical Journal}  \bvol{880}~(1),  \pg{L2},  \arxiv{arXiv: 1905.11685}.

\bibitem[{Squire} {\em et~al.\/}(2019){Squire}, {Schekochihin}, {Quataert} \&
  {Kunz}]{squire19}
{\sc \au{{Squire}, J.}, \au{{Schekochihin}, A.~A.}, \au{{Quataert}, E.} \&
  \au{{Kunz}, M.~W.}} \yr{2019}  \at{{Magneto-immutable turbulence in weakly
  collisional plasmas}}.  \jt{Journal of Plasma Physics}  \bvol{85}~(1),
  \pg{905850114},  \arxiv{arXiv: 1811.12421}.

\bibitem[{Tenerani} {\em et~al.\/}(2017){Tenerani}, {Velli} \&
  {Hellinger}]{tenerani17}
{\sc \au{{Tenerani}, A.}, \au{{Velli}, M.} \& \au{{Hellinger}, P.}} \yr{2017}
  \at{{The Parametric Instability of Alfv{\'e}n Waves: Effects of Temperature
  Anisotropy}}.  \jt{\apj}  \bvol{851},  \pg{99}.

\bibitem[{Tu} \& {Marsch}(1995)]{tumarsch95}
{\sc \au{{Tu}, C.} \& \au{{Marsch}, E.}} \yr{1995}  \at{{MHD structures, waves
  and turbulence in the solar wind: Observations and theories}}.  \jt{Space
  Science Reviews}  \bvol{73},  \pg{1--210}.

\bibitem[{Usmanov} {\em et~al.\/}(2014){Usmanov}, {Goldstein} \&
  {Matthaeus}]{usmanov14}
{\sc \au{{Usmanov}, A.~V.}, \au{{Goldstein}, M.~L.} \& \au{{Matthaeus}, W.~H.}}
  \yr{2014}  \at{{Three-fluid, Three-dimensional Magnetohydrodynamic Solar Wind
  Model with Eddy Viscosity and Turbulent Resistivity}}.  \jt{\apj}
  \bvol{788},  \pg{43}.

\bibitem[{van Ballegooijen} \& {Asgari-Targhi}(2016)]{vanballegooijen16}
{\sc \au{{van Ballegooijen}, A.~A.} \& \au{{Asgari-Targhi}, M.}} \yr{2016}
  \at{{Heating and Acceleration of the Fast Solar Wind by Alfv{\'e}n Wave
  Turbulence}}.  \jt{\apj}  \bvol{821},  \pg{106},  \arxiv{arXiv: 1602.06883}.

\bibitem[{van Ballegooijen} \& {Asgari-Targhi}(2017)]{vanballegooijen17}
{\sc \au{{van Ballegooijen}, A.~A.} \& \au{{Asgari-Targhi}, M.}} \yr{2017}
  \at{{Direct and Inverse Cascades in the Acceleration Region of the Fast Solar
  Wind}}.  \jt{\apj}  \bvol{835},  \pg{10},  \arxiv{arXiv: 1612.02501}.

\bibitem[{van Ballegooijen} {\em et~al.\/}(2011){van Ballegooijen},
  {Asgari-Targhi}, {Cranmer} \& {DeLuca}]{vanballegooijen11}
{\sc \au{{van Ballegooijen}, A.~A.}, \au{{Asgari-Targhi}, M.}, \au{{Cranmer},
  S.~R.} \& \au{{DeLuca}, E.~E.}} \yr{2011}  \at{{Heating of the Solar
  Chromosphere and Corona by Alfv{\'e}n Wave Turbulence}}.  \jt{\apj}
  \bvol{736},  \pg{3--+},  \arxiv{arXiv: 1105.0402}.

\bibitem[{van der Holst} {\em et~al.\/}(2014){van der Holst}, {Sokolov},
  {Meng}, {Jin}, {Manchester}, {T{\'o}th} \& {Gombosi}]{vanderholst14}
{\sc \au{{van der Holst}, B.}, \au{{Sokolov}, I.~V.}, \au{{Meng}, X.},
  \au{{Jin}, M.}, \au{{Manchester}, IV, W.~B.}, \au{{T{\'o}th}, G.} \&
  \au{{Gombosi}, T.~I.}} \yr{2014}  \at{{Alfv{\'e}n Wave Solar Model (AWSoM):
  Coronal Heating}}.  \jt{\apj}  \bvol{782},  \pg{81},  \arxiv{arXiv:
  1311.4093}.

\bibitem[{Vasquez} \& {Hollweg}(1996)]{vasquez96}
{\sc \au{{Vasquez}, B.~J.} \& \au{{Hollweg}, J.~V.}} \yr{1996}  \at{{Formation
  of arc-shaped Alfv{\'e}n waves and rotational discontinuities from oblique
  linearly polarized wave trains}}.  \jt{\jgr}  \bvol{101},  \pg{13527--13540}.

\bibitem[{Vech} {\em et~al.\/}(2018){Vech}, {Mallet}, {Klein} \&
  {Kasper}]{vech18}
{\sc \au{{Vech}, D.}, \au{{Mallet}, A.}, \au{{Klein}, K.~G.} \& \au{{Kasper},
  J.~C.}} \yr{2018}  \at{{Magnetic Reconnection May Control the Ion-scale
  Spectral Break of Solar Wind Turbulence}}.  \jt{\apjl}  \bvol{855},
  \pg{L27},  \arxiv{arXiv: 1803.00065}.

\bibitem[{Velli}(1993)]{velli93}
{\sc \au{{Velli}, M.}} \yr{1993}  \at{{On the propagation of ideal, linear
  Alfven waves in radially stratified stellar atmospheres and winds}}.
  \jt{\aap}  \bvol{270},  \pg{304--314}.

\bibitem[{Velli} {\em et~al.\/}(1989){Velli}, {Grappin} \& {Mangeney}]{velli89}
{\sc \au{{Velli}, M.}, \au{{Grappin}, R.} \& \au{{Mangeney}, A.}} \yr{1989}
  \at{{Turbulent cascade of incompressible unidirectional Alfven waves in the
  interplanetary medium}}.  \jt{Physical Review Letters}  \bvol{63},
  \pg{1807--1810}.

\bibitem[{Verdini} {\em et~al.\/}(2012){Verdini}, {Grappin}, {Pinto} \&
  {Velli}]{verdini12}
{\sc \au{{Verdini}, A.}, \au{{Grappin}, R.}, \au{{Pinto}, R.} \& \au{{Velli},
  M.}} \yr{2012}  \at{{On the Origin of the 1/f Spectrum in the Solar Wind
  Magnetic Field}}.  \jt{\apjl}  \bvol{750},  \pg{L33},  \arxiv{arXiv:
  1203.6219}.

\bibitem[{Verdini} \& {Velli}(2007)]{verdini07}
{\sc \au{{Verdini}, A.} \& \au{{Velli}, M.}} \yr{2007}  \at{{A}lfv{\'e}n waves
  and turbulence in the solar atmosphere and solar wind}.  \jt{\apj}
  \bvol{662},  \pg{669--676},  \arxiv{arXiv: arXiv:astro-ph/0702205}.

\bibitem[{Verdini} {\em et~al.\/}(2009){Verdini}, {Velli} \&
  {Buchlin}]{verdini09a}
{\sc \au{{Verdini}, A.}, \au{{Velli}, M.} \& \au{{Buchlin}, E.}} \yr{2009}
  \at{{Turbulence in the Sub-Alfv{\'e}nic Solar Wind Driven by Reflection of
  Low-Frequency Alfv{\'e}n Waves}}.  \jt{\apjl}  \bvol{700},  \pg{L39--L42},
  \arxiv{arXiv: 0905.2618}.

\bibitem[{Verdini} {\em et~al.\/}(2010){Verdini}, {Velli}, {Matthaeus},
  {Oughton} \& {Dmitruk}]{verdini10}
{\sc \au{{Verdini}, A.}, \au{{Velli}, M.}, \au{{Matthaeus}, W.~H.},
  \au{{Oughton}, S.} \& \au{{Dmitruk}, P.}} \yr{2010}  \at{{A Turbulence-Driven
  Model for Heating and Acceleration of the Fast Wind in Coronal Holes}}.
  \jt{\apjl}  \bvol{708},  \pg{L116--L120},  \arxiv{arXiv: 0911.5221}.

\bibitem[{Wicks} {\em et~al.\/}(2013a){Wicks}, {Mallet}, {Horbury}, {Chen},
  {Schekochihin} \& {Mitchell}]{wicks13a}
{\sc \au{{Wicks}, R.~T.}, \au{{Mallet}, A.}, \au{{Horbury}, T.~S.}, \au{{Chen},
  C.~H.~K.}, \au{{Schekochihin}, A.~A.} \& \au{{Mitchell}, J.~J.}} \yr{2013a}
  \at{{Alignment and Scaling of Large-Scale Fluctuations in the Solar Wind}}.
  \jt{Physical Review Letters}  \bvol{110}~(2),  \pg{025003},  \arxiv{arXiv:
  1209.5362}.

\bibitem[{Yoon} \& {Fang}(2009)]{yoon09}
{\sc \au{{Yoon}, P.~H.} \& \au{{Fang}, T.-M.}} \yr{2009}  \at{{Proton heating
  by parallel Alfv{\'e}n wave cascade}}.  \jt{Physics of Plasmas}
  \bvol{16}~(6),  \pg{062314}.

\bibitem[{Zank} {\em et~al.\/}(2018){Zank}, {Adhikari}, {Hunana}, {Tiwari},
  {Moore}, {Shiota}, {Bruno} \& {Telloni}]{zank18}
{\sc \au{{Zank}, G.~P.}, \au{{Adhikari}, L.}, \au{{Hunana}, P.}, \au{{Tiwari},
  S.~K.}, \au{{Moore}, R.}, \au{{Shiota}, D.}, \au{{Bruno}, R.} \&
  \au{{Telloni}, D.}} \yr{2018}  \at{{Theory and Transport of Nearly
  Incompressible Magnetohydrodynamic Turbulence. IV. Solar Coronal
  Turbulence}}.  \jt{\apj}  \bvol{854},  \pg{32}.

\bibitem[{Zhou} \& {Matthaeus}(1989)]{zhou89}
{\sc \au{{Zhou}, Y.} \& \au{{Matthaeus}, W.~H.}} \yr{1989}  \at{{Non-WKB
  evolution of solar wind fluctuations: A turbulence modeling approach}}.
  \jt{\grl}  \bvol{16},  \pg{755--758}.

\bibitem[{Zhou} \& {Matthaeus}(1990)]{zhou90}
{\sc \au{{Zhou}, Y.} \& \au{{Matthaeus}, W.~H.}} \yr{1990}  \at{{Transport and
  turbulence modeling of solar wind fluctuations}}.  \jt{\jgr}  \bvol{95},
  \pg{10291--10311}.

\end{thebibliography}
\bibliographystyle{jpp}

\end{document}